\documentclass[APA,STIX1COL]{WileyNJD-v2}

\articletype{Article Type}%

\received{9 March 2026}
\revised{9 March 2026}
\accepted{9 March 2026}

\begin{document}

\title{Uncertainty quantification for critical energy systems during compound extremes via BMW-GAM}

\author[1]{Mitchell L. Krock}

\author[2]{W.\ Neal Mann}

\author[2]{Zhi Zhou}

\authormark{KROCK \textsc{et al}}

\address[1]{\orgdiv{Mathematics and Computer Science}, \orgname{Argonne National Laboratory}, \orgaddress{\state{Illinois}, \country{USA}}}
\address[2]{\orgdiv{Energy Systems and Infrastructure Assessment}, \orgname{Argonne National Laboratory}, \orgaddress{\state{Illinois}, \country{USA}}}

\corres{\email{mk52n@missouri.edu}}

\abstract[Summary]{
    Extreme weather poses a large risk to critical energy systems \citep{Ekisheva2021,levin2022}.
    Uncertainty quantification of negative impacts is important for developing resilience, especially during compound extreme weather events involving multiple climate variables. 
    We leverage BMW-GAM \citep{economou2022}, a copula workflow that relies on fitting marginal distributions with Bayesian generalized additive models in moving windows---an embarrassingly parallel task. 
    The Gaussian copula has separable multivariate space-time correlation, allowing for efficient emulation and likelihood fitting with big datasets.
    Overall, the formulation is interpretable and provides uncertainty quantification through probabilistic simulations of weather variables during extreme events. 
    Our method is illustrated in an analysis of temperature, wind speed, and global horizontal irradiance from Argonne National Laboratory's high-fidelity climate model output ADDA.
}

\maketitle

\section{Introduction}
\label{sec:introduction}

Natural gas networks, transmission networks, and electricity supply and demand are more closely coupled than ever.
All are subject to threats from extreme weather, particularly when the weather event involves multiple climate variables.
The industry sees this as the greatest risk to system reliability, but understudied.
\citet{SILLMANN201765} discuss several challenges of modeling weather extremes; in this work we analyze short-duration extremes of climate model output in Northeast U.S.\ 
Our target is to create a workflow that investigates the simultaneous risks of extreme weather to the interdependent electricity and natural gas network systems.
The contribution of this paper is constructing a probabilistic model for uncertainty quantification during compound weather extremes.
Simulations from our model can be used as synthetic input to the electricity and gas networks, once the connection is established, to conduct a weather-impacted grid reliability assessment \citep{lee2024,adhikari2025}.
Linking gas and electricity networks is the subject of previous research \citep{li2008,Gillessen2019,Shahidehpour2005,Martinez2012,clegg2017} but remains a challenge.
Researchers from Argonne National Laboratory are developing a soft-coupled workflow between electricity dispatch model A-LEAF \citep{aleaf} and natural gas network operation model NGTransient \citep{ngtransient} to find simultaneous risks to these interdependent systems.
Ultimately, one could consider extremes not just in the weather, but also in grid and gas networks, such as common mode or cascade events.

A simple approach to studying extremes with climate model output is empirically counting occurrences of the extreme event of interest.
However, this assumes the number of climate simulations is enough to adequately quantify uncertainty. 
One solution, if possible, is using more high-performance computing resources to run additional large climate ensembles; this is not always viable due to limited computational budget.
Instead, a computationally efficient statistical model can emulate climate model output outside the range of the original data, providing uncertainty quantification for impacts from rare events.

\section{Bayesian Moving Window Generalized Additive Model (BMW-GAM)}
\label{sec:bmwgam}

Our probabilistic model follows the BMW-GAM framework proposed by \citet{economou2022}.
It begins with windowed Bayesian inference for generalized additive models, then a Gaussian copula endows multivariate spatio-temporal dependence.
GAMs are a flexible, interpretable nonparametric regression technique with state-of-the-art performance in energy load forecasting competitions \citep{GAILLARD20161038}.
\citet{han2009} use GAMs to study power grid outage caused by extreme weather.
Windowed estimation has a rich history in geostatistics \citep{wiens2021,ver2004flexible,risser2015local,haas1990kriging,haas1990lognormal}.
Similar ideas of encoding local estimates to spatial process models are found in \citet{nychka2018} and \citet{wiens2020}.

\subsection{Marginal Distribution}

Consider a single weather variable $Y_{s,t} \in \mathbb{R}$ at location $s \in \mathbb{R}^2$ and time $t \in \mathbb{R}$. 
BMW-GAM writes
\begin{align}
    \label{eq:gam}
\begin{split}
    Y_{s,t} &\sim p(y_{s,t}; \mu_{s,t}, \phi) \\
    g(\mu_{s,t}) &= f(s,t)
\end{split}
\end{align}
where $p(\cdot)$ is a probability distribution with mean $\mu_{s,t}$, parameters independent of $(s,t)$ are represented by $\phi$ (e.g.,\ $\phi = \sigma^2>0$, the distribution's scale parameter), $g(\cdot)$ is a link function, and $f(s,t)$ is a regression spline.
In 1D, $Y_{i} \sim p(y_{i}; \mu(x_{i}), \phi)$ and $g(\mu(x_{i})) = f(x_i) = \sum_{j=1}^J \beta_j b_j(x_i)$ and $\boldsymbol \beta = (\beta_1,\dots,\beta_J)^{\mathrm{T}}$ contains coefficients of a cubic spline basis.
In higher dimensions, tensor-product splines are a natural option for $f$, but come at heavy computational cost.
We instead opt for a thin-plane spline \citep{wahba1990}, balancing computational efficiency with an assumption of isotropy.
By using localized fitting, this assumption needs only to be true for a small window and can be checked for violations.
Parameters are estimated by minimizing
\begin{equation}
    \label{eq:penalizedlikelihood}
-\ell(\boldsymbol \beta, \phi ; \mathbf{y}, \mathbf{X}) + \lambda \boldsymbol \beta^{\mathrm{T}} \mathbf{S}  \boldsymbol \beta,
\end{equation}
where $\ell(\boldsymbol \beta, \phi ; \mathbf{y}, \mathbf{X})$ denotes the log-likelihood
for data $(\mathbf{X},\mathbf{y})$ in a nearest-neighbor moving window centered at $y_{s,t}$, 
and $\mathbf{S}$ is a penalty matrix (constructed before fitting) with regularization parameter $\lambda>0$ selected through generalized cross-validation.
The R package \texttt{mgcv} \citep{GAMbook} automates these steps.
Hyperparameters are the window size, in both space and time, as well as the number of knots in the spline function.
Following recommendation from \cite{economou2022}, we employ 50 knots and $30$ nearest neighbors in space.
The temporal window size of $9$ nearest neighbors (including the central timepoint) corresponds to 24 hours\footnote{Data are at a three-hour frequency, so 8 readings per day.}.
To simulate $Y_{s,t}$, we use Bayesian Monte Carlo sampling.
The posterior predictive distribution of the response $\tilde{y}$ given data $\mathbf{y}$ is
\begin{equation}
    \label{eq:posterior}
    p(\tilde{y} \mid \mathbf{y}) = \int_{\boldsymbol \beta} p(\tilde{y} \mid \boldsymbol \beta, \hat \lambda, \hat \phi) p(\boldsymbol \beta \mid \mathbf{y}) \; d \boldsymbol \beta,
\end{equation}
where point estimates are substituted for $\lambda$ and $\phi$ rather than performing the full Bayesian hierarchy.
Assume the prior $\boldsymbol \beta \sim N(\mathbf{0},  \sigma^2 \mathbf{S}^\dagger/\lambda)$ with $\mathbf{S}^\dagger$ the pseudo-inverse of $\mathbf{S}$.
If the response variable is Normal\footnote{If not, large sample approximations result in a multivariate Normal posterior.}, the posterior is $\boldsymbol\beta \mid \mathbf{y} = N(\hat{\boldsymbol \beta}, (\mathbf{X}^{\mathrm{T}} \mathbf{X} + \lambda \mathbf{S})^{-1} \sigma^2)$; the posterior mean $\hat{\boldsymbol{\beta}}$ minimizes \eqref{eq:penalizedlikelihood}, and the posterior variance matrix is easily extracted from \texttt{mgcv}.
Marginal inference can be accelerated with parallel computations across windows.

\subsection{Copula}

Simulations under the posterior \eqref{eq:posterior} are independently collected at each central point in the moving window, separately for the weather variables.
To account for dependence, we leverage the copula \citep{Nels06,Joe14}, in particular, the Gaussian copula.
Recent developments in Gaussian copula literature include sensitivity analysis to unobserved confounding \citep{balgi2025sensitivityanalysisunobservedconfounding}, and estimation of correlation parameters under differential privacy \citep{wang2026differentiallyprivatebayesianinference}.
We do not analyze precipitation, but related ideas in stochastic weather generators \citep{verdin2015, VERDIN2018835, verdin2019, LI2021100474} may be of independent interest.
\cite{bracken2016} embed the Gaussian copula in a Bayesian hierarchical model to study precipitation extremes over a large spatial domain.
Gaussian copula is also used for wind and PV scenario generation \citep{dumas2021,pinson2009,GOLESTANEH201680},
modeling extreme cold and weak-wind events over Europe \citep{Tedesco2023}, sparse multi-task learning \citep{goncalves2016}, and importance sampling of rare events \citep{rossman2025}.
Copula are popular in forecasting and time series analysis \citep{bessa2011,bessa2012,copulamarkovbook}.
\cite{ZHANG2025125369} use Gaussian copula to forecast load, wind, and solar power.
\cite{moradi2026copulabasedaggregationcontextawareconformal} combine Gaussian copula with conformal prediction to forecast solar power in three key U.S.\ energy markets.
\cite{yang2025adaptiveensemblelearninggaussian} perform expectation-maximization with Gaussian copula to handle missing data, then predictions from multiple machine learning algorithms are combined via adaptive ensemble for load forecasting.
\cite{pokou2026predictiveaccuracyversusinterpretability} study whether structural econometric models based on copula rival machine learning in energy market forecasting.

Let $\mathbf{Y}_{i} \in \mathbb{R}^{NT}$ be the vector corresponding to the $i$th weather variable obtained from concatenating columns of the $N \times T$ data matrix whose rows and columns are indexed by space and time, respectively, for $i=1,\dots,P$.
Our model for the random vector $\mathbf{Y} = (\mathbf{Y}_{1}^{\mathrm{T}},\dots,\mathbf{Y}_{P}^{\mathrm{T}})^{\mathrm{T}} \in \mathbb{R}^{PNT}$ is $f^{-1}(\Phi(\mathbf{Z}))$, where $\mathbf{Z} \sim N(0,\Sigma)$ is mean-zero multivariate Normal with correlation matrix $\Sigma$, $\Phi(\cdot)$ is the standard Normal cdf, and $f^{-1}(\cdot)$ is the empirical quantile function obtained from simulations of the posterior \eqref{eq:posterior}.
We follow inference-for-marginals to separate the marginal and dependence steps in copula modeling.
This approach is asymptotically efficient and often viable when the full maximum likelihood is not \citep{joe2005}.
The copula correlation matrix $\Sigma$ is estimated by minimizing the Gaussian negative log-likelihood
\begin{equation}
    \label{eq:gaussiannegloglikelihood}
-2 L (\Sigma) =  \log \det \Sigma + \mathbf{z}^{\mathrm{T}} \Sigma^{-1} \mathbf{z} 
\end{equation}
over $\Sigma$ positive semidefinite.
We assume a separable multivariate-space-time correlation function \citep{genton2007}, yielding the Kronecker product representation $\Sigma = \Sigma_p \otimes \Sigma_t \otimes \Sigma_n$, equivalent to a Tensor-Normal distribution on $\mathbf{Z}$ \citep{MANCEUR201337}, for which maximum likelihood computations are efficient.
Another advantage of the Kronecker representation is that $\Sigma$ never has to be stored in memory, only $\Sigma_p$ and $\Sigma_t$ and $\Sigma_n$.
Matérn correlation functions \citep{stein1999} parameterize $\Sigma_n$ and $\Sigma_t$, depending on the distance between locations (in space or time):
\begin{align}
    \label{eq:Matérn}
\begin{split}
    C(d_{\text{space}}) &= \text{Matérn}_{\nu}(d_{\text{space}}; V_{\text{space}}),  \\
    C(d_{\text{time}}) &= \text{Matérn}_{\nu}(d_{\text{time}}; V_{\text{time}}).
\end{split}
\end{align}
The Matérn smoothness parameter is fixed at $\nu = 5/2$, a typical choice to produce smooth spatial fields; in this case, the Bessel function in the Matérn correlation function has a closed form.
$\quad V_{\text{space}} > 0$ and $\quad V_{\text{time}} > 0$ are range parameters controlling the strength of spatial and temporal correlations.
Stepping away from restrictive assumptions of isotropy and separability is desirable but the subject of cutting-edge research in spatio-temporal statistics \citep{porcu2018,alegria2019,porcu2021}.
Finally, $\Sigma_p$ is an arbitrary $3 \times 3$ correlation matrix with $3$ unknown parameters in the interval $(-1,1)$.
Five parameters are estimated via maximum likelihood using a standard L-BFGS optimizer.

Simulating $\mathbf{Y}$ is efficient due to the Kronecker product structure of $\Sigma$.
Let $L_p, L_t,$ and $L_n$ denote the lower Cholesky triangle of $\Sigma_p, \Sigma_t$, and $\Sigma_n$, respectively.
The Cholesky decomposition of $\Sigma$ equals $(L_p \otimes L_t \otimes L_n) (L_p \otimes L_t \otimes L_n)^{\mathrm{T}}$, so we can simulate $\mathbf{Z}$ by multiplying $(L_p \otimes L_t \otimes L_n)$ into a Gaussian noise vector $\boldsymbol{\epsilon}$. 
Such multiplication is done with repeated applications of the formula $(B^{\mathrm{T}} \otimes A) \boldsymbol{\epsilon} = \text{vec}(A \text{mat}(\boldsymbol{\epsilon}) B)$, where $\text{mat}(\boldsymbol{\epsilon})$ reshapes $\boldsymbol{\epsilon}$ into a matrix of appropriate size and $\text{vec}(\cdot)$ is the columnwise vectorization of a matrix.
Transforming from $\mathbf{Z}$ to $\mathbf{Y}$ follows immediately from componentwise applications of $\Phi(\cdot)$ then $f^{-1}(\cdot)$.

\section{ADDA Analysis}

Argonne Downscaled Data Archive (ADDA), hosted at ClimRR hub \citep{climrr}, leverages the regional weather model WRF to dynamically downscale output from three global climate models (HadGEM, GFDL, CCSM) under various RCP emission scenarios \citep{zobel2018,wang2015,zobel2017}.
Our analysis considers HadGEM data under RCP 8.5, without loss of generality; see Section \ref{sec:discussion} for more discussion on this matter.
We study the northeastern United States, targeting a five-day synthetic weather event, a winter storm with exceptionally low temperature.
One could target multivariate extreme events with more sophisticated methods \citep{xu2025,sharma2025variationalautoencodersbaseddetectionextremes,coulaud2026traknnefficienttrajectoryaware}.
For our weather variables, data are observed at $N = 2841$ locations and $T=33$ time points.
The following selections of \eqref{eq:gam} are made:
\begin{align}
    \label{eq:gamchoice}
\begin{split}
    \texttt{temperature: } Y_{s,t} \sim \text{Normal}(\mu_{s,t},\sigma^2) \texttt{ and } g(x) = x \\
    \texttt{wind speed: } Y_{s,t} \sim \text{Gamma}(\mu_{s,t},\phi) \texttt{ and } g(x) = \sqrt{x} \\
    \texttt{(nonzero) global horizontal irradiance: } Y_{s,t} \sim \text{Gamma}(\mu_{s,t},\phi) \texttt{ and } g(x) = \sqrt{x} \\
\end{split}
\end{align}
\citet{economou2022} model temperature in identical fashion.
Wind speed requires a positively-supported probability distribution $p(\cdot)$ and a link function $g(\cdot)$ whose input space is the domain $(0,\infty)$.
Gamma distribution is a practical choice for wind speed modeling \citep{ARIES201878}.
Global horizontal irradiance is most difficult due to the distribution's point mass at zero.
Our concession is to only model the nonzeros of the irradiance; discussion about this concludes our paper.
For simplicity, we match the setup of wind speed and global horizontal irradiance.
Selecting square-root rather than log link function is crucial to tamper unrealistic values of global horizontal irradiance, and, to a lesser extent, wind speed.

\begin{figure}[h!]
    \centering
    \includegraphics[scale=.28]{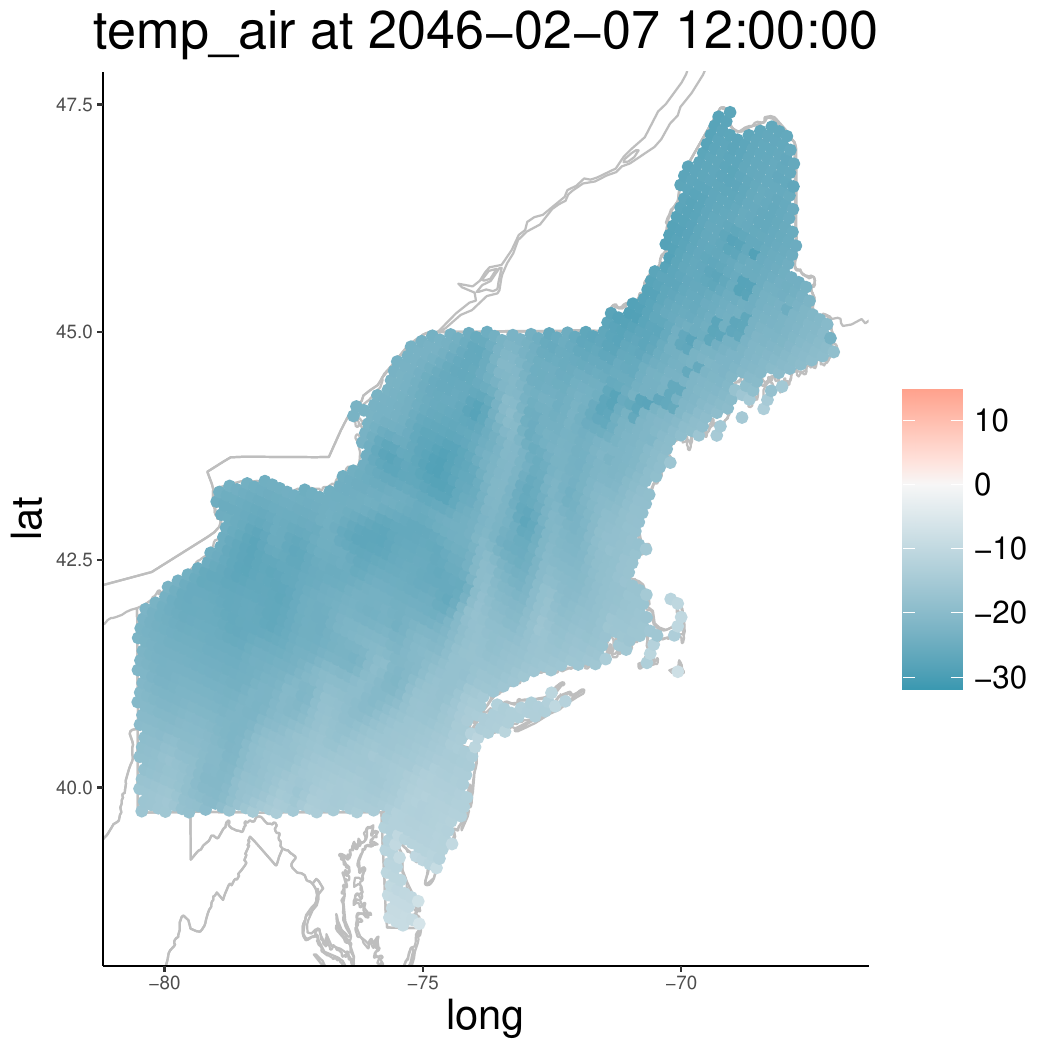}
    \includegraphics[scale=.28]{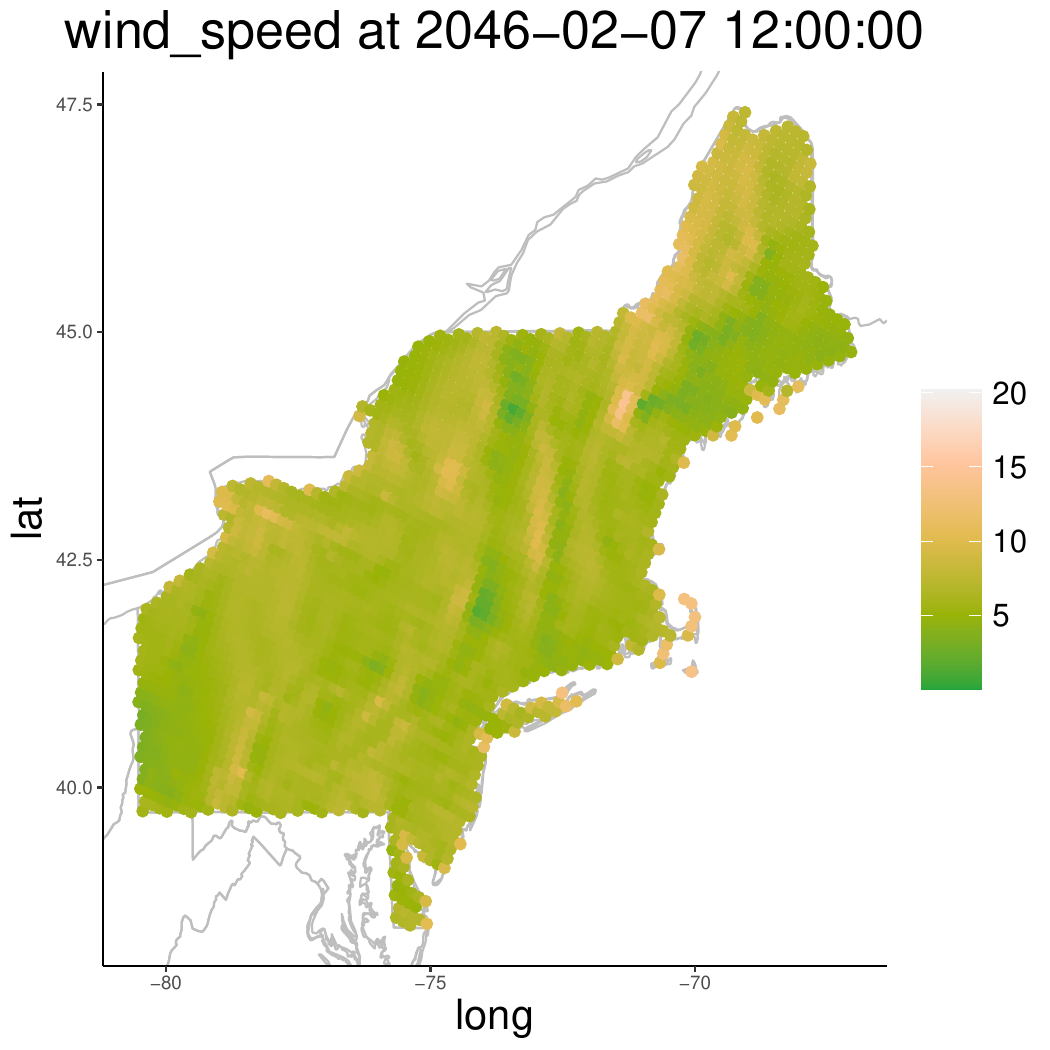}
    \includegraphics[scale=.28]{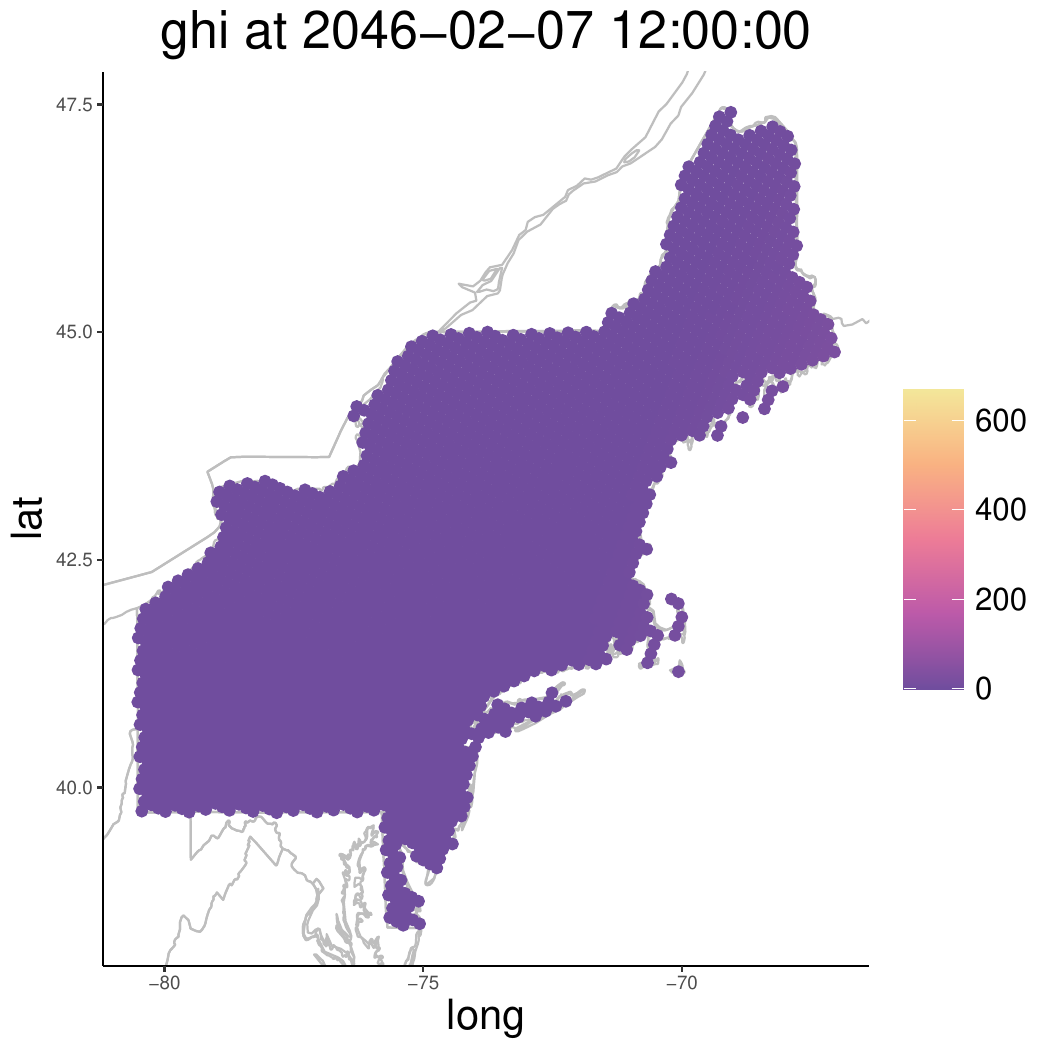}
    
    \includegraphics[scale=.28]{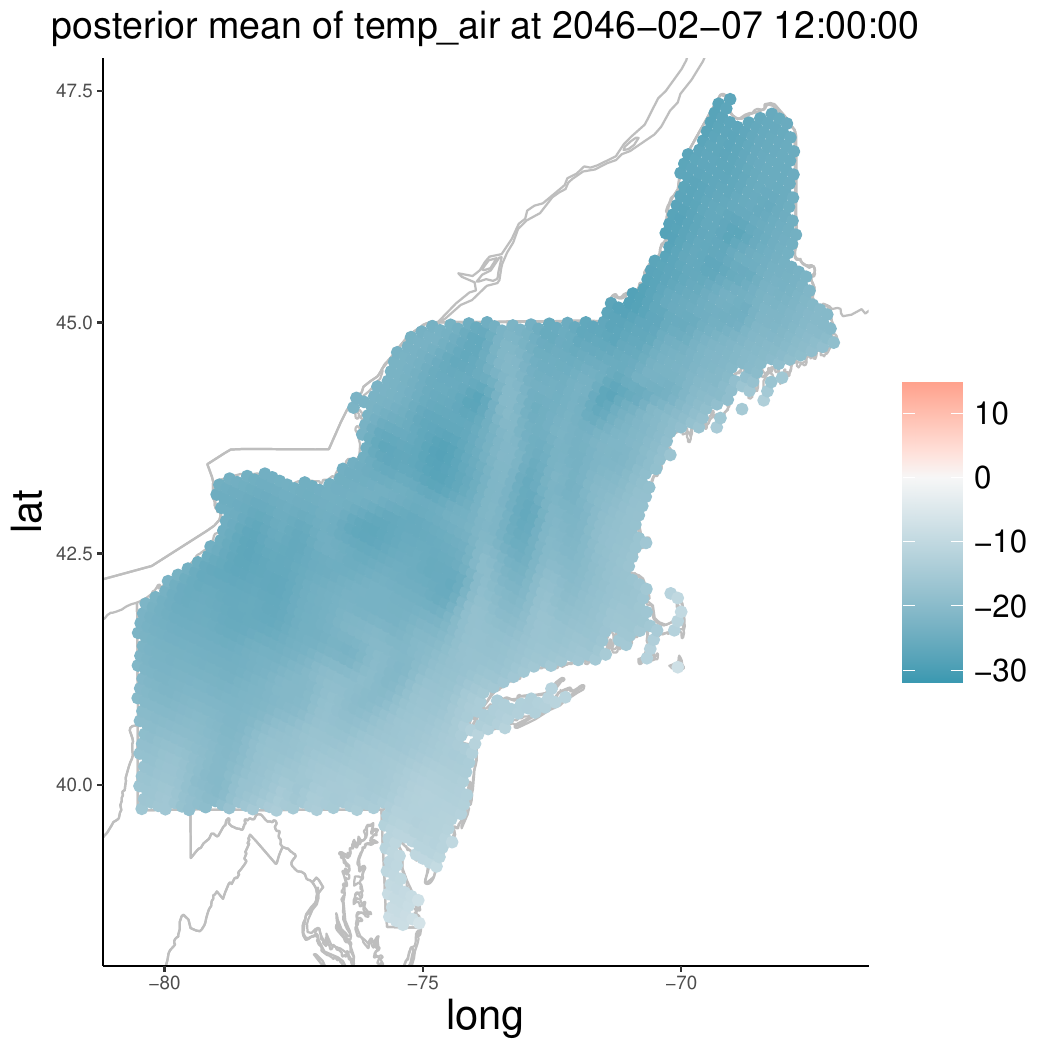}
    \includegraphics[scale=.28]{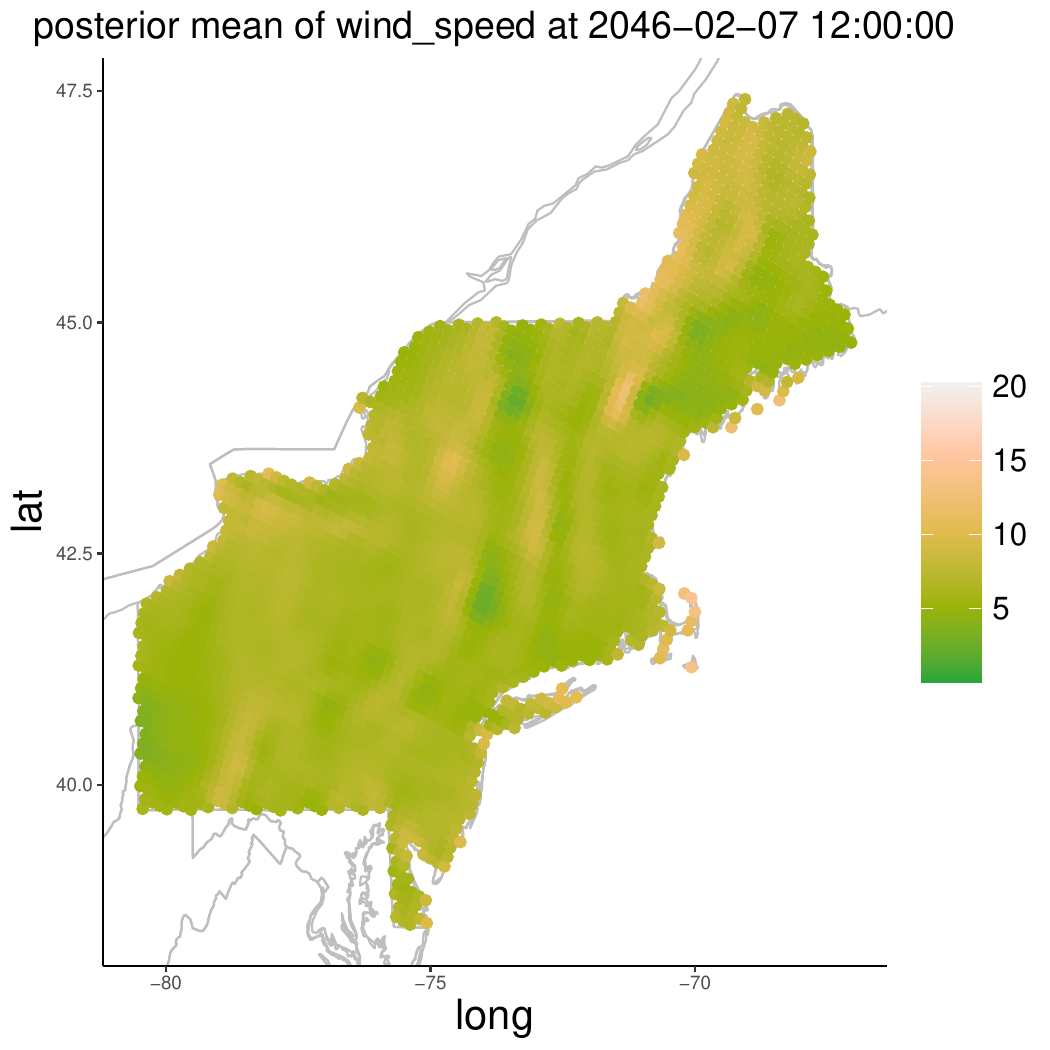}
    \includegraphics[scale=.28]{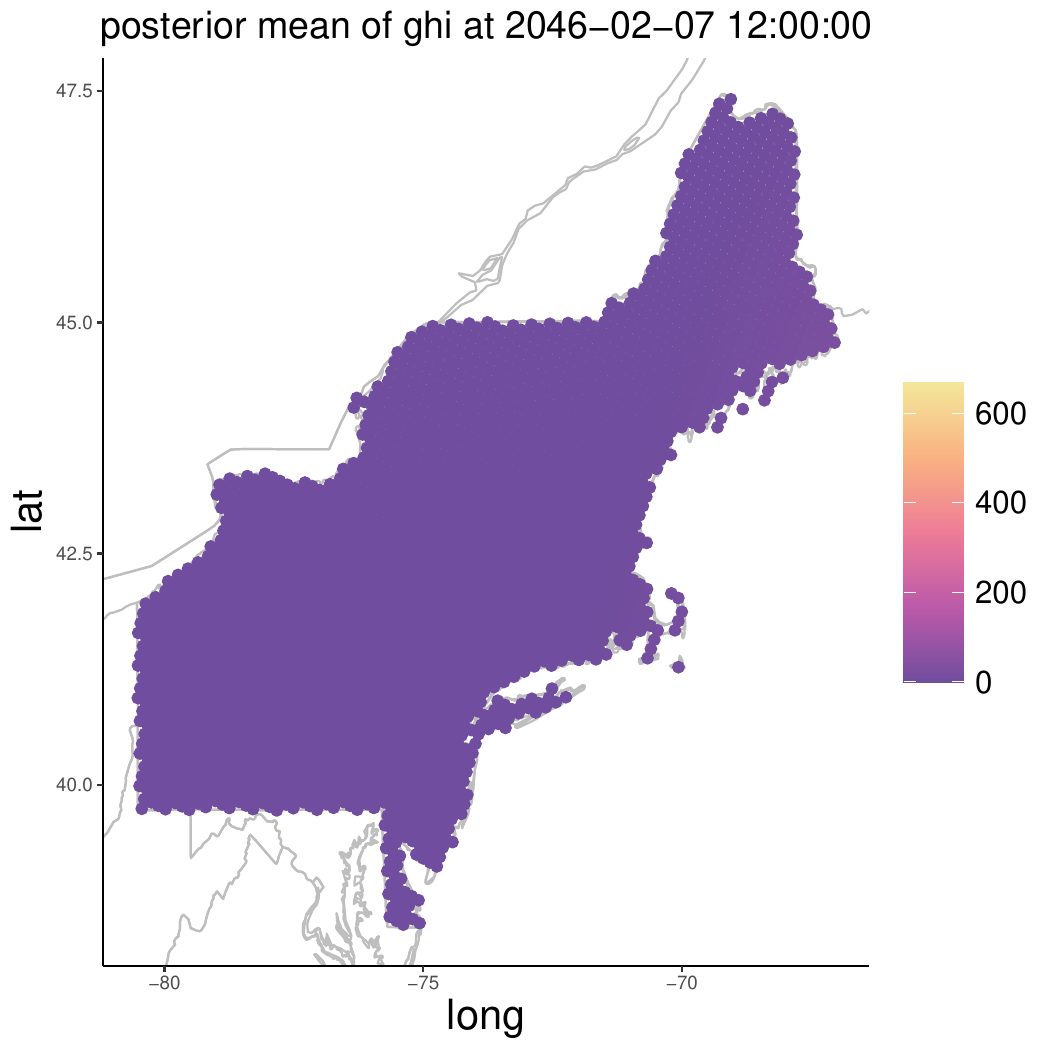}

    \includegraphics[scale=.28]{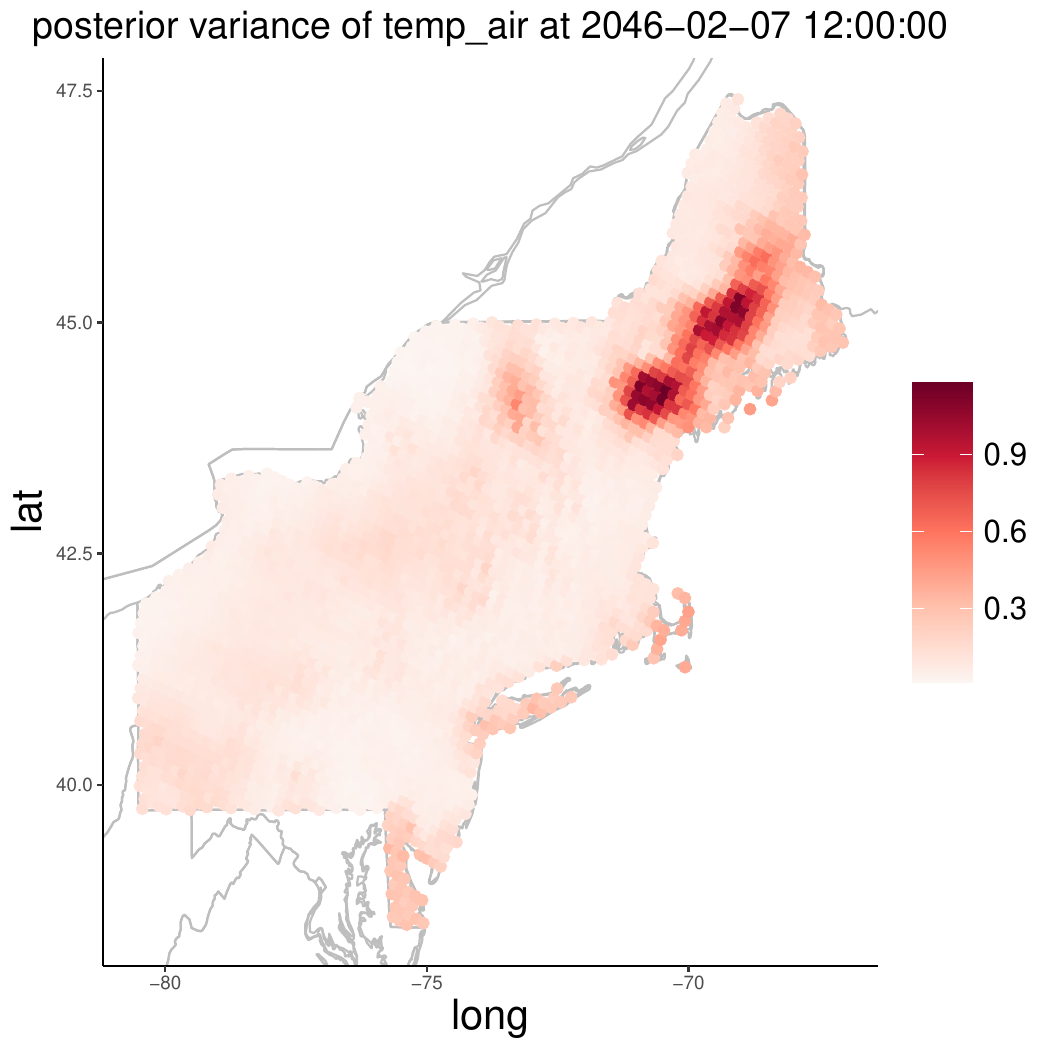}
    \includegraphics[scale=.28]{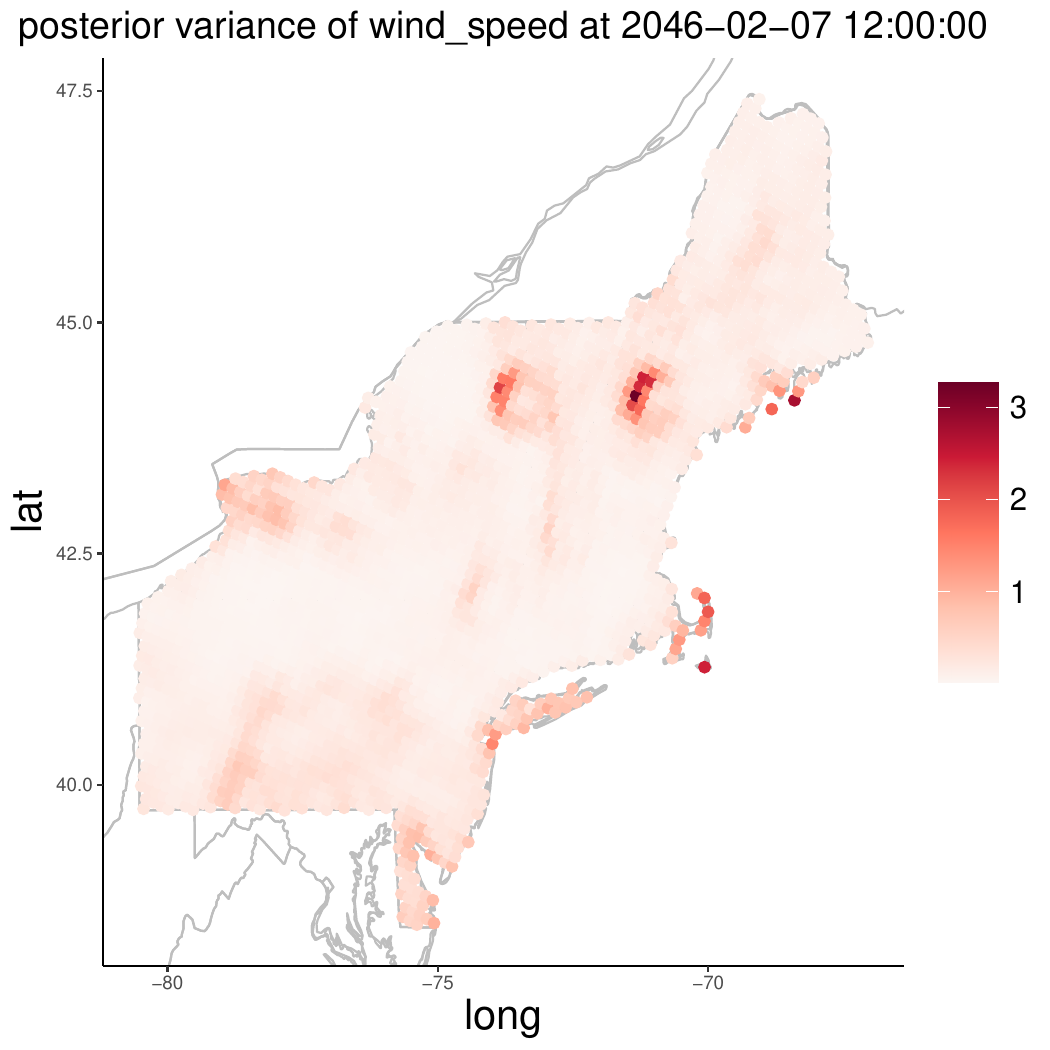}
    \includegraphics[scale=.28]{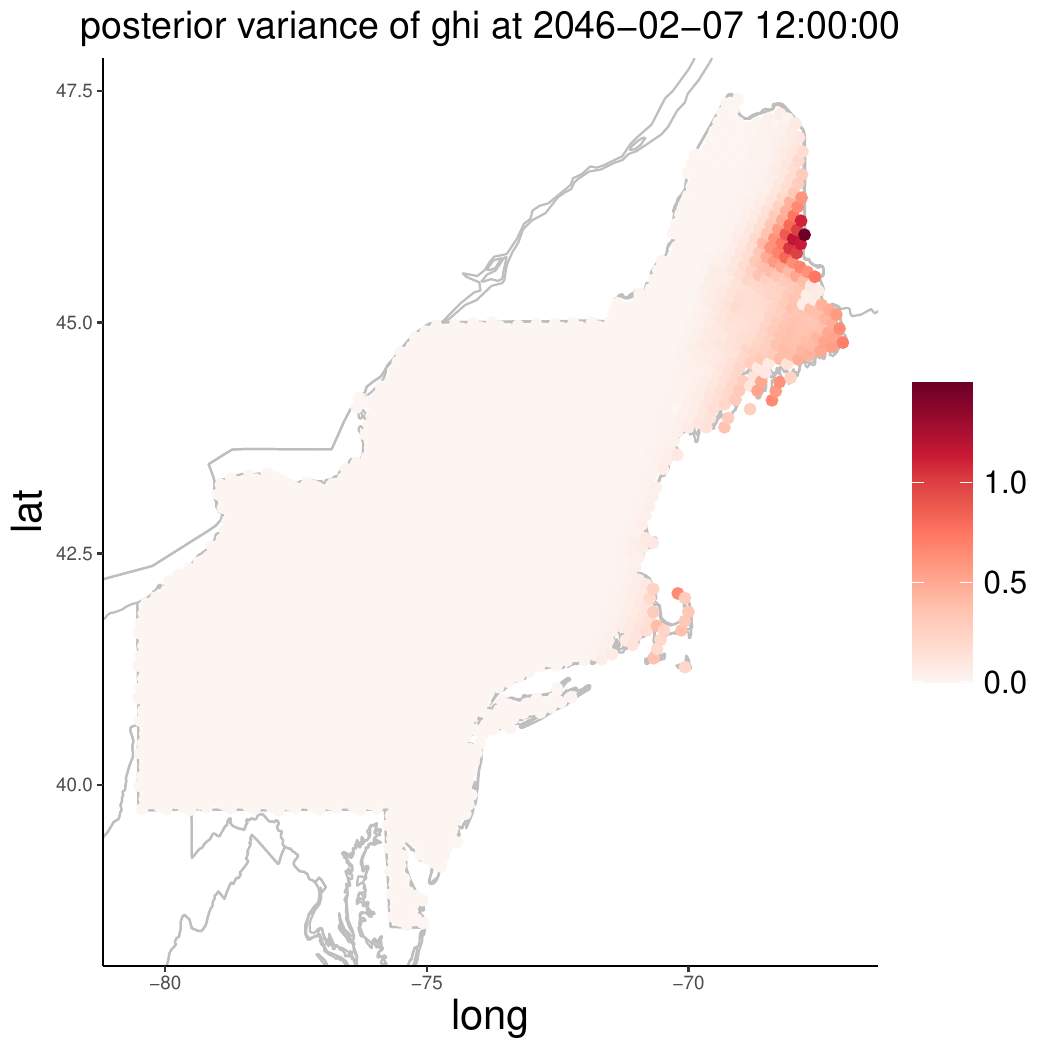}

    \caption{First row: historical data. Second row: posterior mean of BMW-GAM simulations. Third row: posterior variance of BMW-GAM simulations.}
    \label{fig:date1plot}
\end{figure}

\begin{figure}[h!]
    \centering
    \includegraphics[scale=.28]{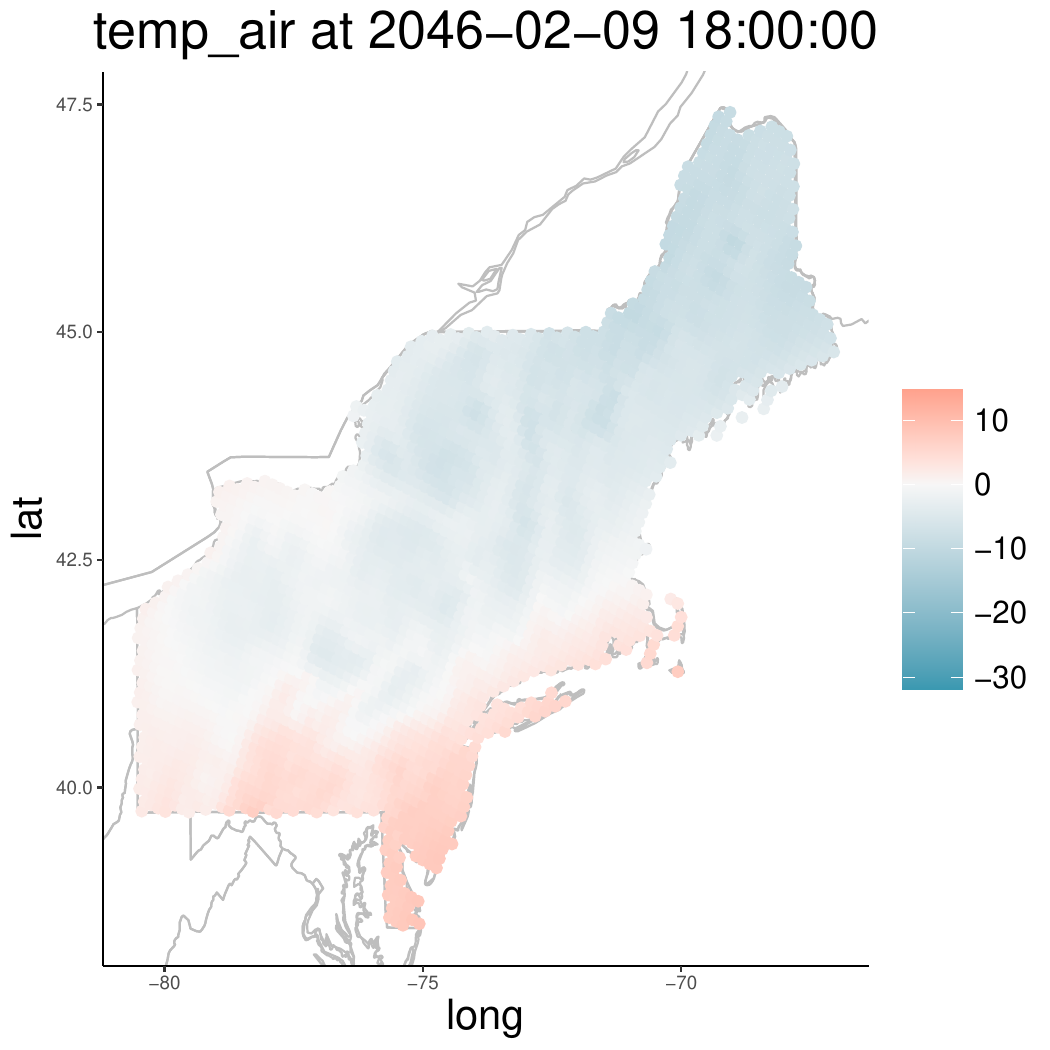}
    \includegraphics[scale=.28]{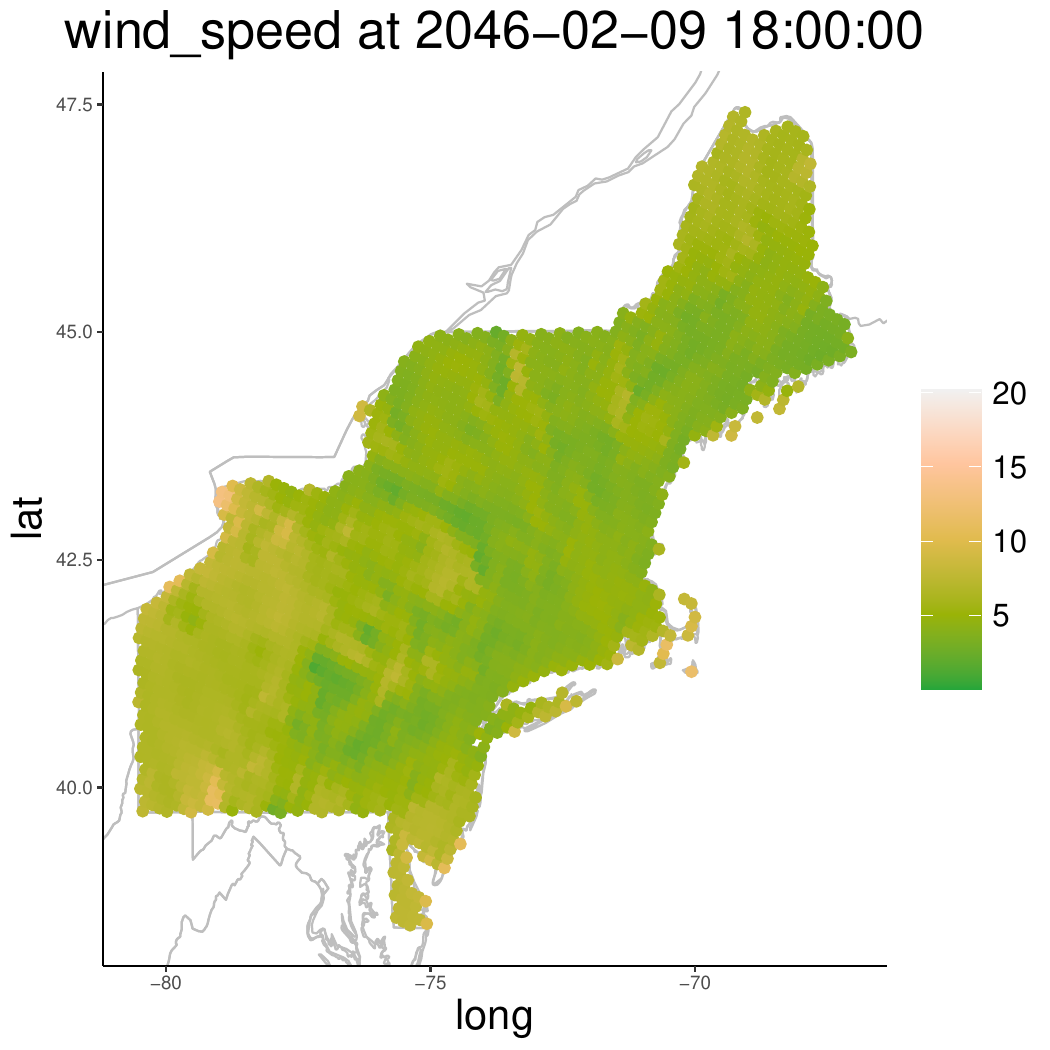}
    \includegraphics[scale=.28]{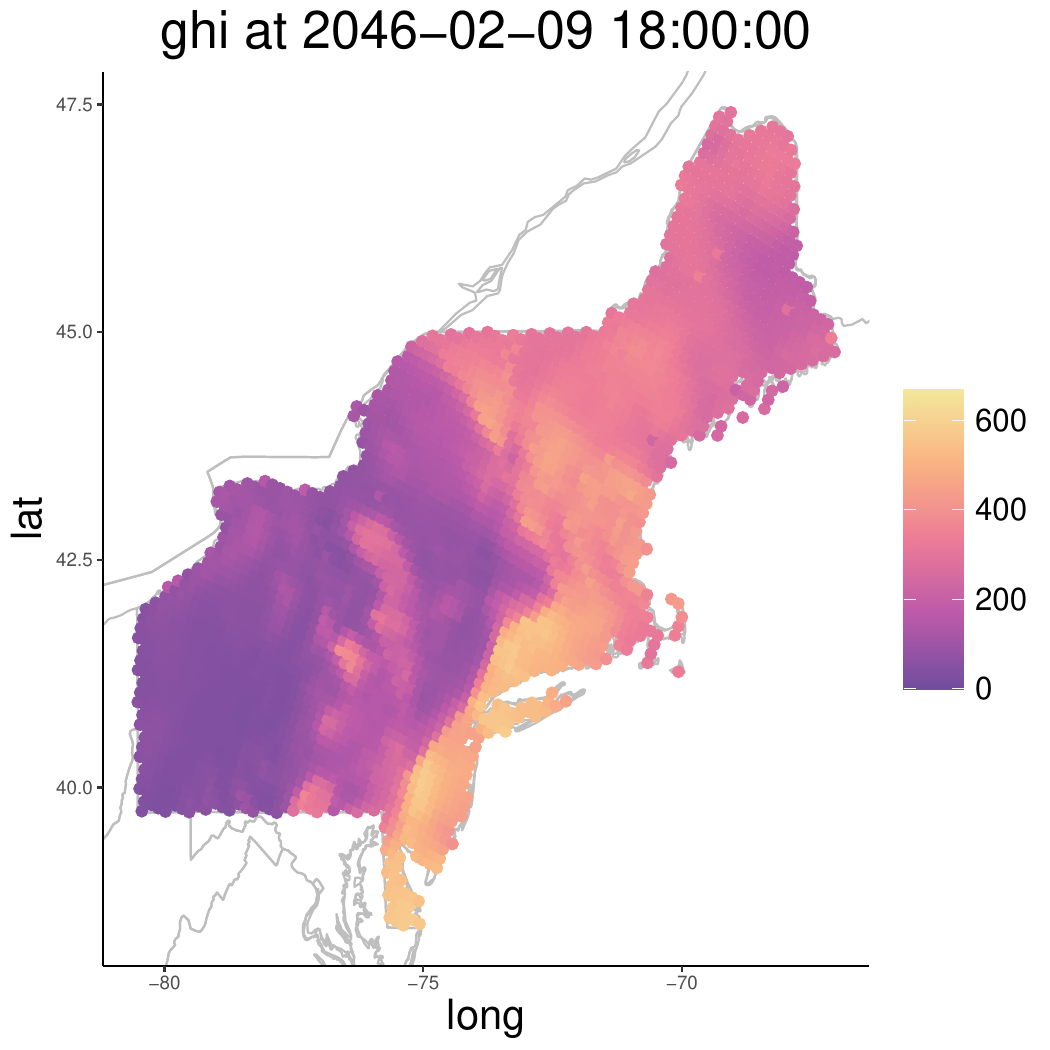}
    
    \includegraphics[scale=.28]{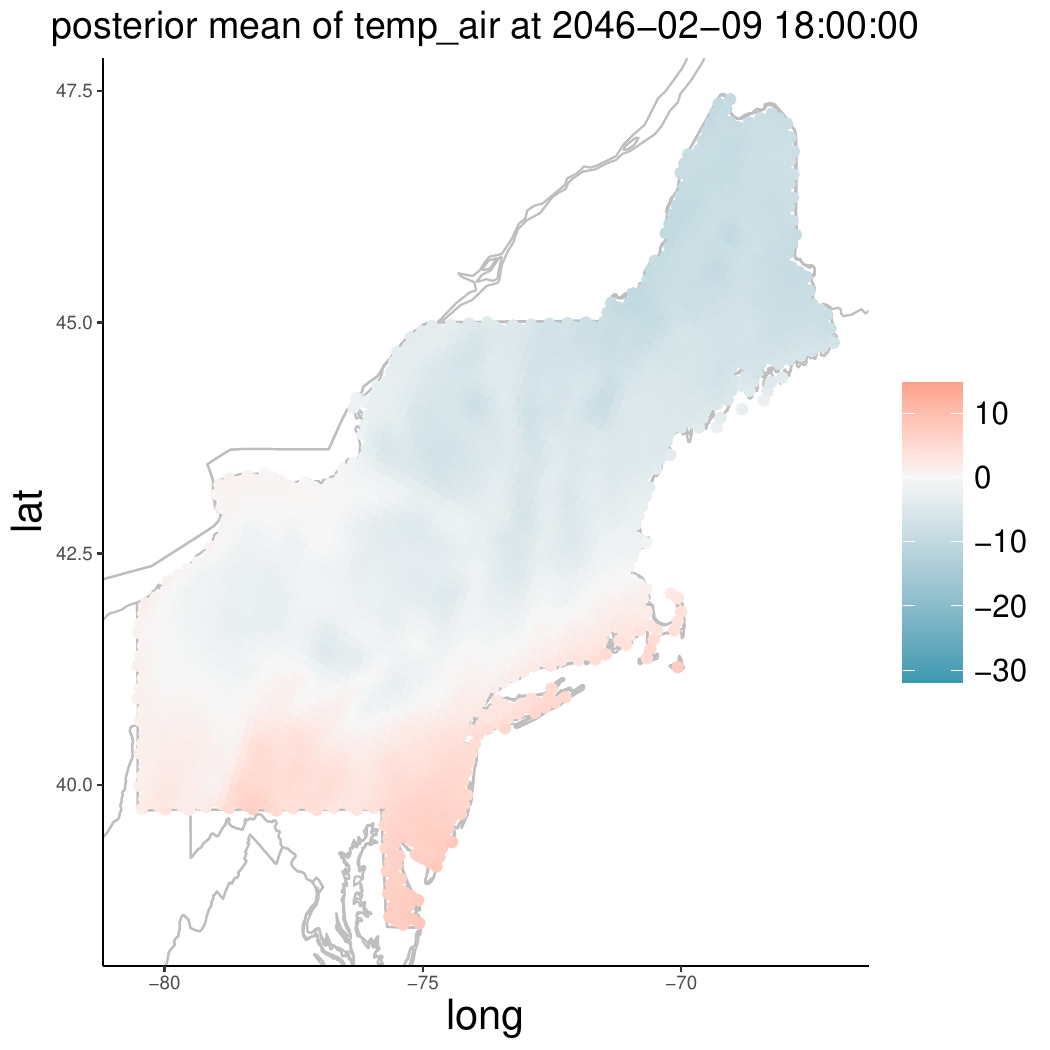}
    \includegraphics[scale=.28]{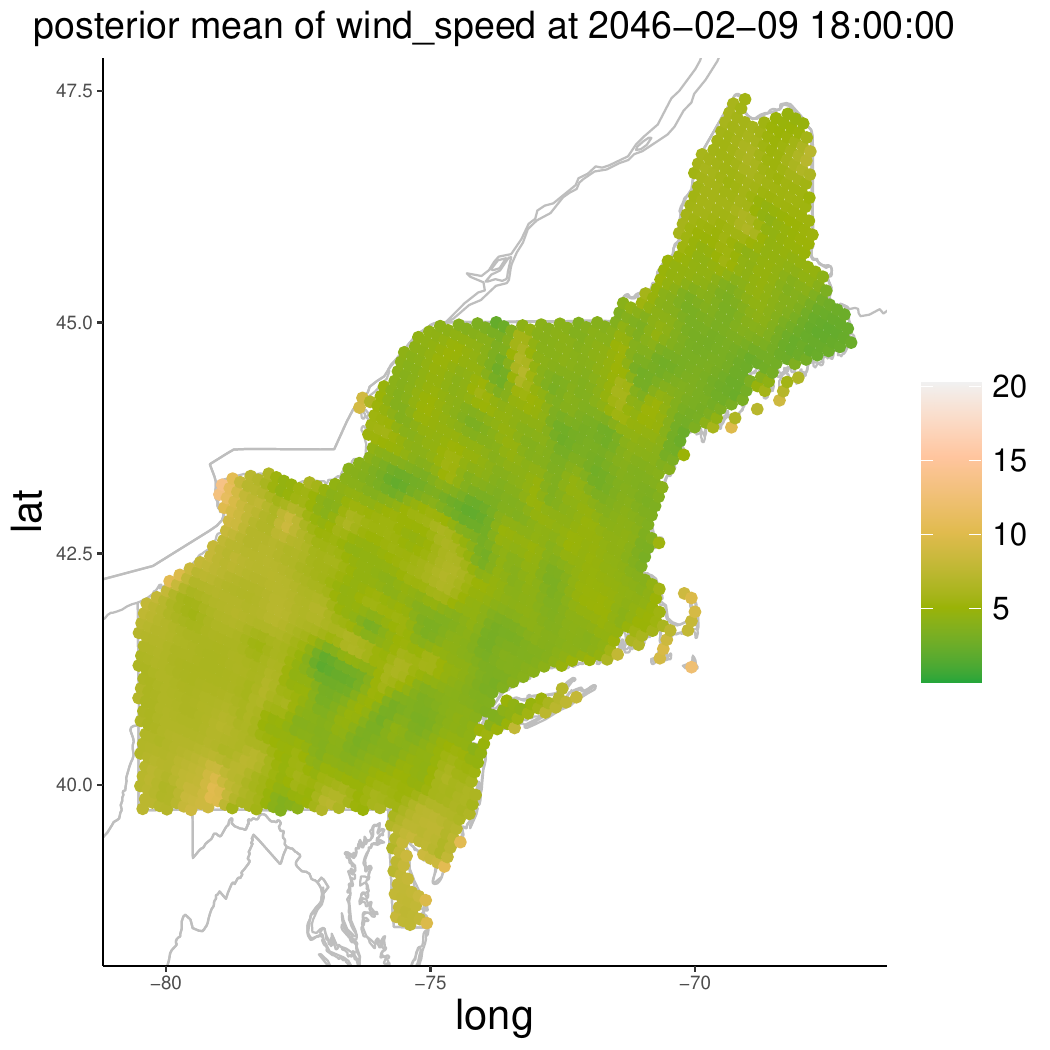}
    \includegraphics[scale=.28]{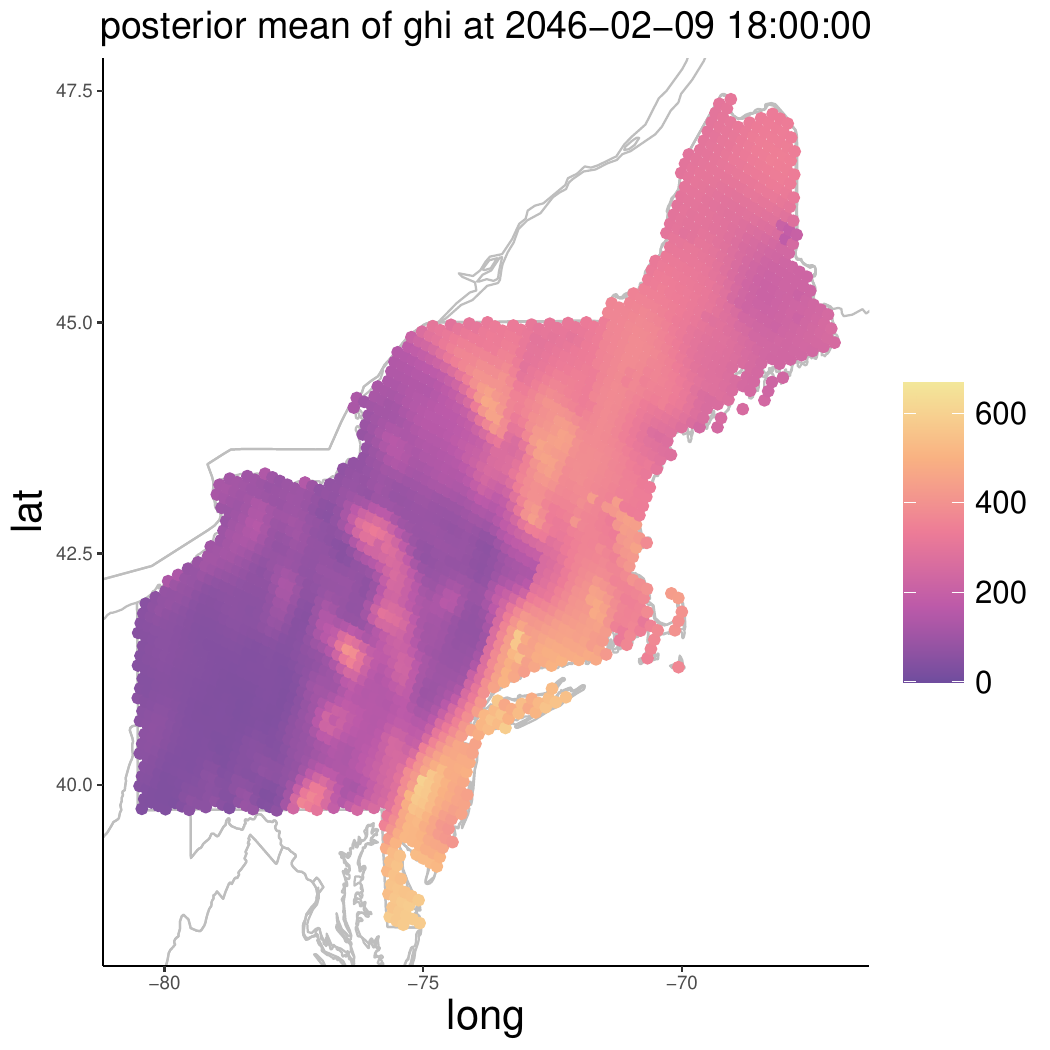}

    \includegraphics[scale=.28]{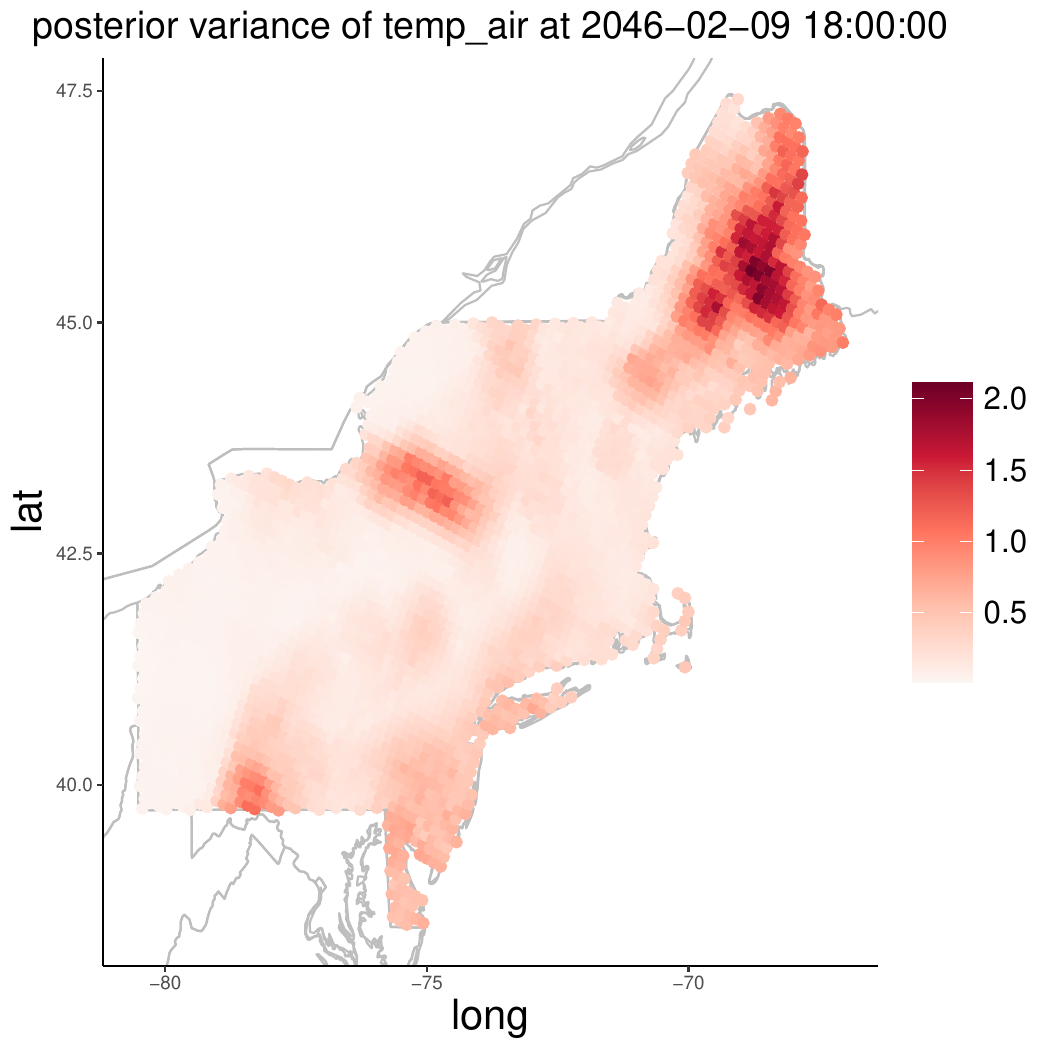}
    \includegraphics[scale=.28]{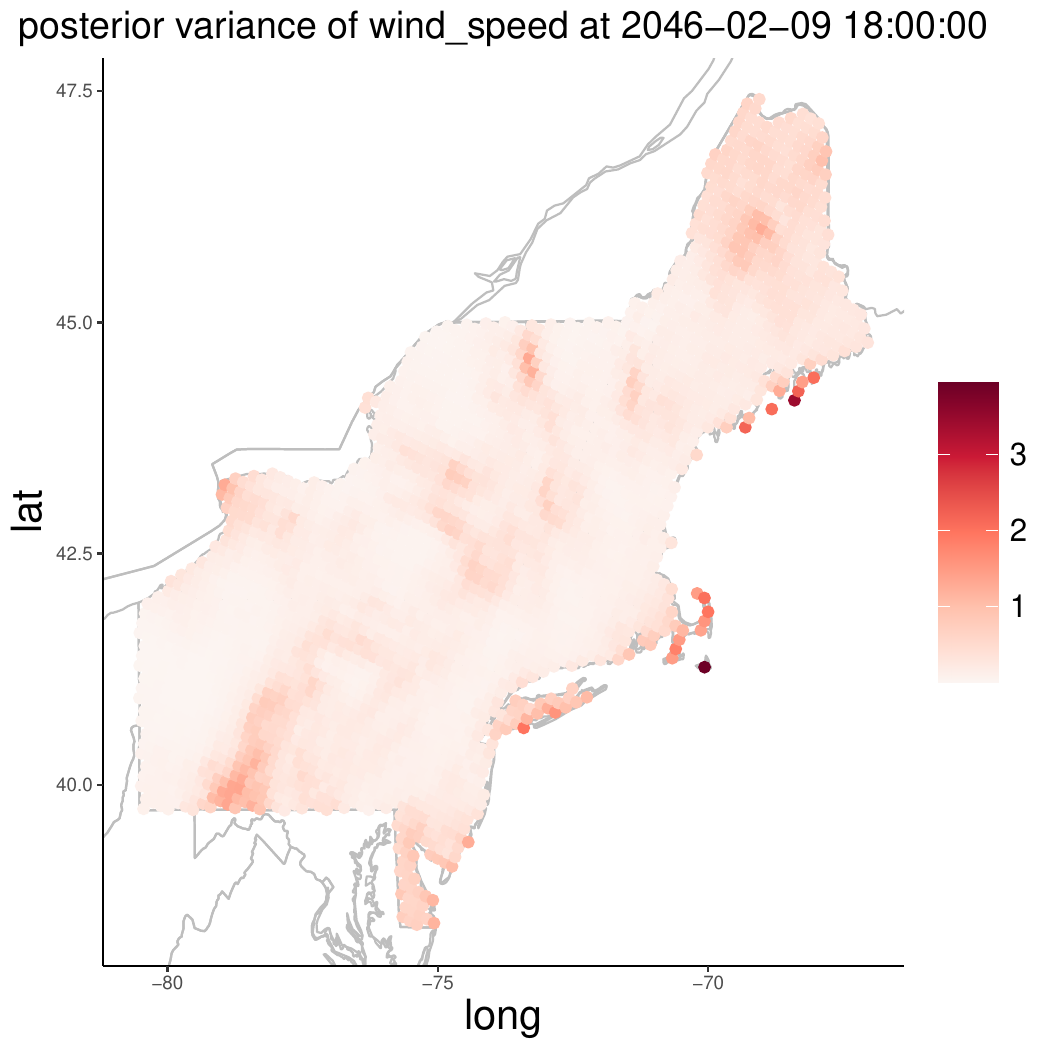}
    \includegraphics[scale=.28]{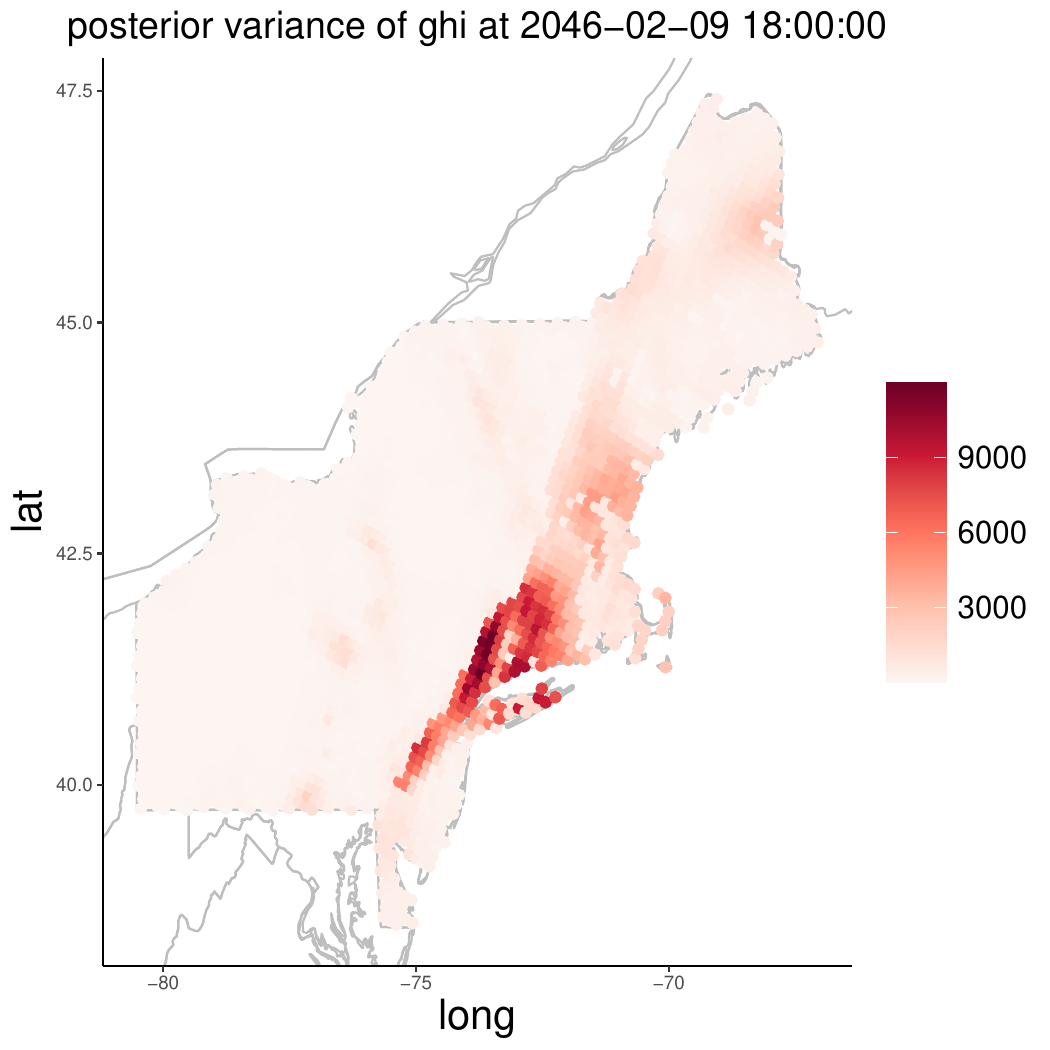}

    \caption{Same configuration as Figure \ref{fig:date1plot}, but during a different date.}
    \label{fig:date2plot}
\end{figure}

\begin{figure}[h!]
    \centering
    \includegraphics[scale=.28]{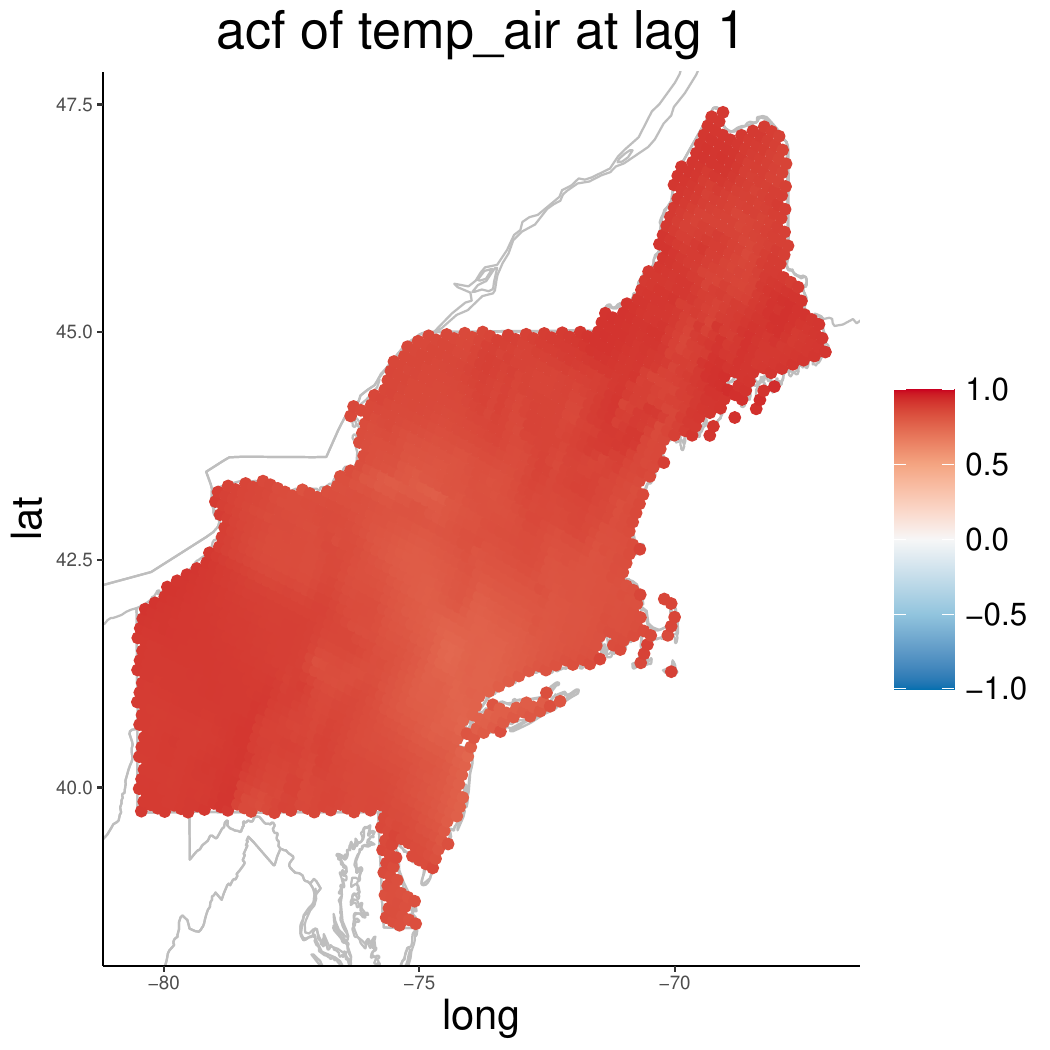}
    \includegraphics[scale=.28]{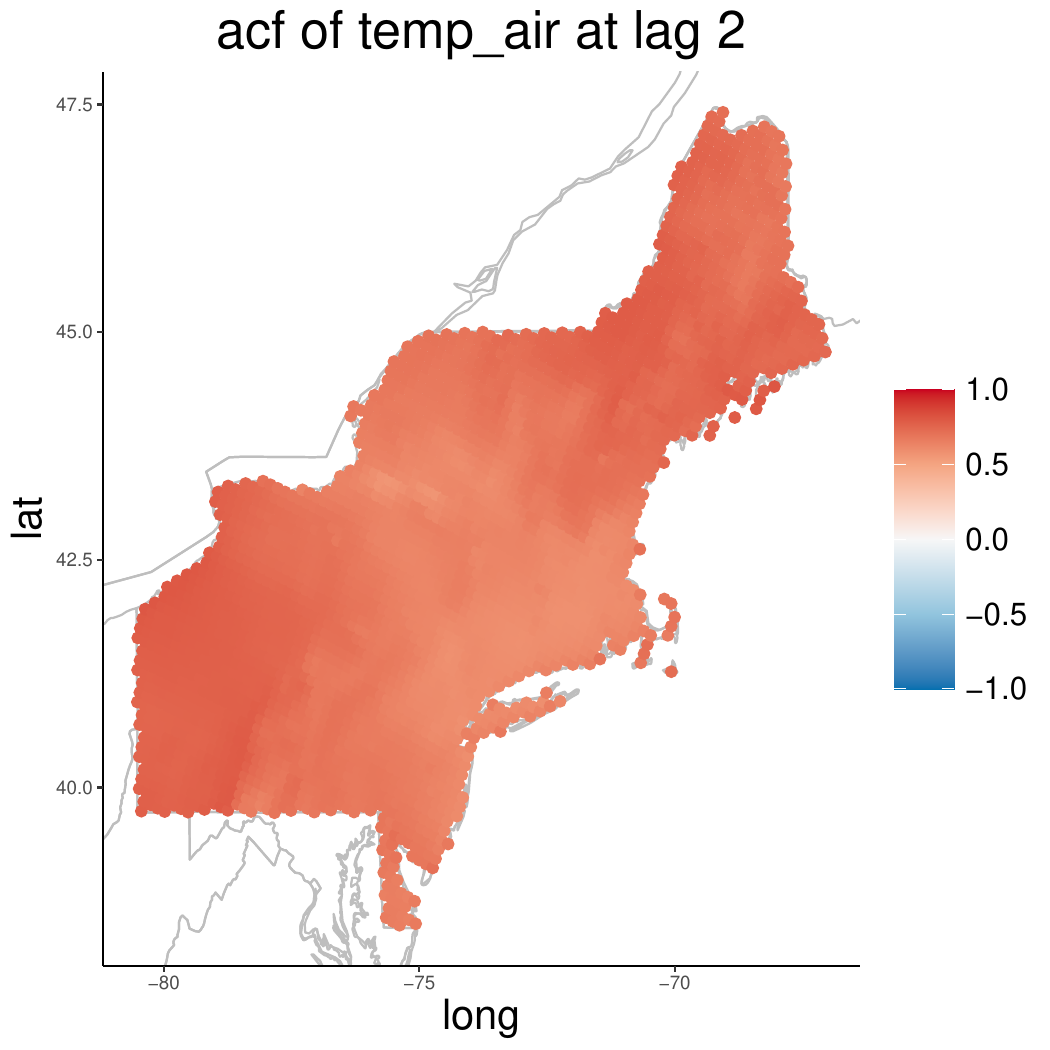}
    \includegraphics[scale=.28]{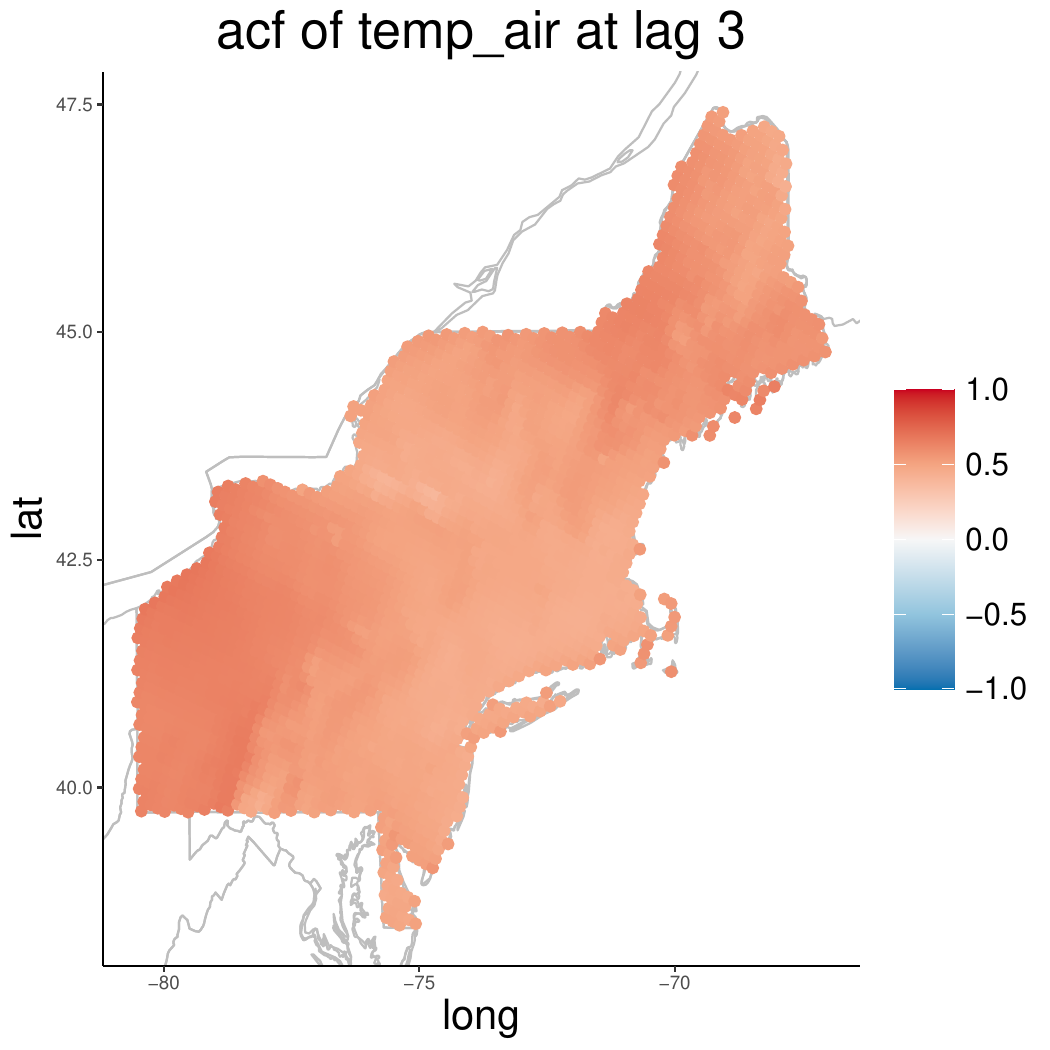}

    \includegraphics[scale=.28]{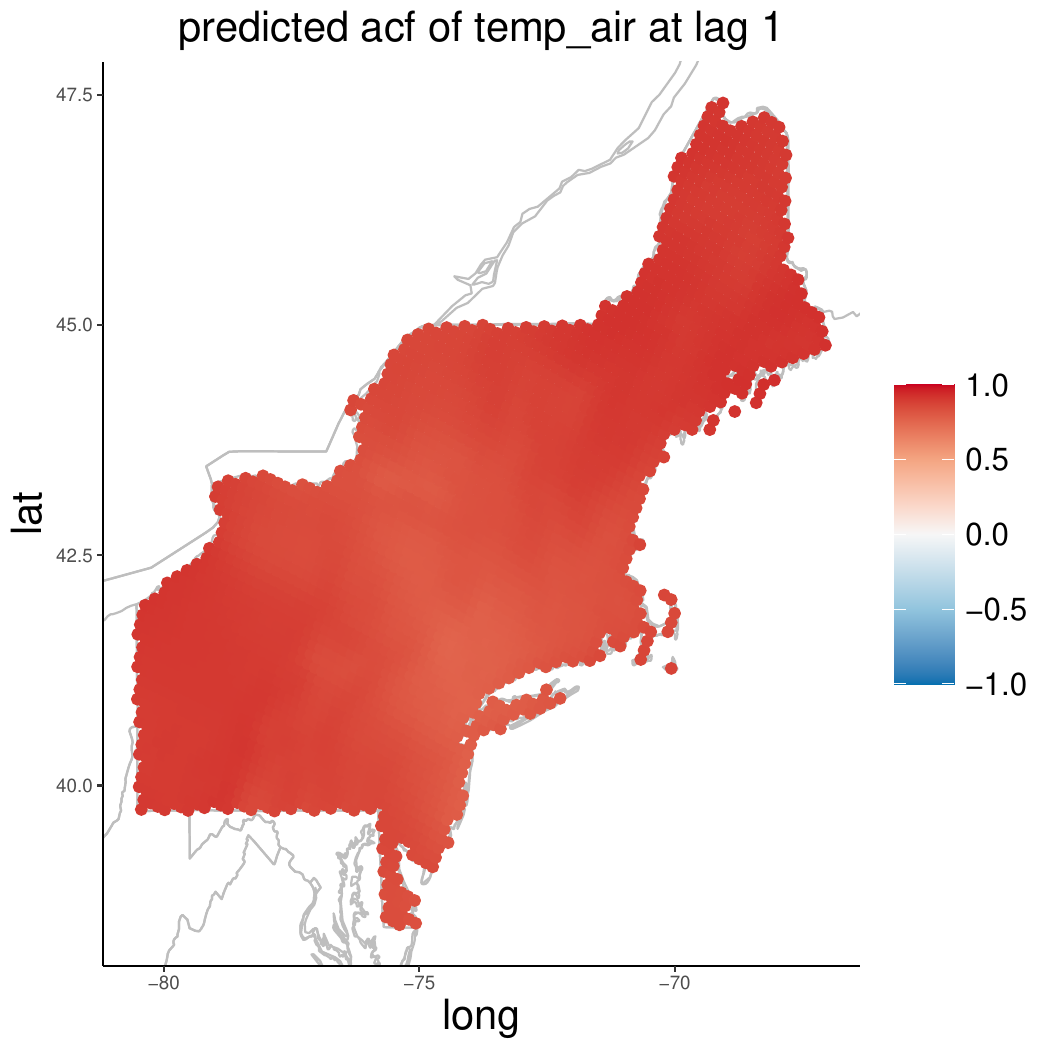}
    \includegraphics[scale=.28]{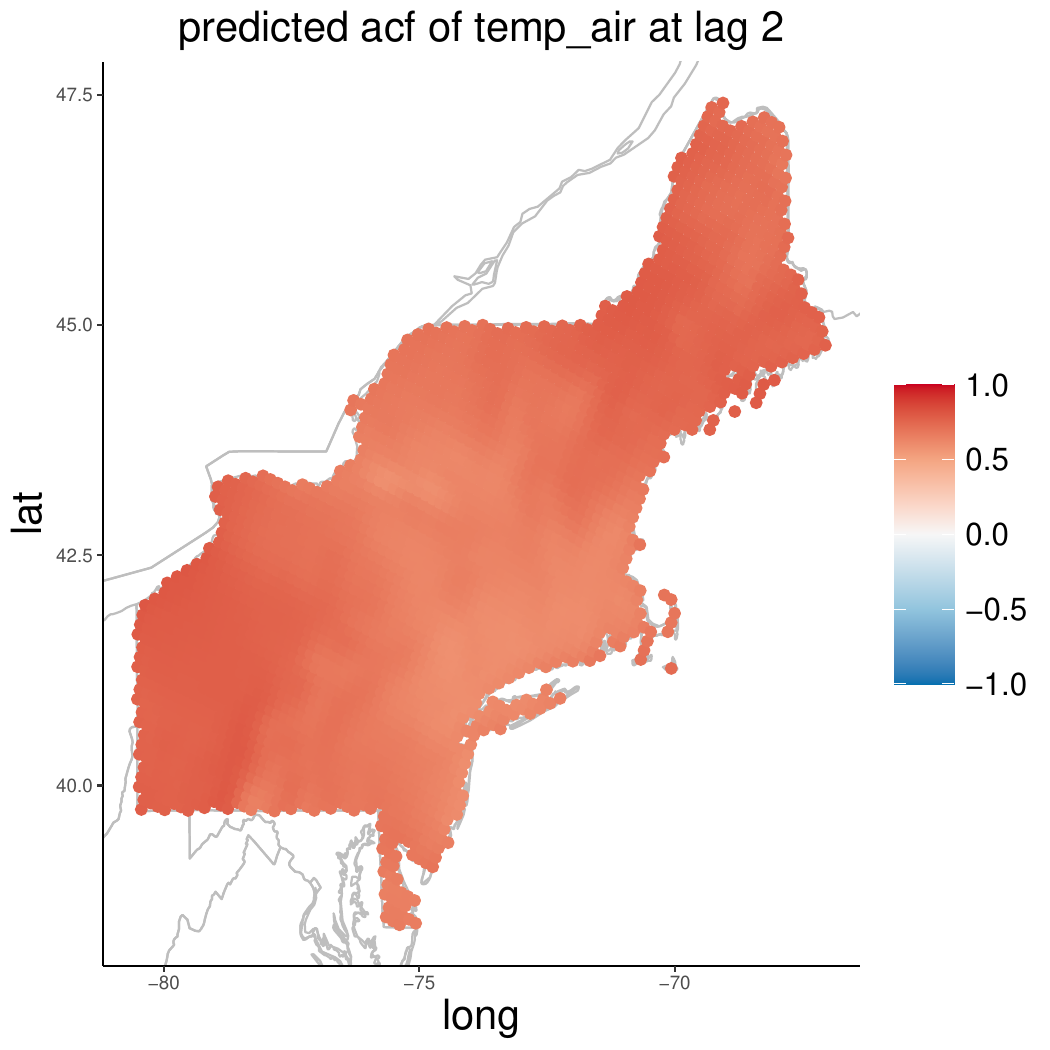}
    \includegraphics[scale=.28]{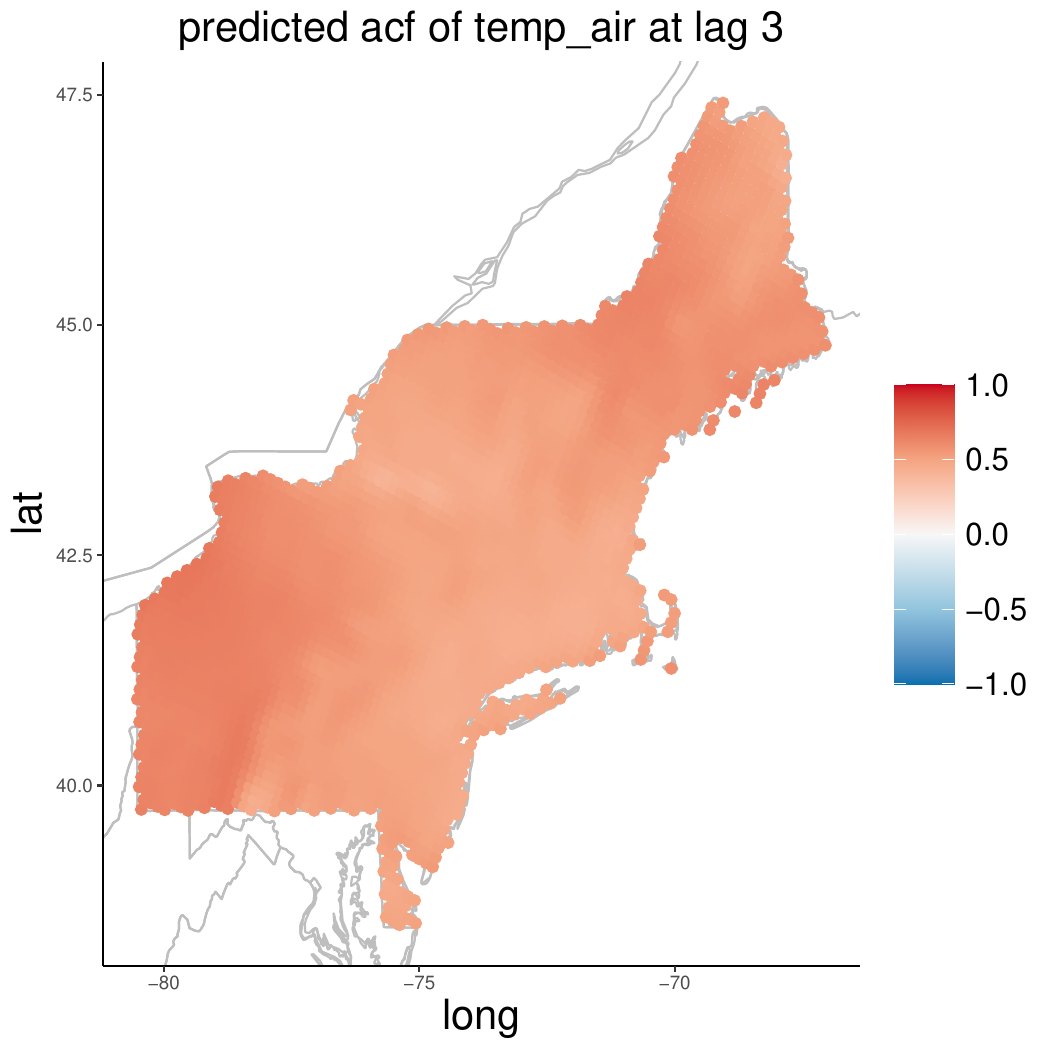}
    
    \includegraphics[scale=.28]{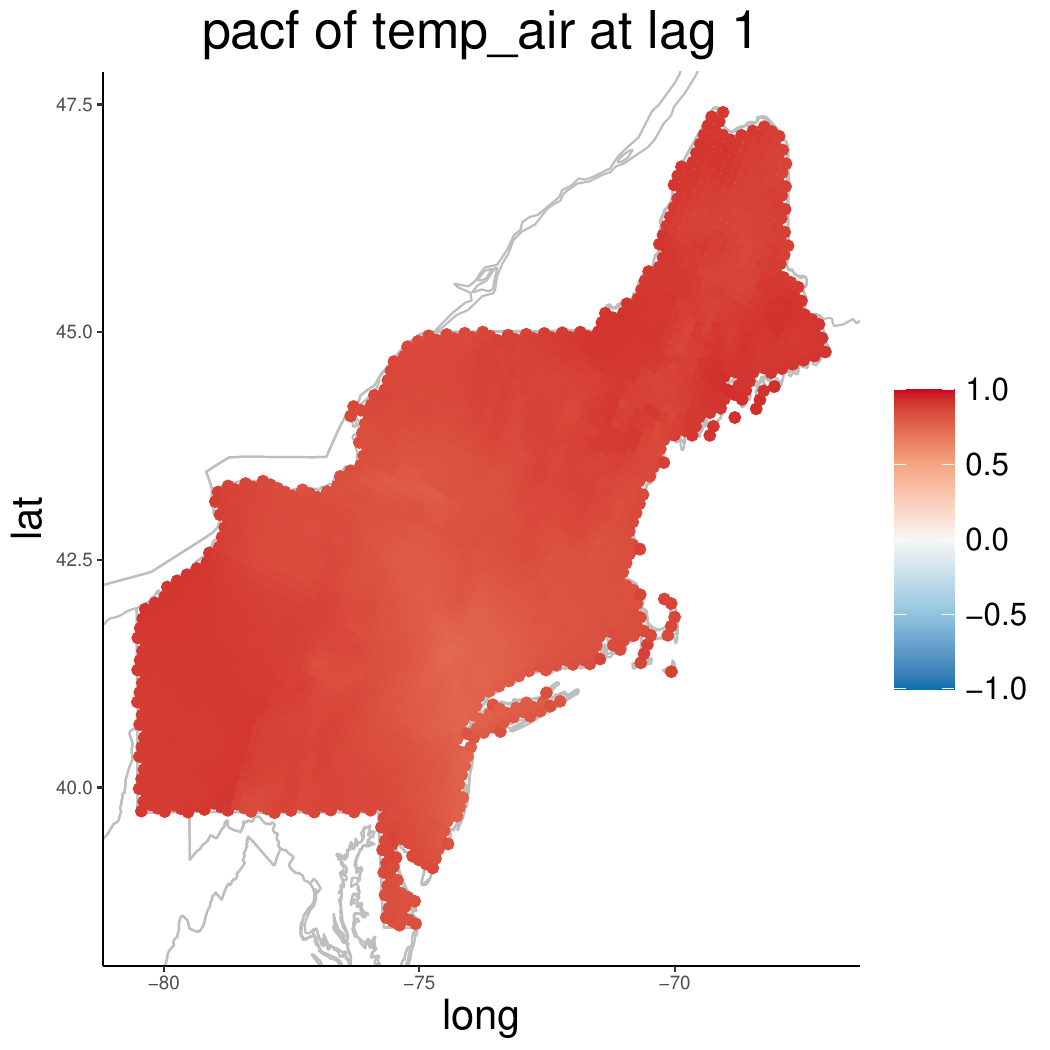}
    \includegraphics[scale=.28]{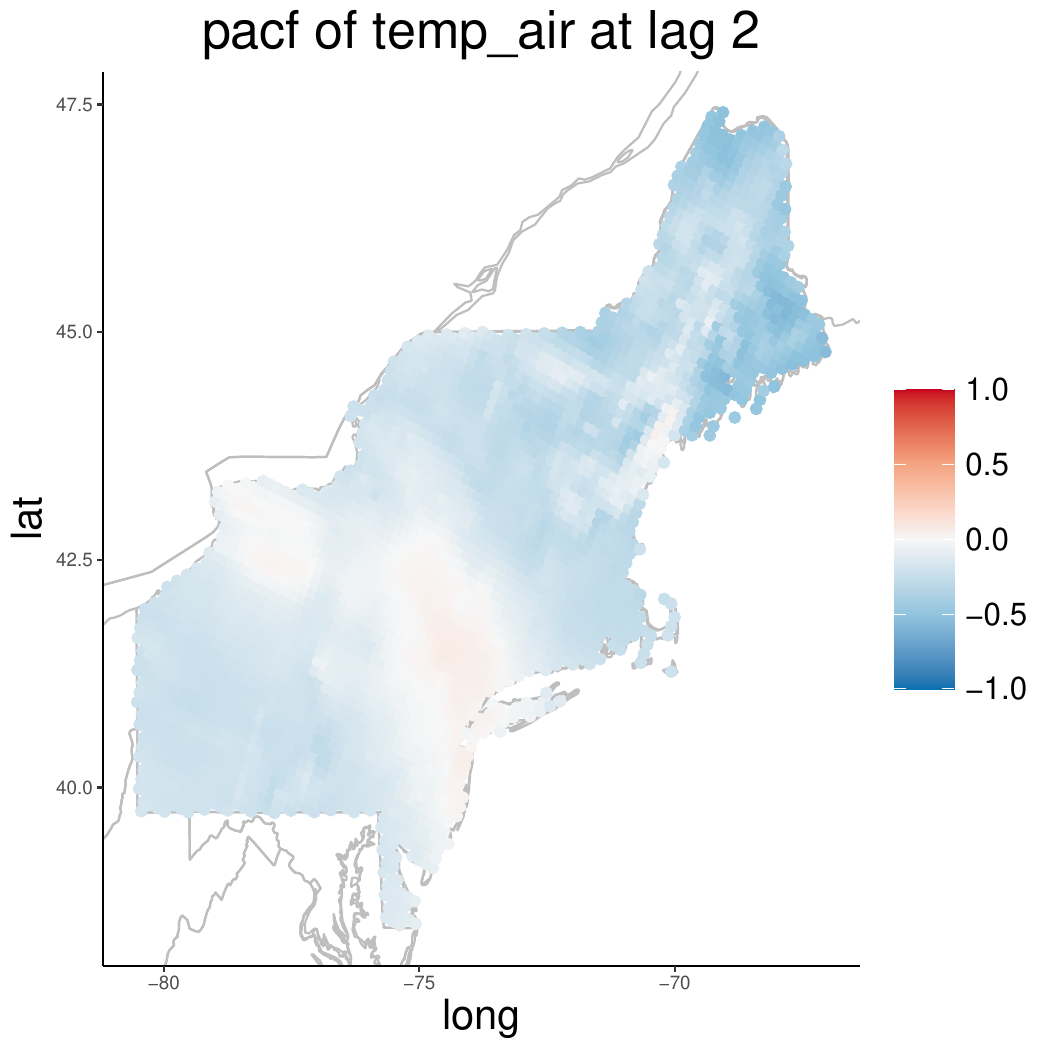}
    \includegraphics[scale=.28]{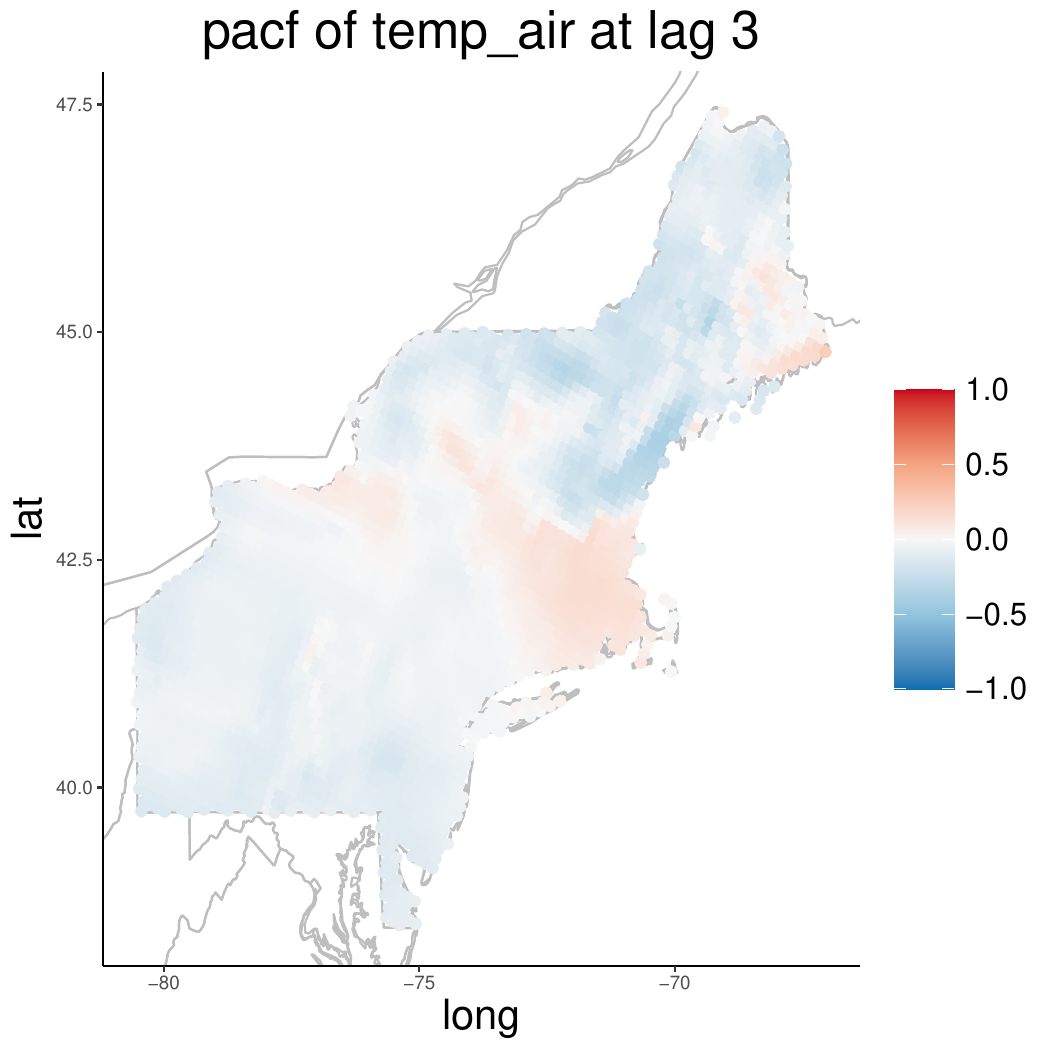}

    \includegraphics[scale=.28]{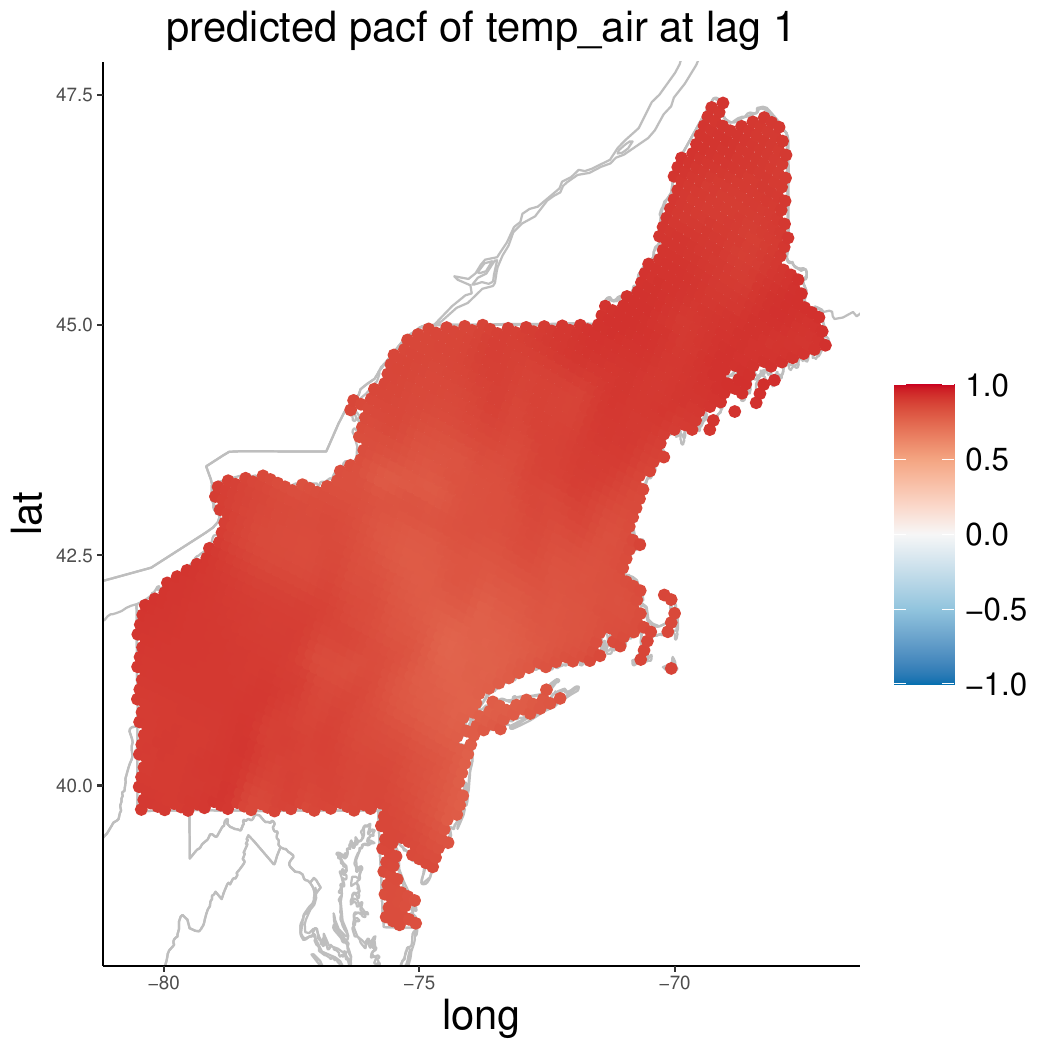}
    \includegraphics[scale=.28]{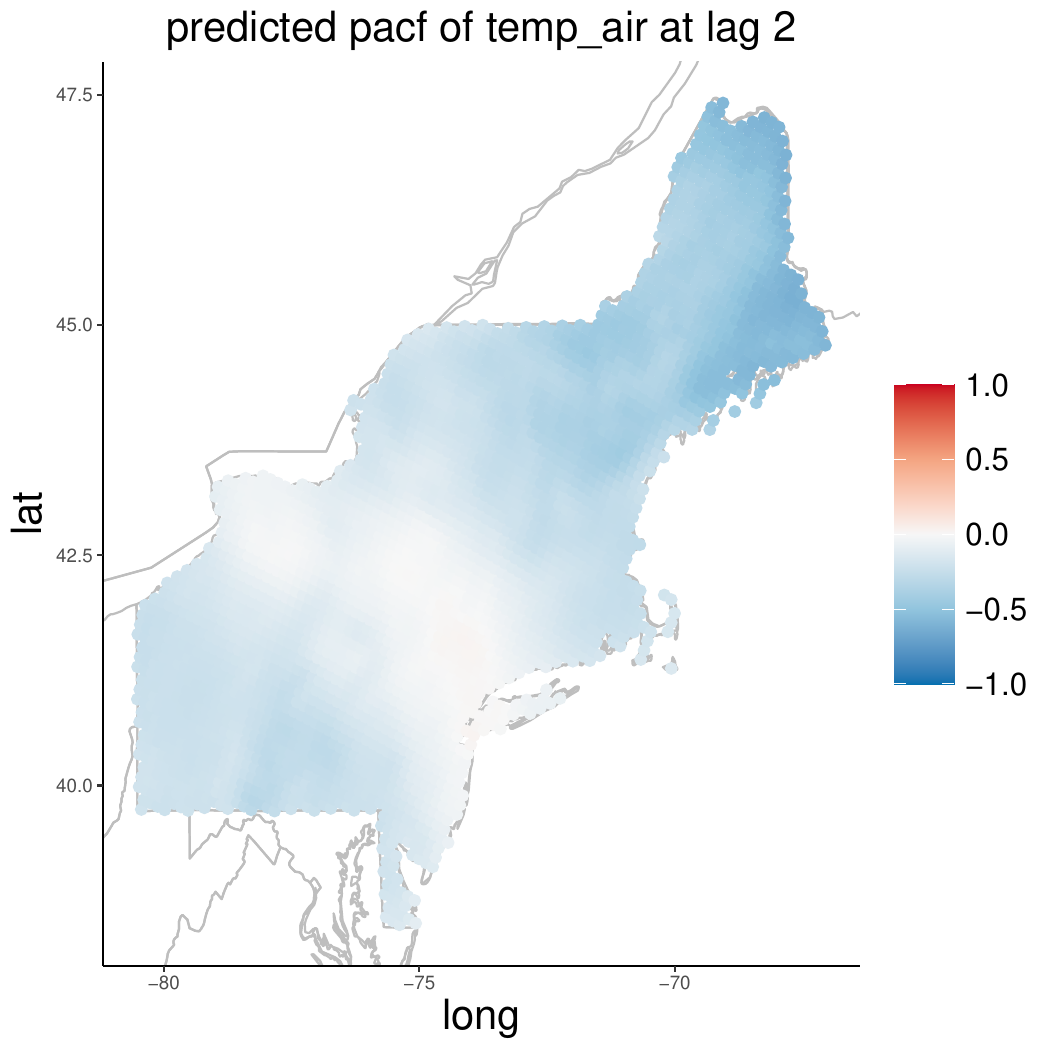}
    \includegraphics[scale=.28]{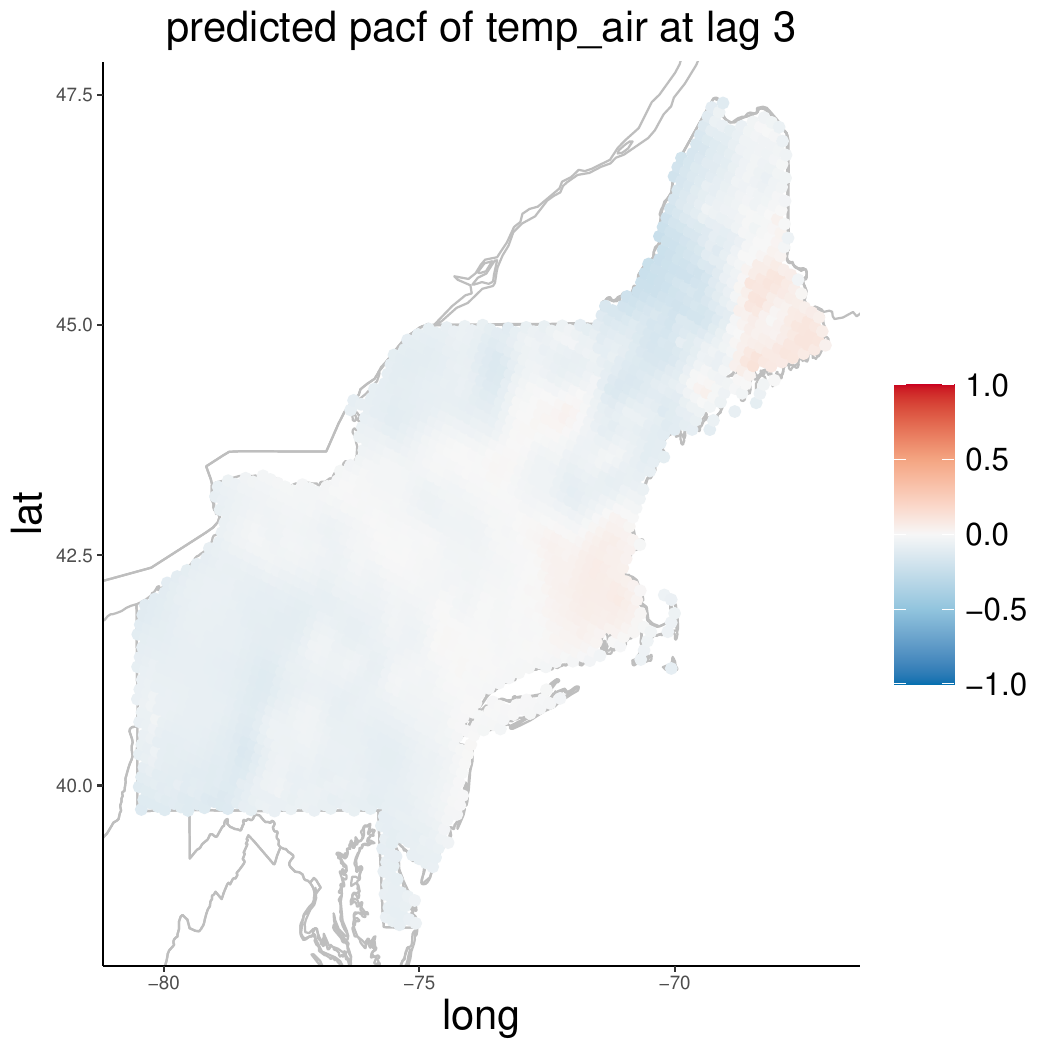}
    
    \caption{Comparison of autocorrelation and partial autocorrelation functions for historical and predicted temperature.}
    \label{fig:acfpacftemp}
\end{figure}

\begin{figure}[h!]
    \centering
    \includegraphics[scale=.28]{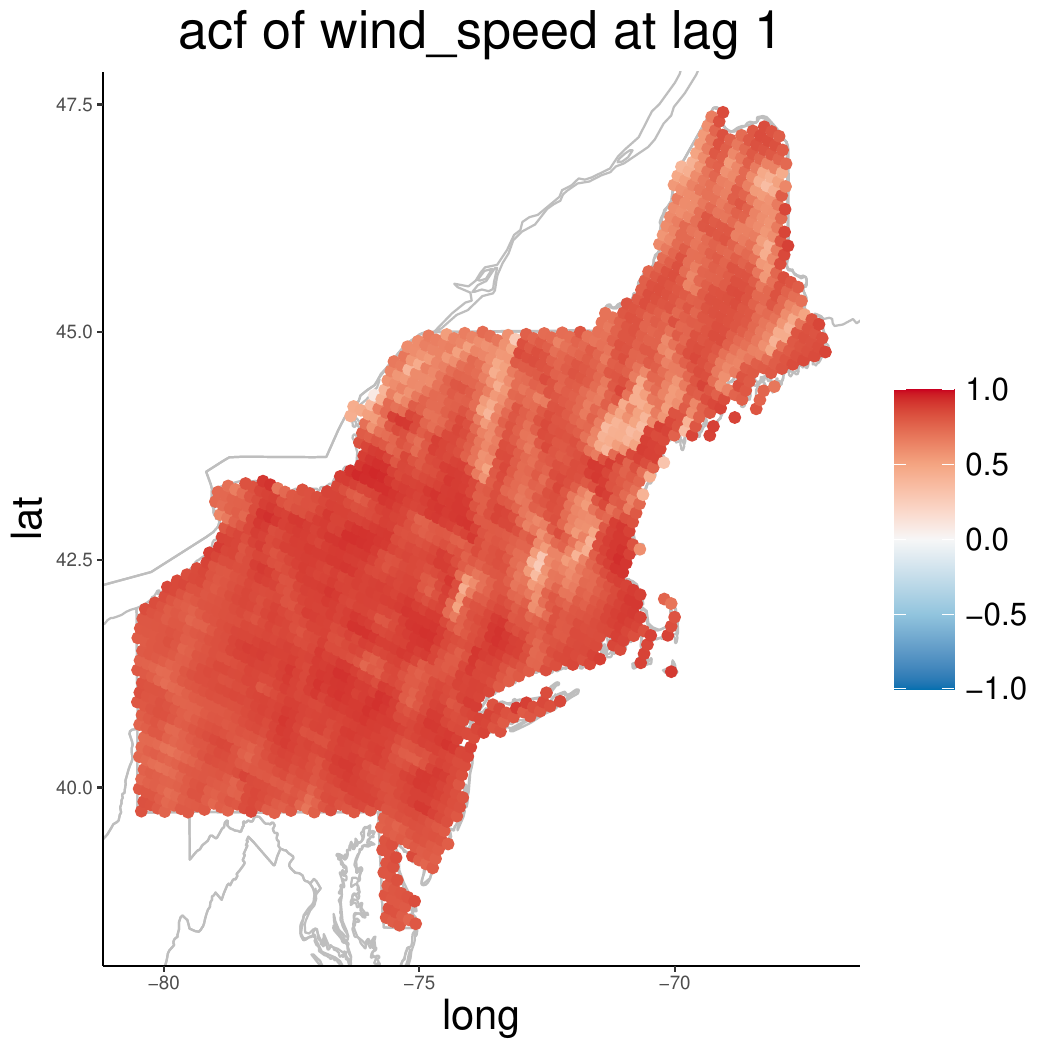}
    \includegraphics[scale=.28]{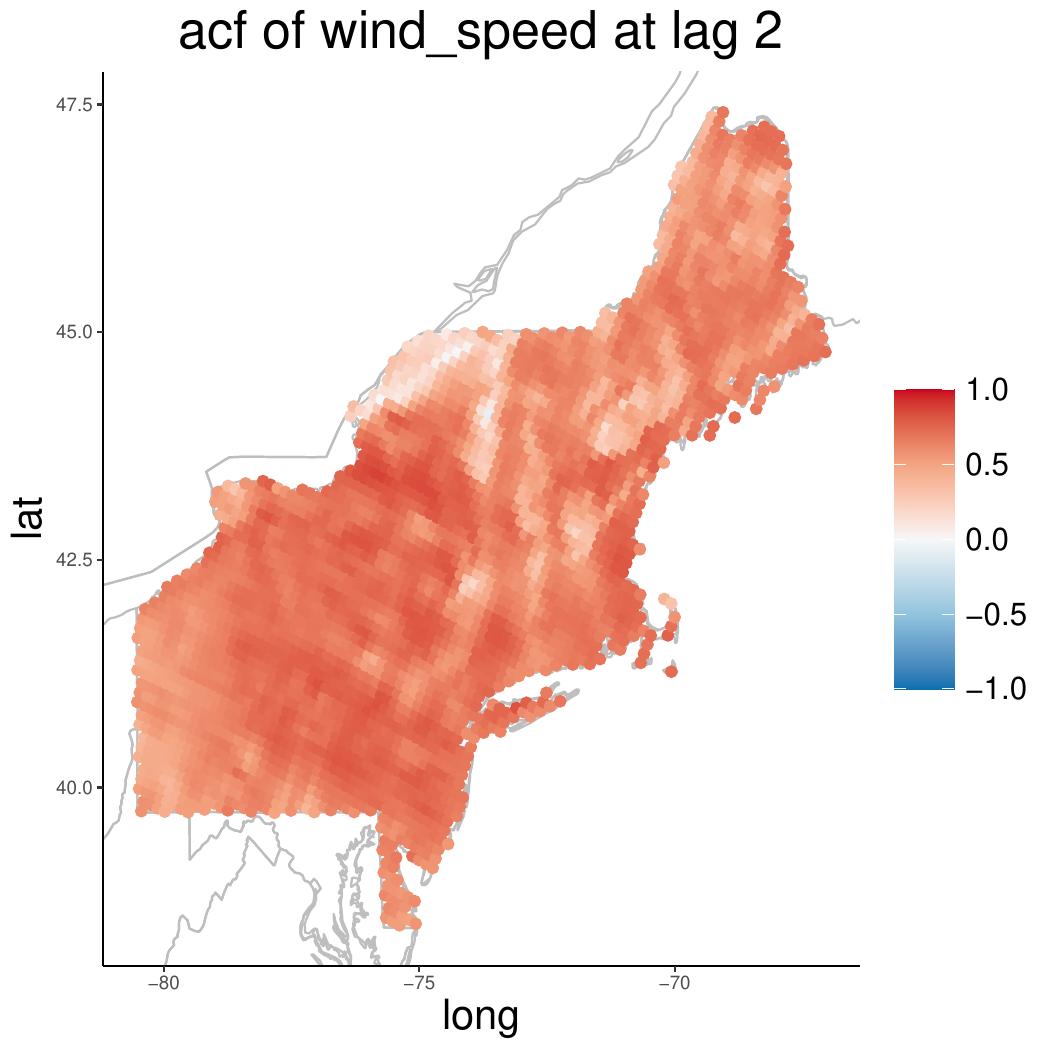}
    \includegraphics[scale=.28]{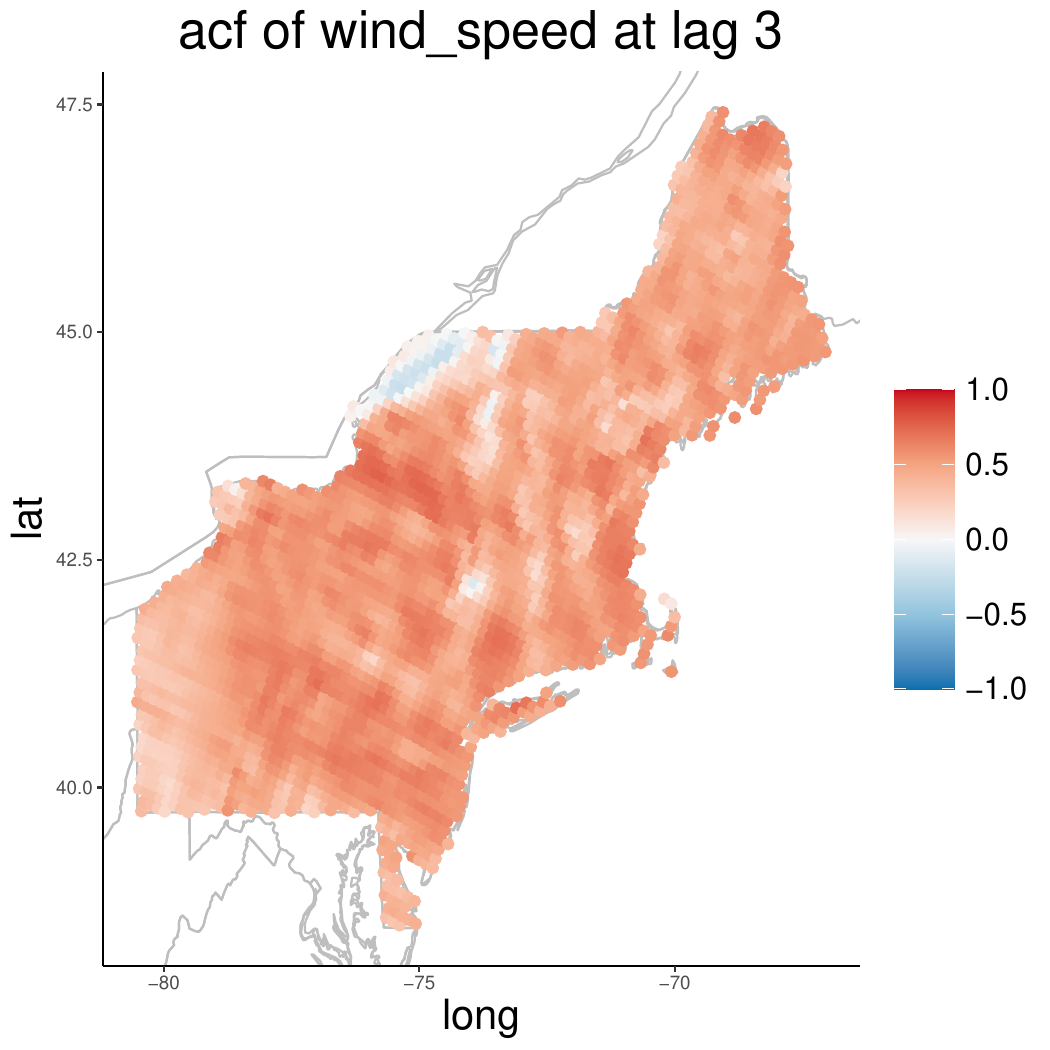}

    \includegraphics[scale=.28]{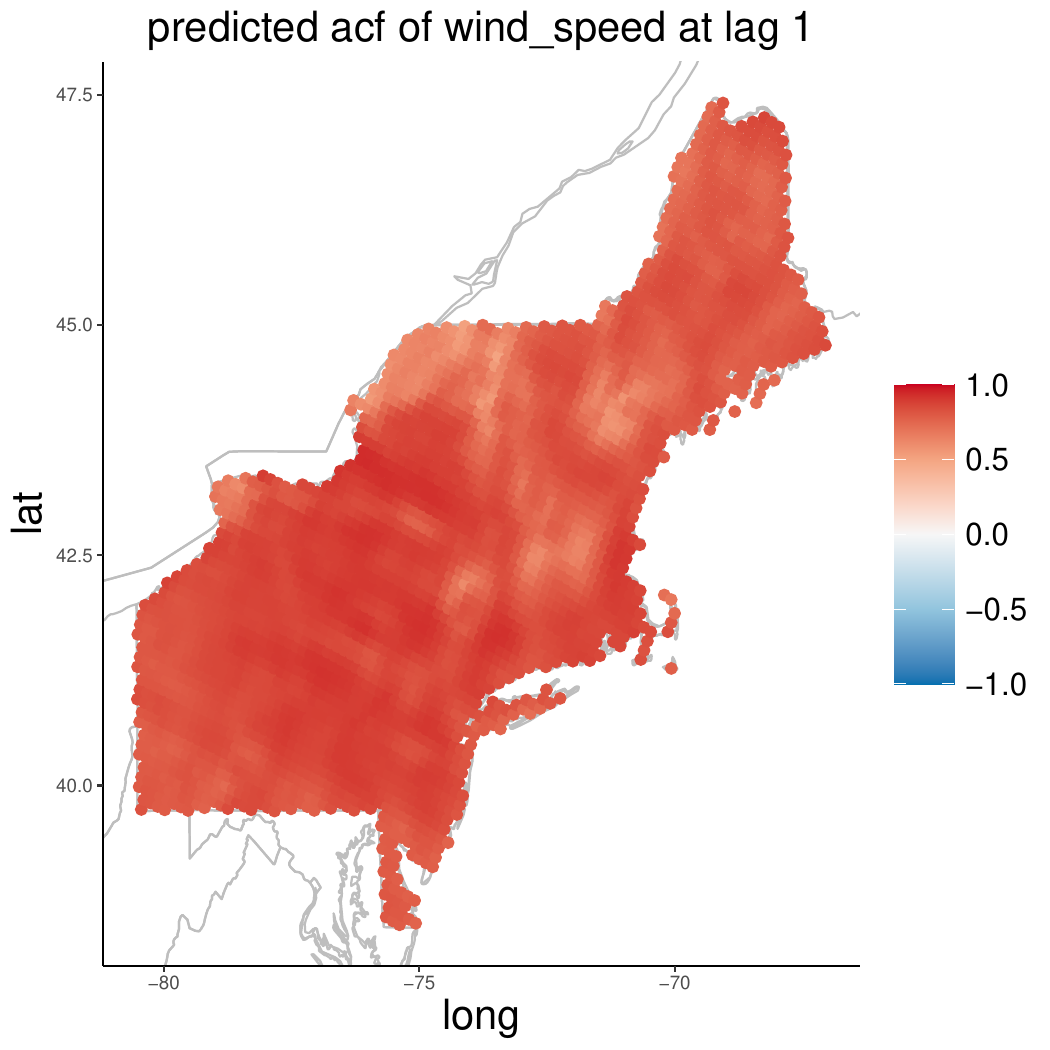}
    \includegraphics[scale=.28]{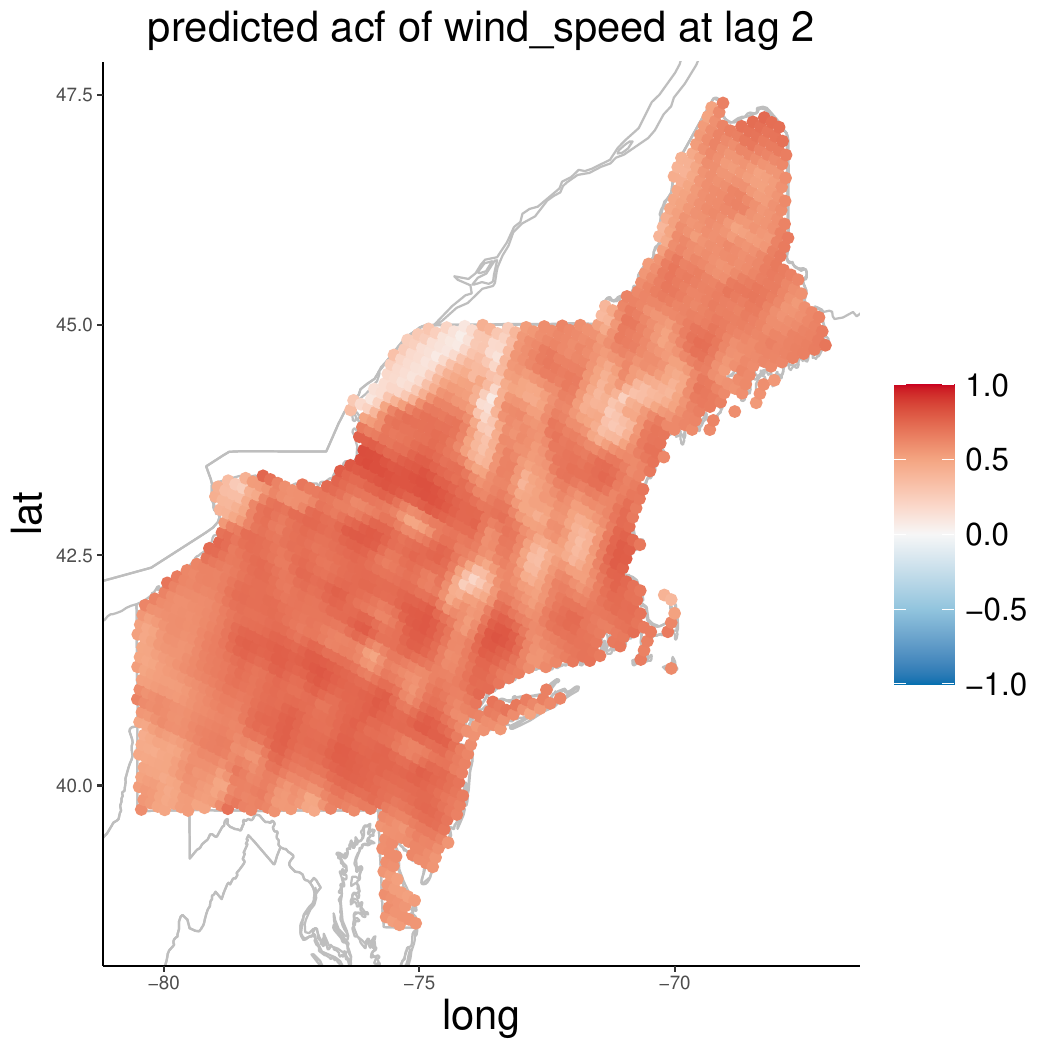}
    \includegraphics[scale=.28]{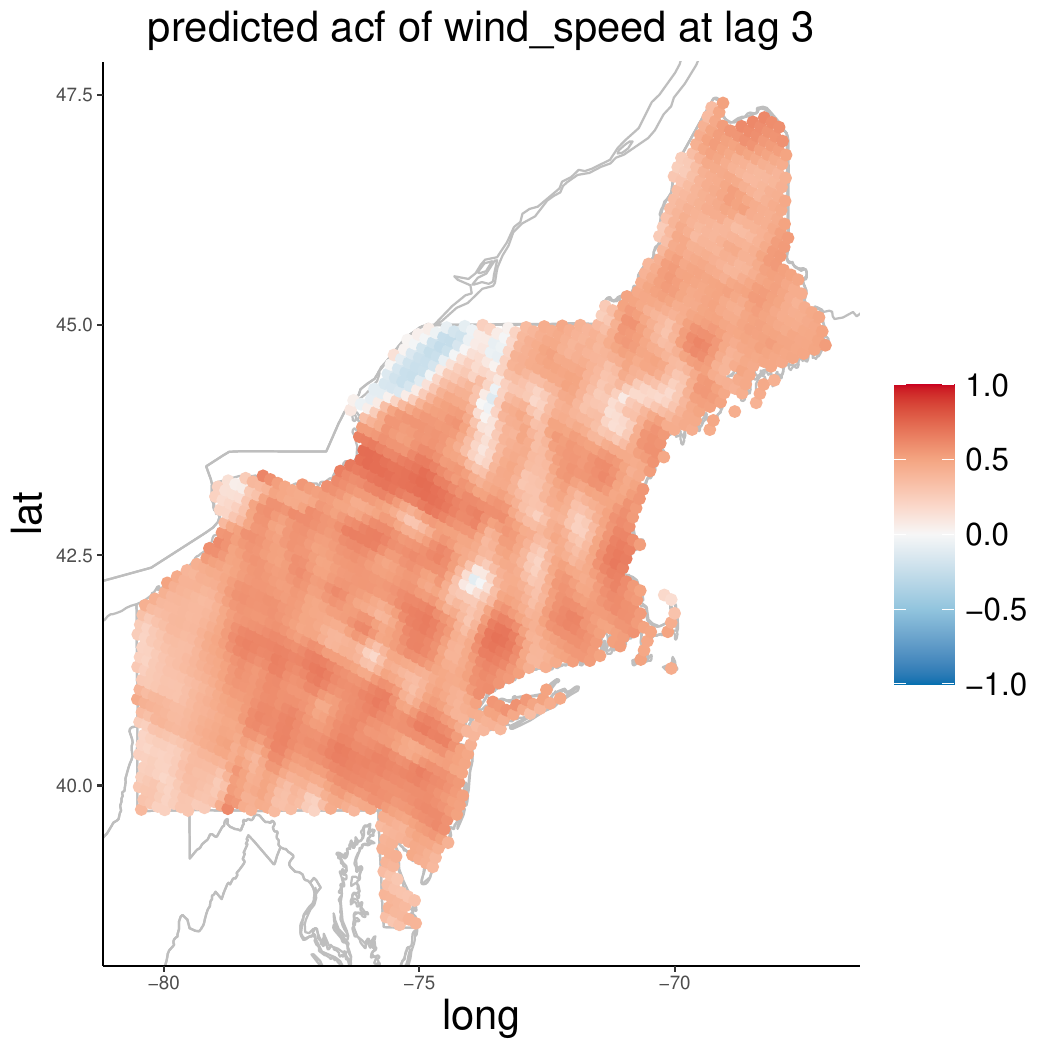}
    
    \includegraphics[scale=.28]{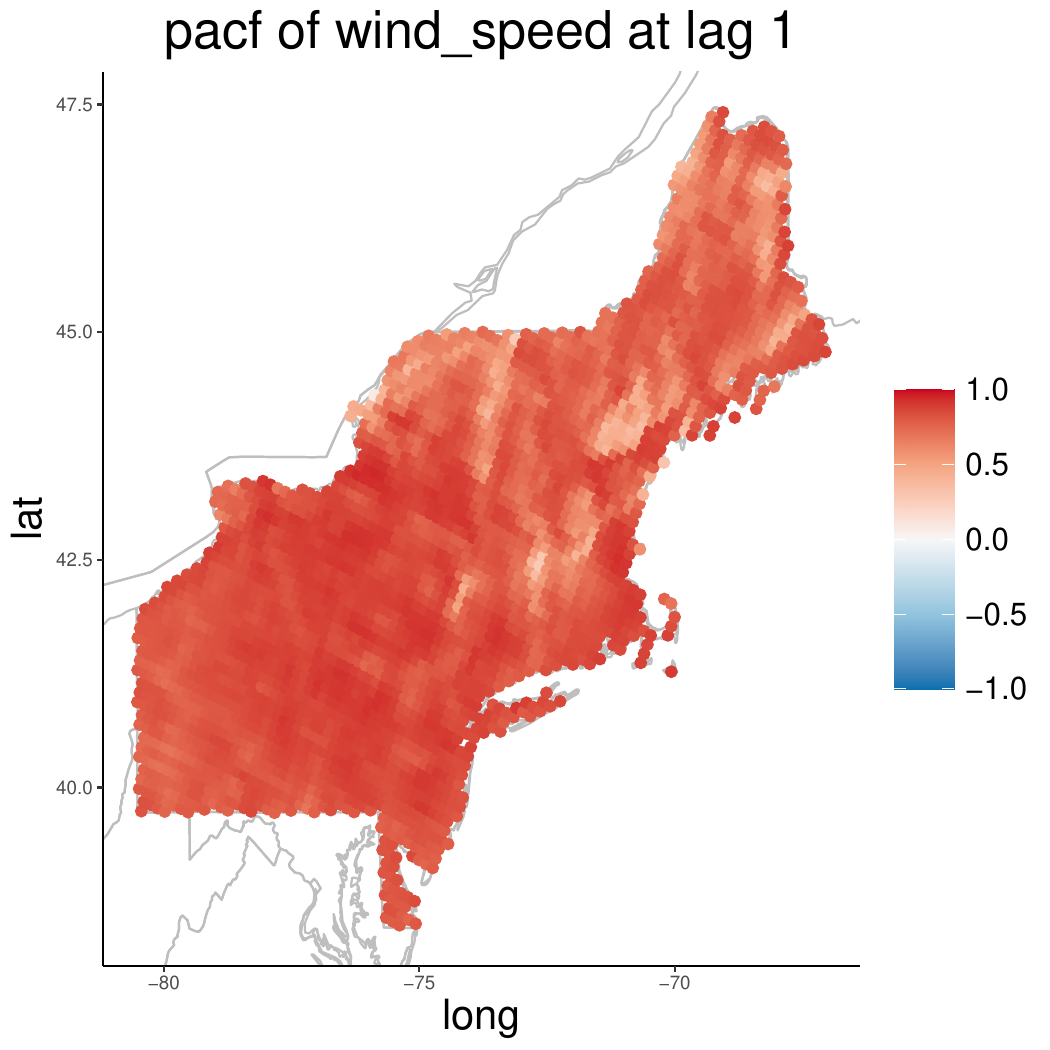}
    \includegraphics[scale=.28]{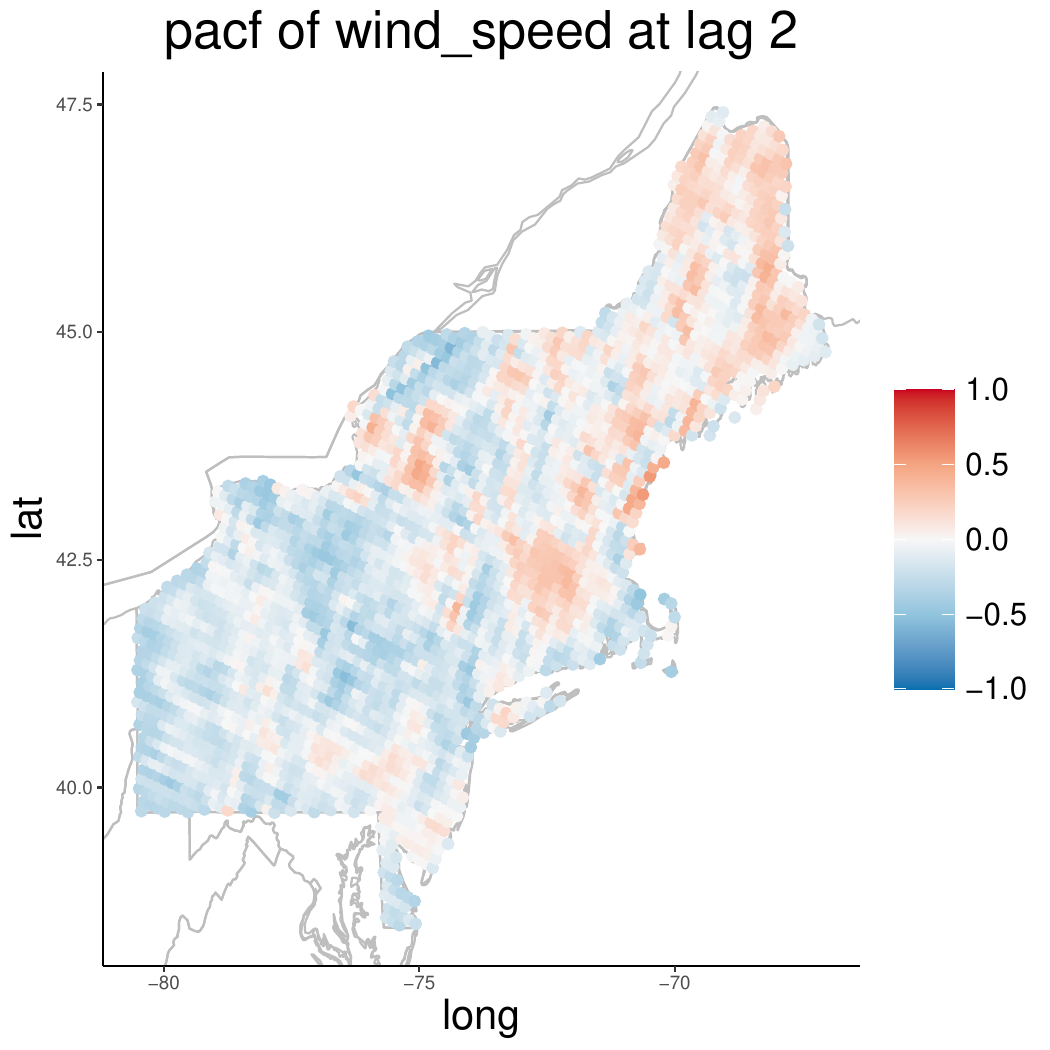}
    \includegraphics[scale=.28]{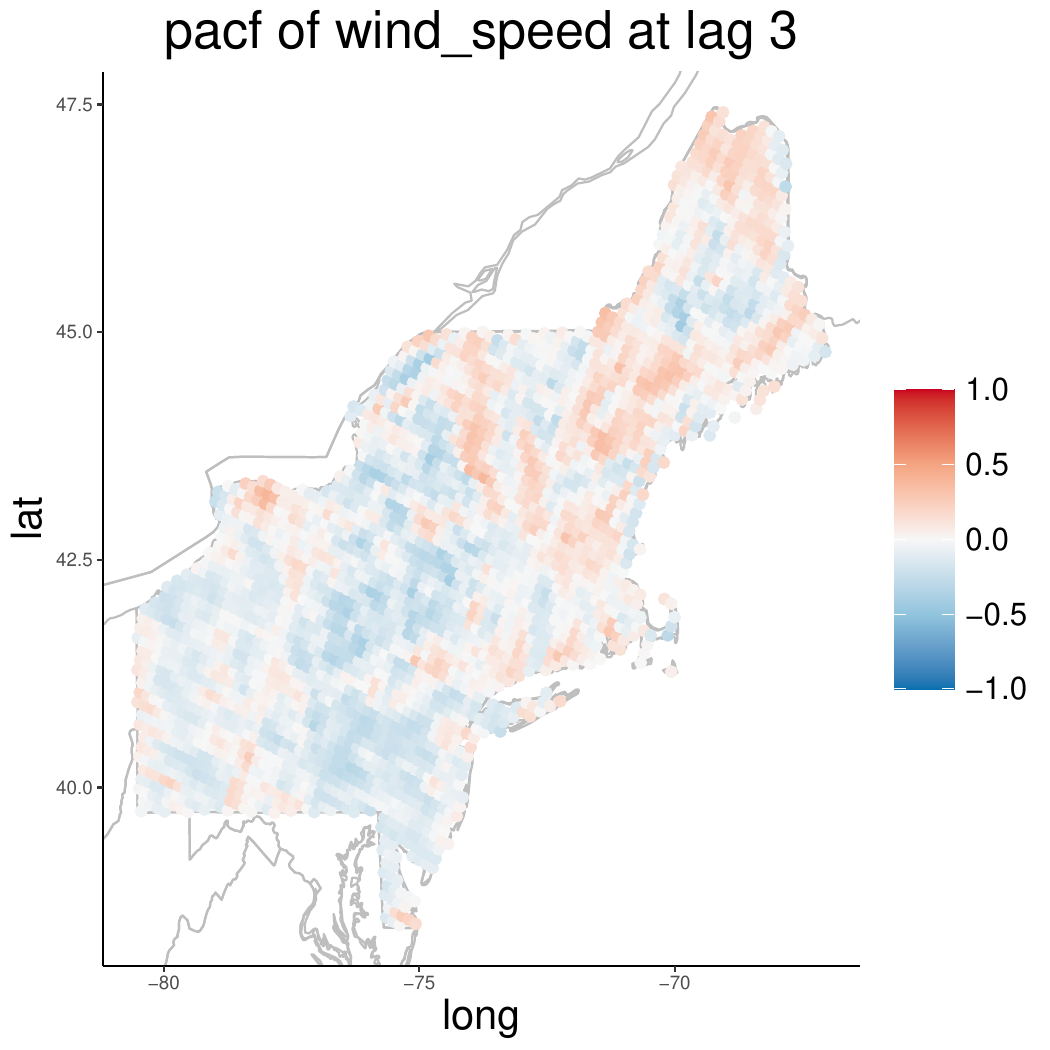}

    \includegraphics[scale=.28]{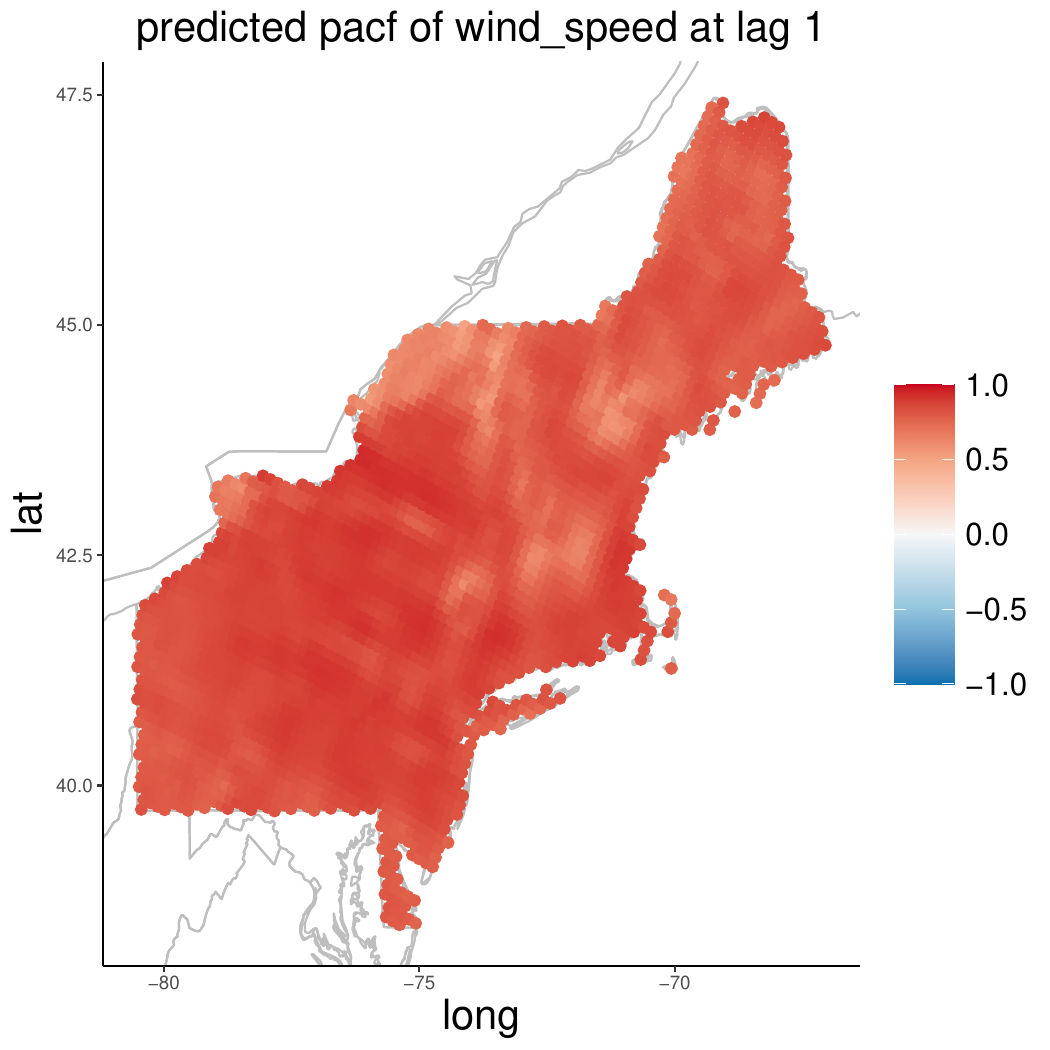}
    \includegraphics[scale=.28]{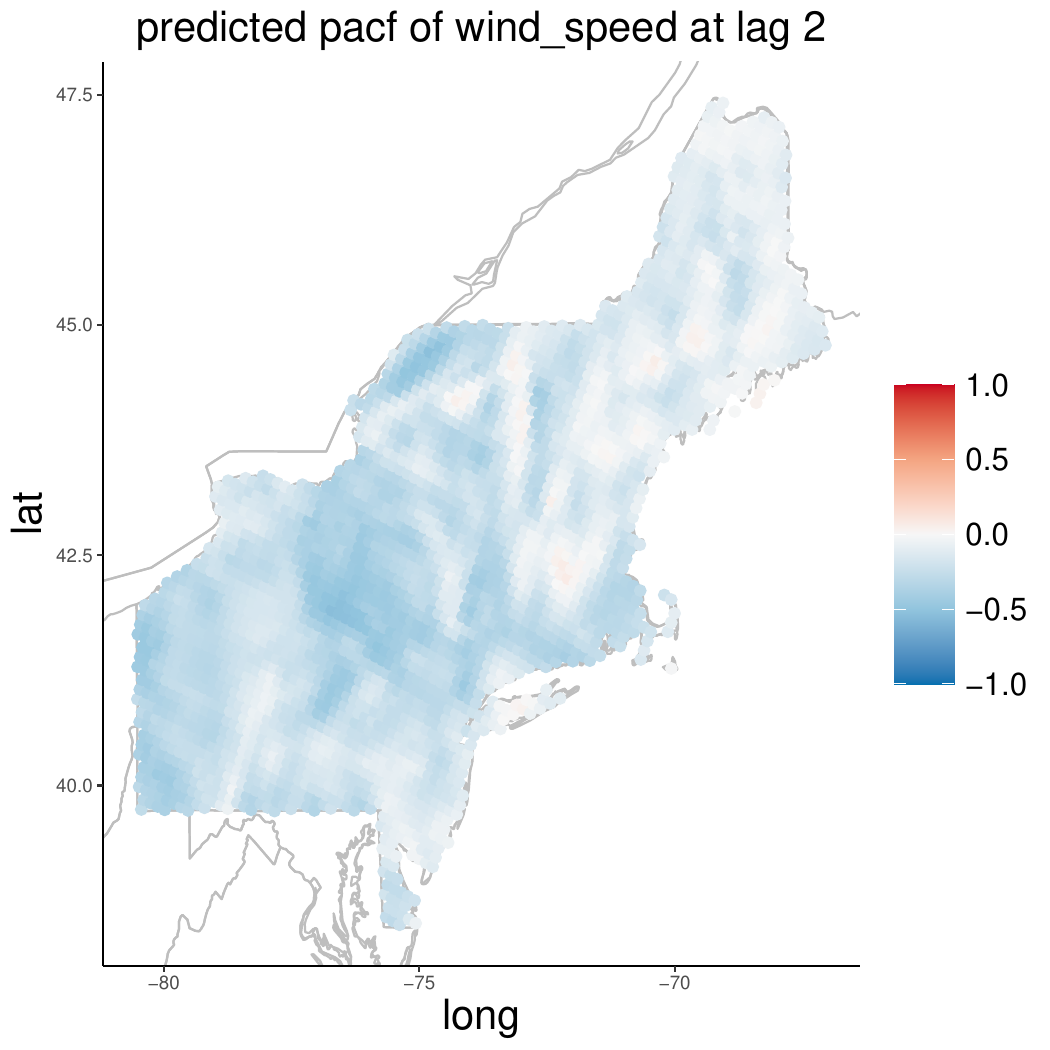}
    \includegraphics[scale=.28]{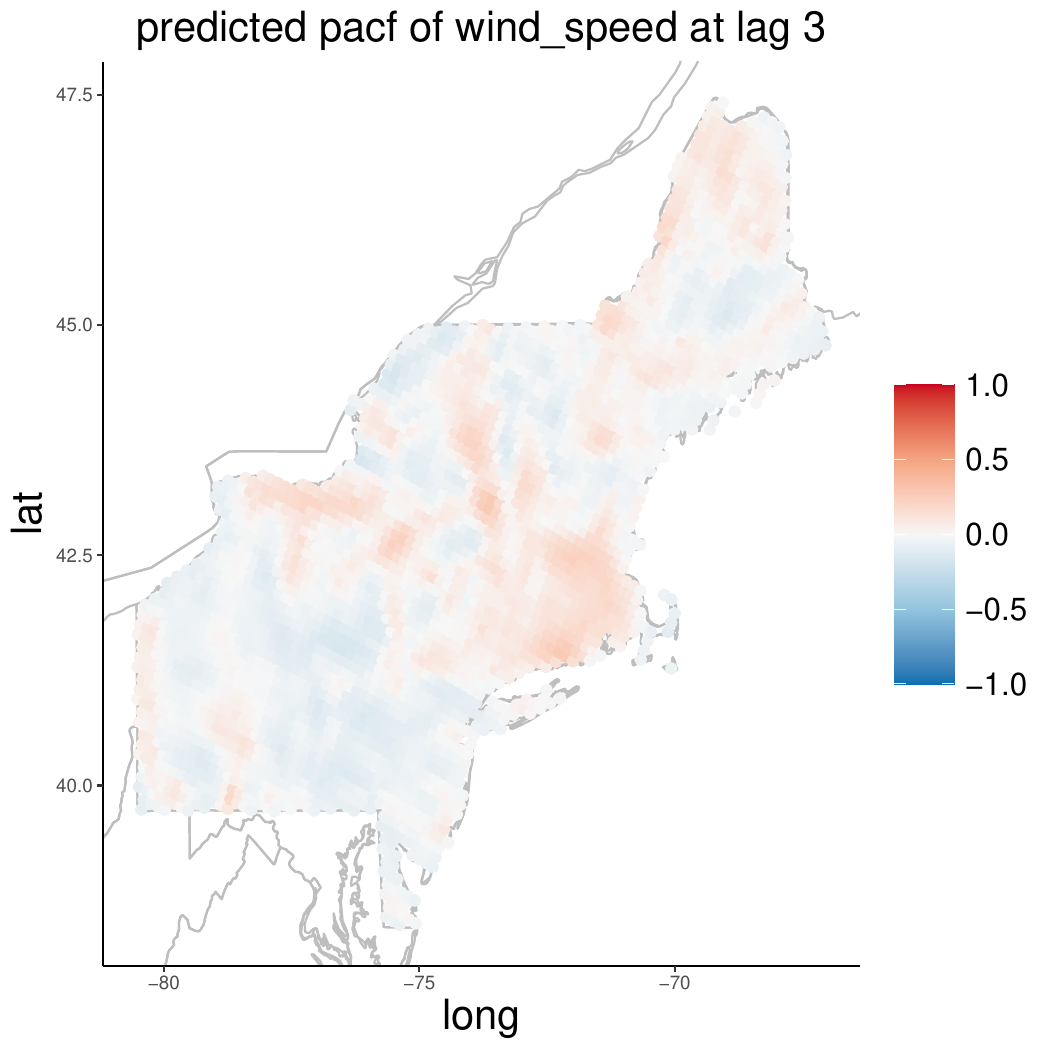}
    
    \caption{Same configuration as Figure \ref{fig:acfpacftemp}, but for wind speed.}
    \label{fig:acfpacfwind}
\end{figure}

\begin{figure}[h!]
    \centering
    \includegraphics[scale=.28]{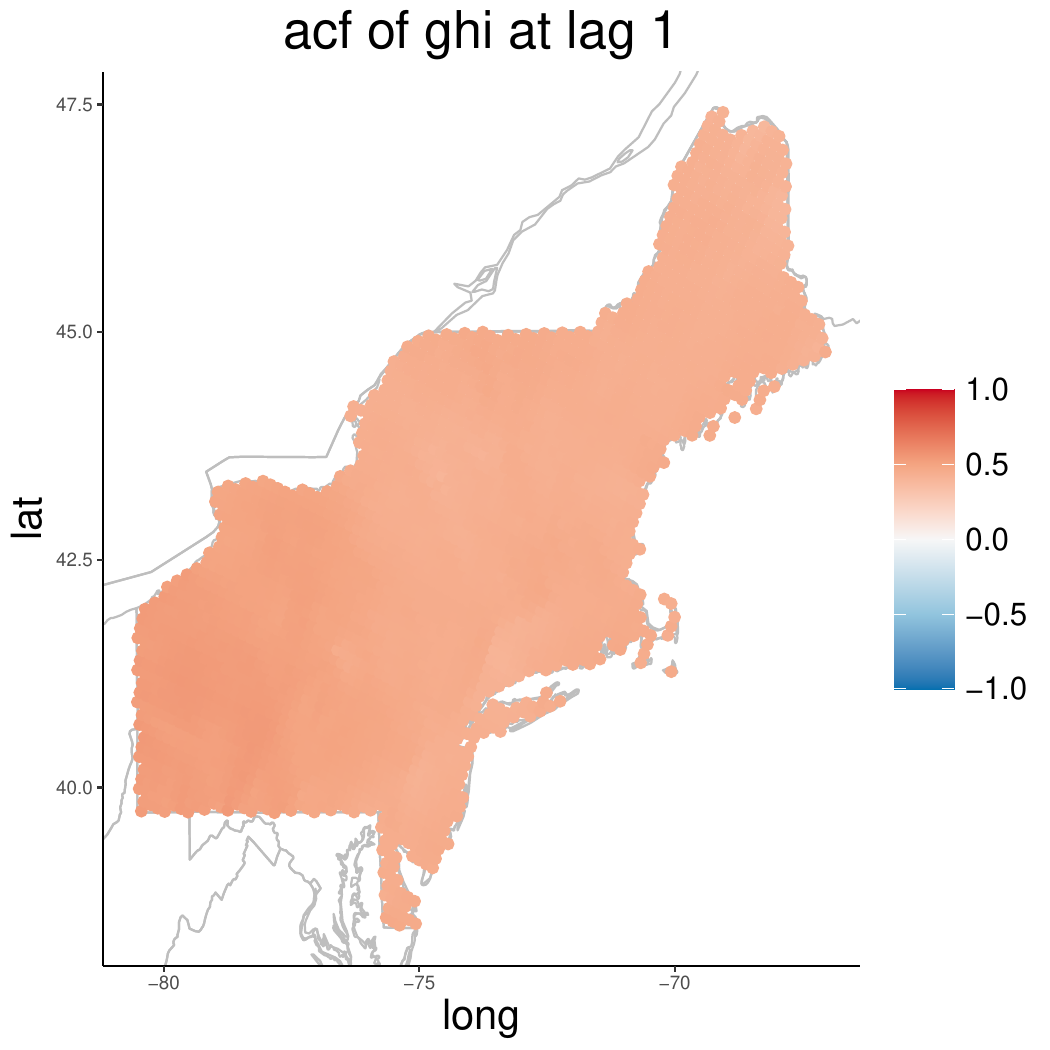}
    \includegraphics[scale=.28]{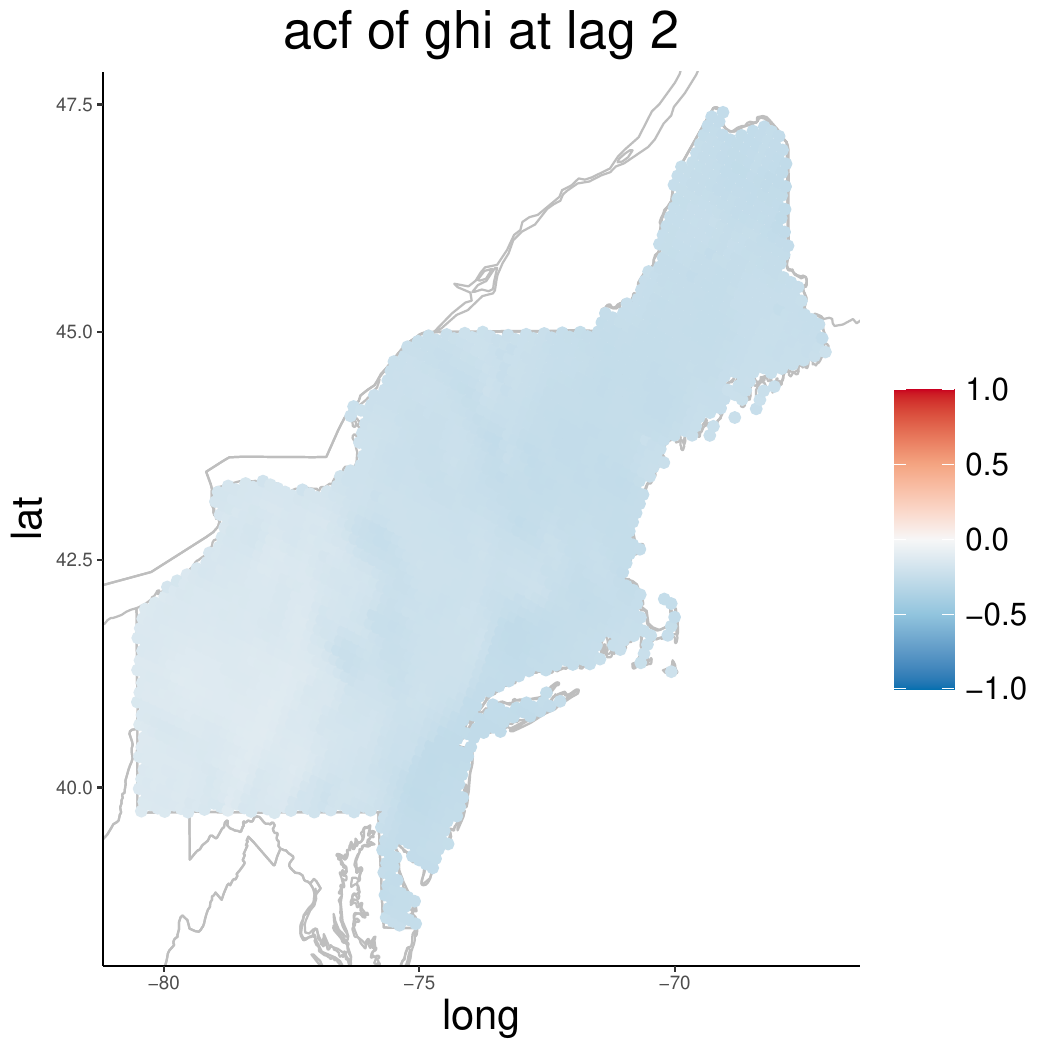}
    \includegraphics[scale=.28]{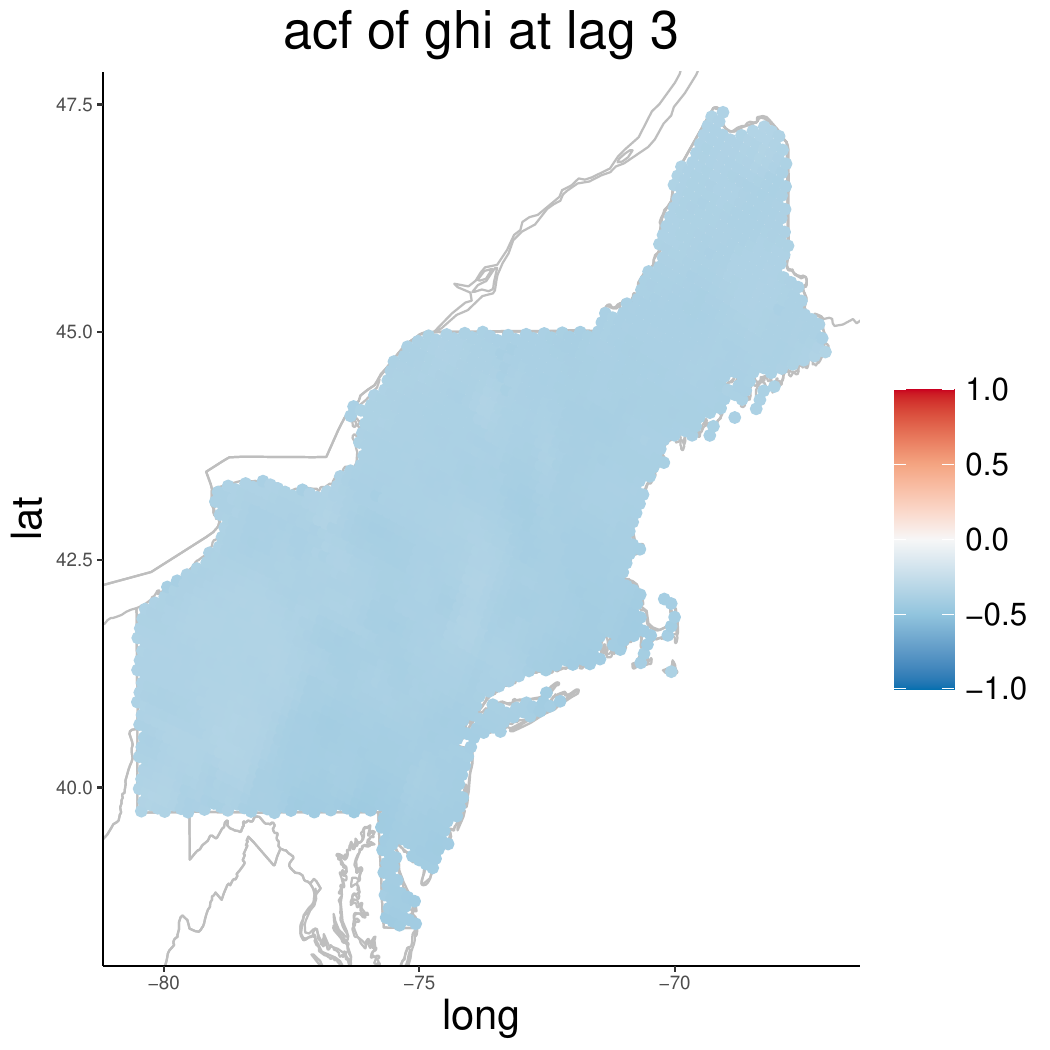}

    \includegraphics[scale=.28]{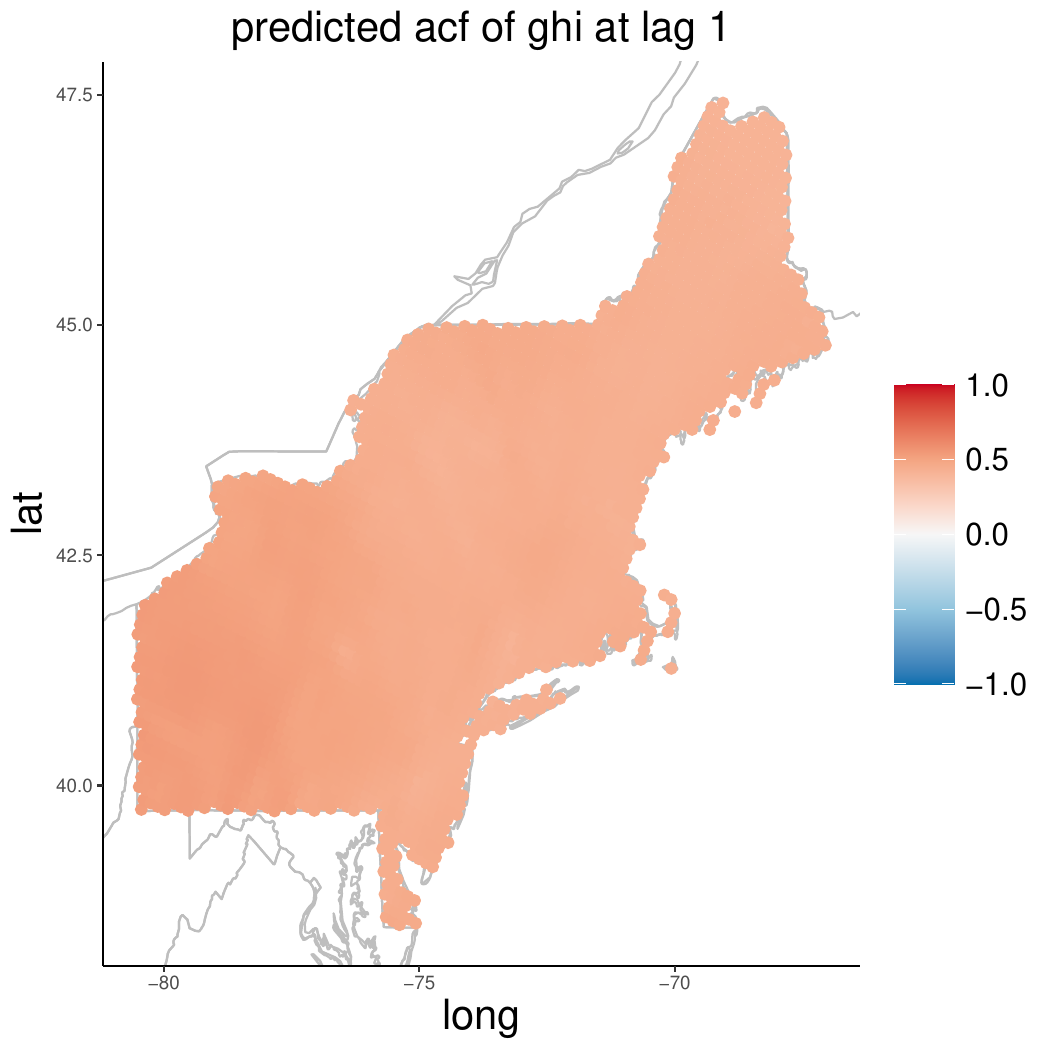}
    \includegraphics[scale=.28]{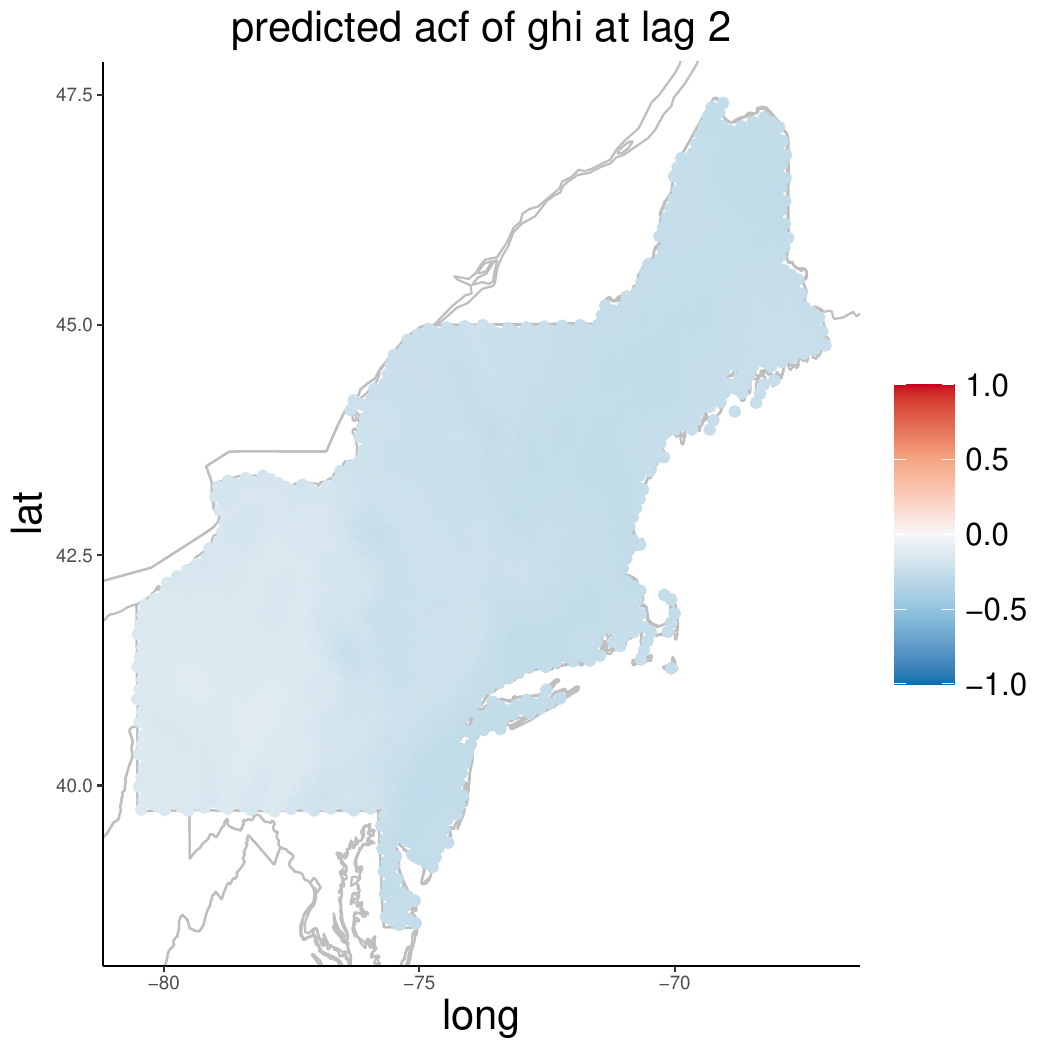}
    \includegraphics[scale=.28]{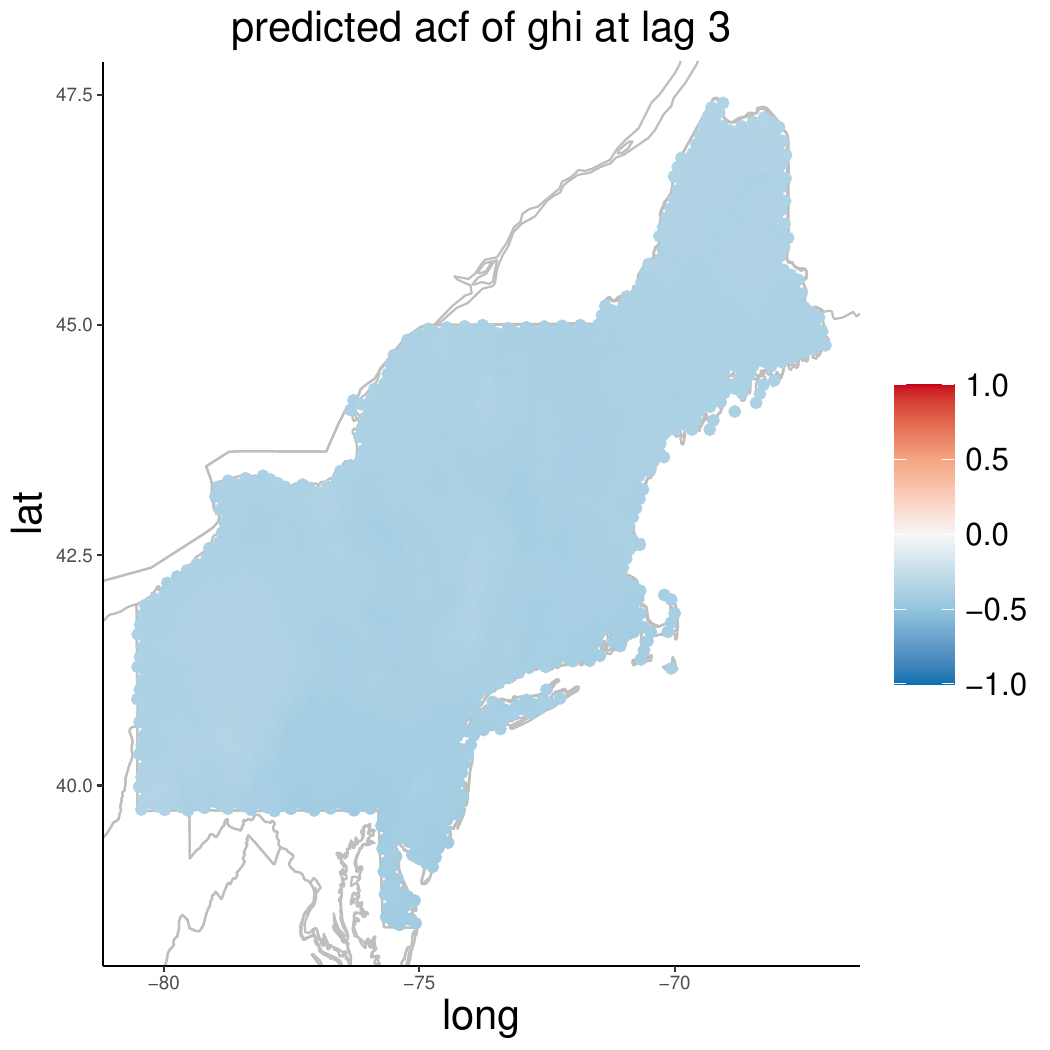}
    
    \includegraphics[scale=.28]{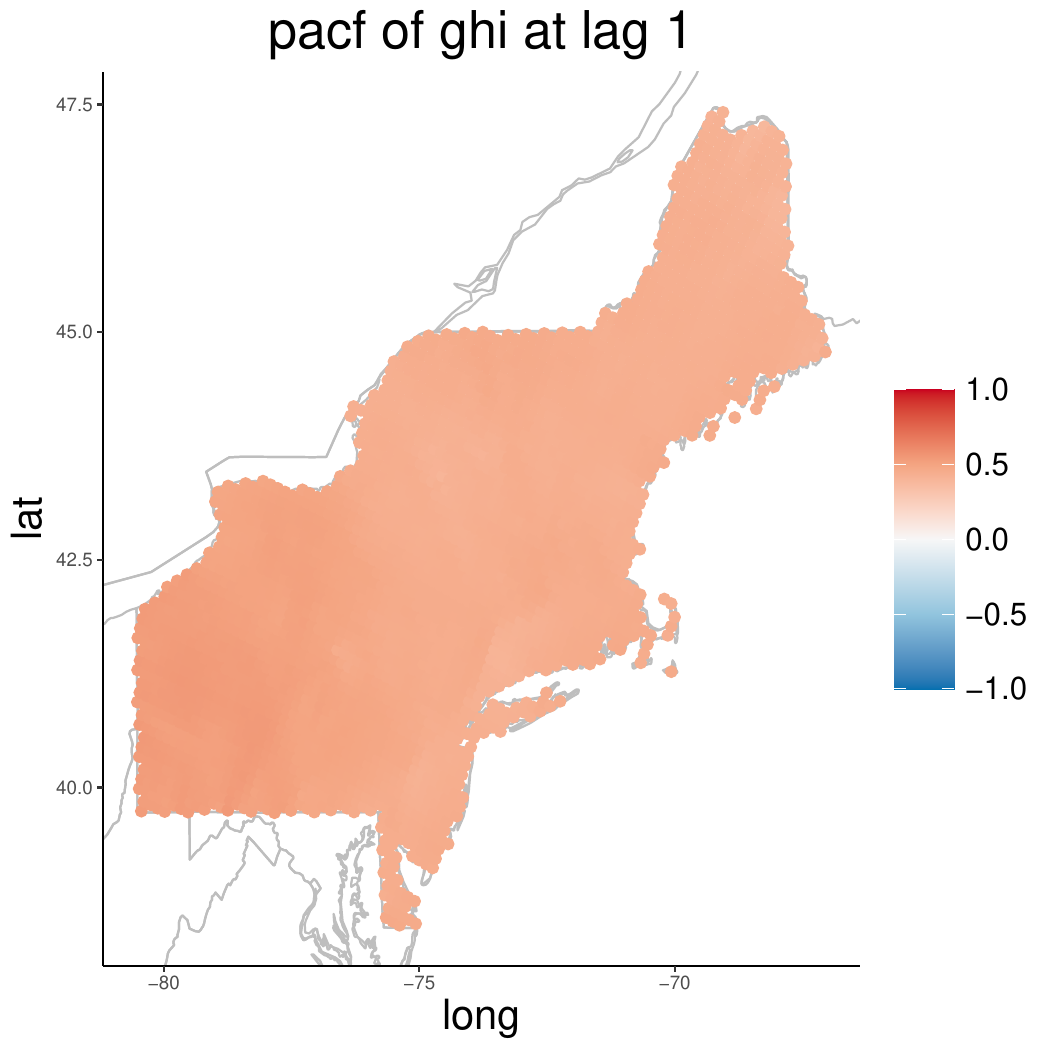}
    \includegraphics[scale=.28]{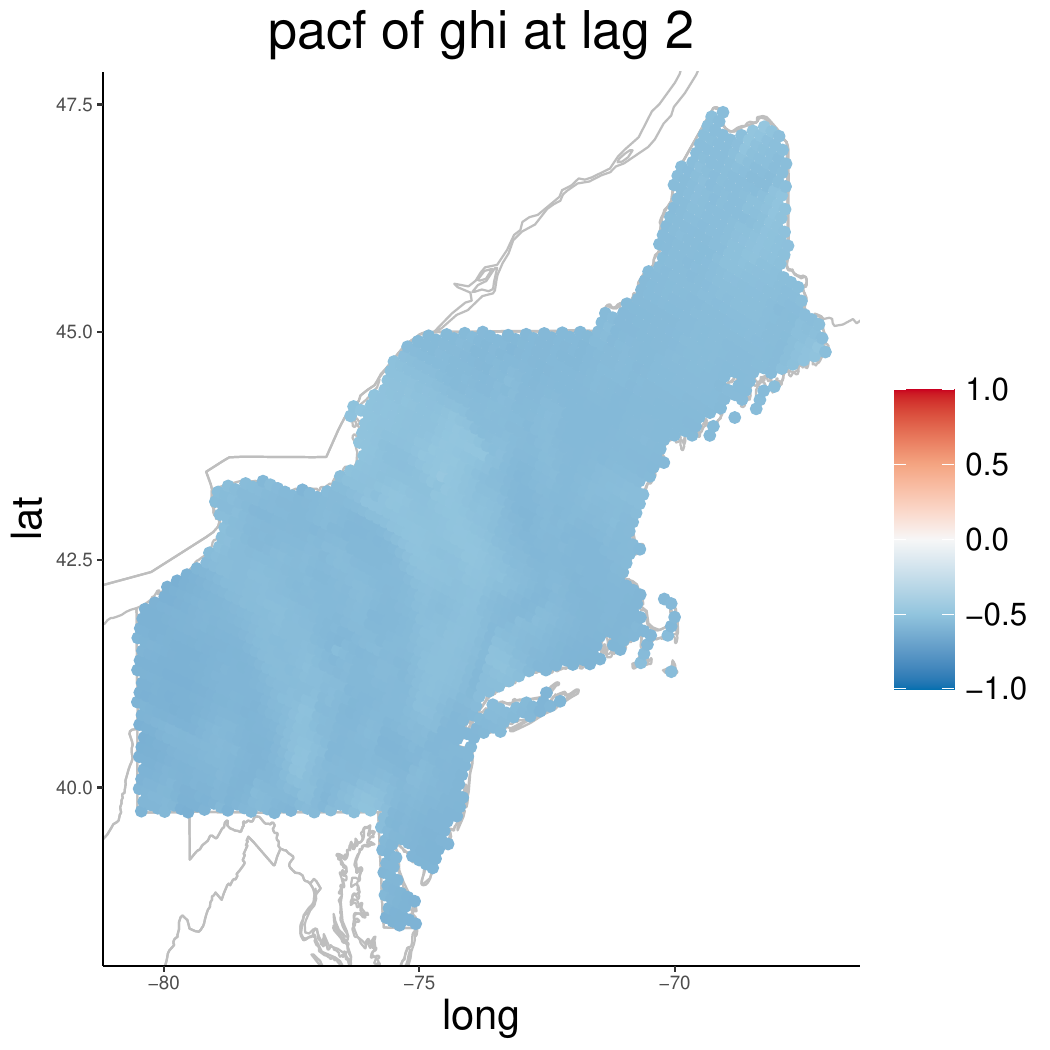}
    \includegraphics[scale=.28]{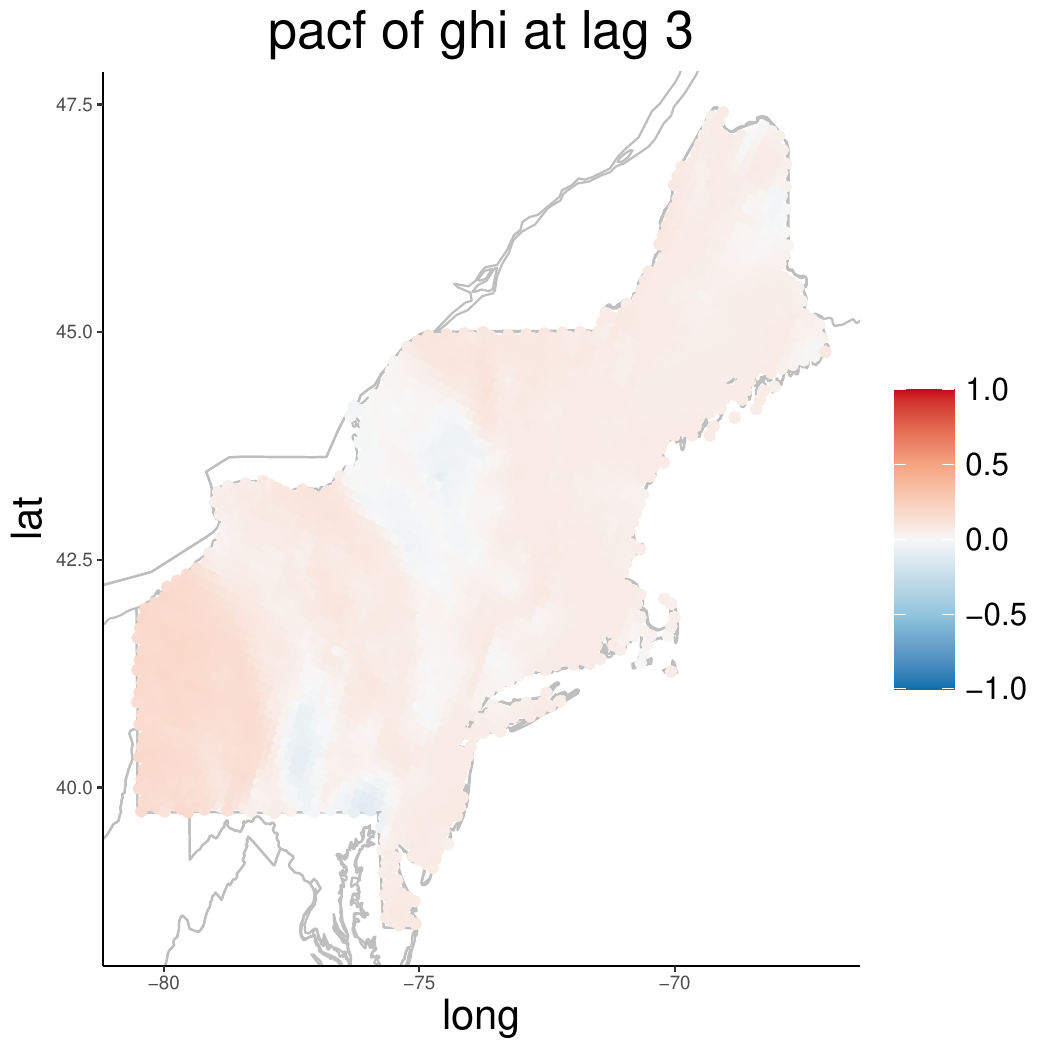}

    \includegraphics[scale=.28]{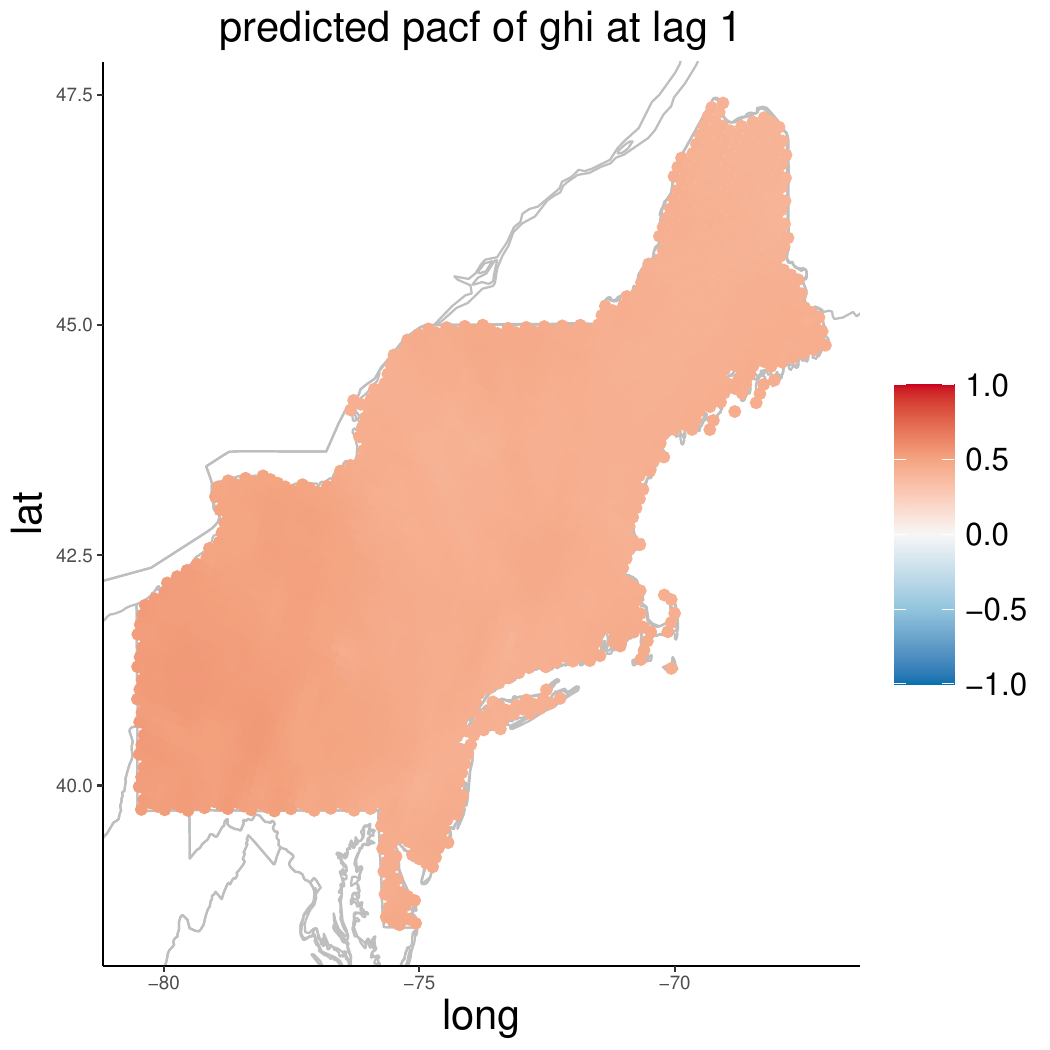}
    \includegraphics[scale=.28]{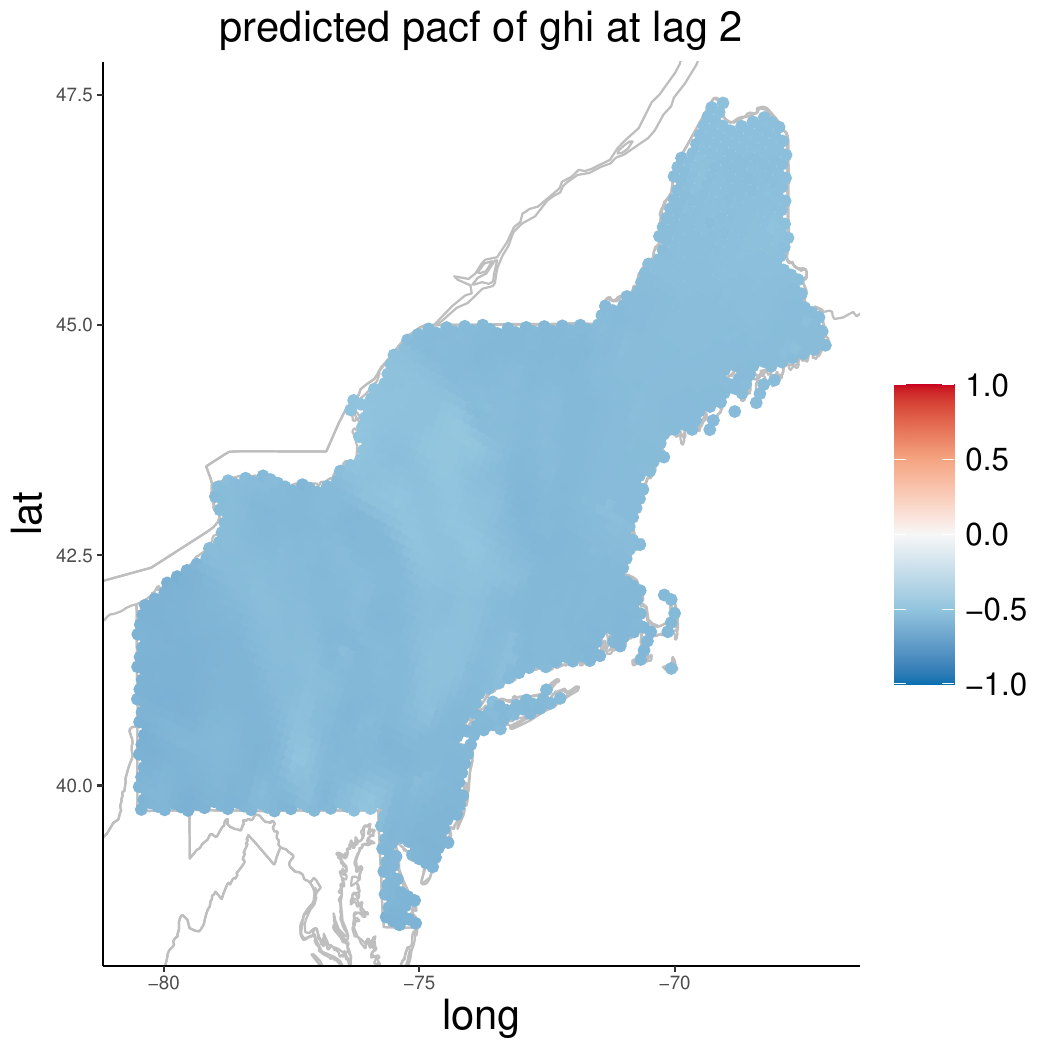}
    \includegraphics[scale=.28]{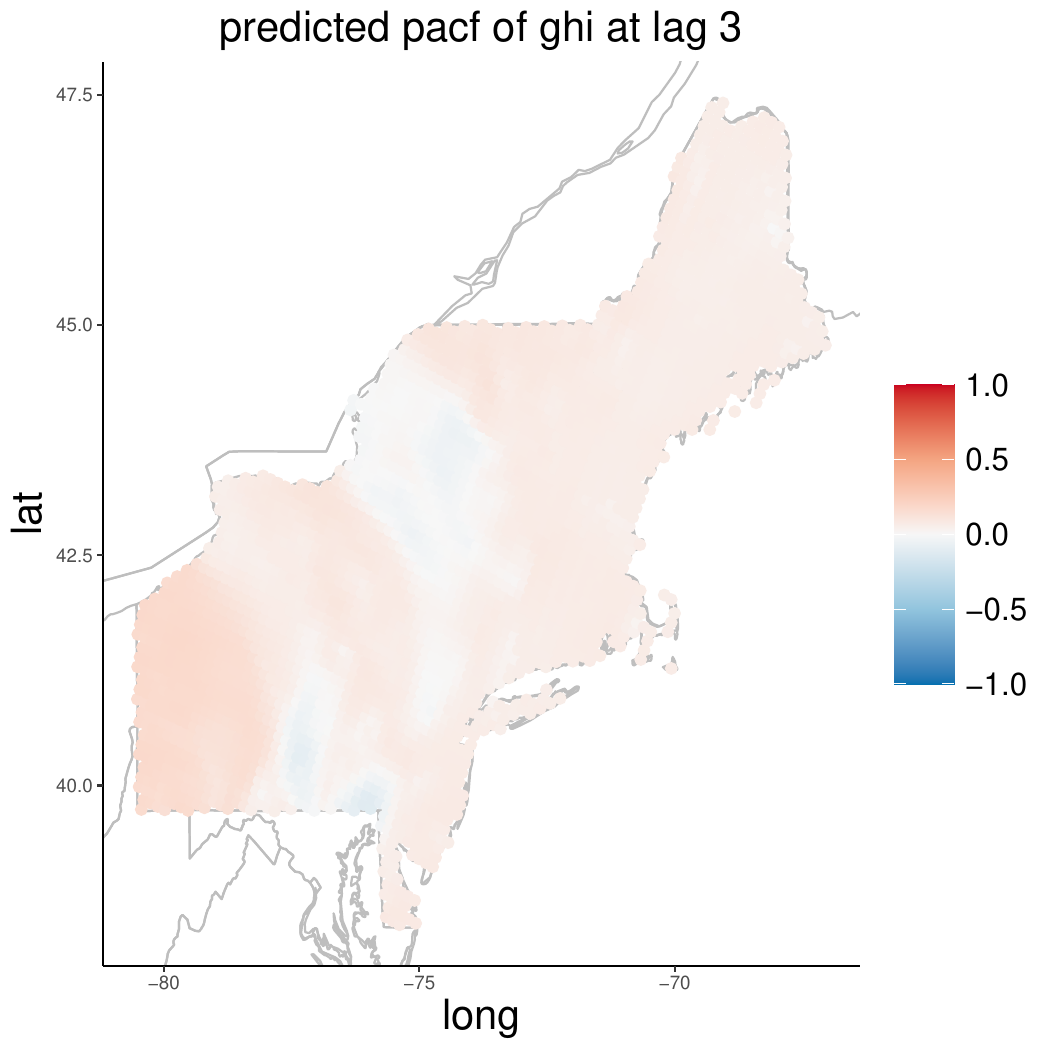}
    
    \caption{Same configuration as Figure \ref{fig:acfpacfwind}, but for global horizontal irradiance.}
    \label{fig:acfpacfghi}
\end{figure}

\begin{figure}[h!]
    \centering
    \includegraphics[scale=.28]{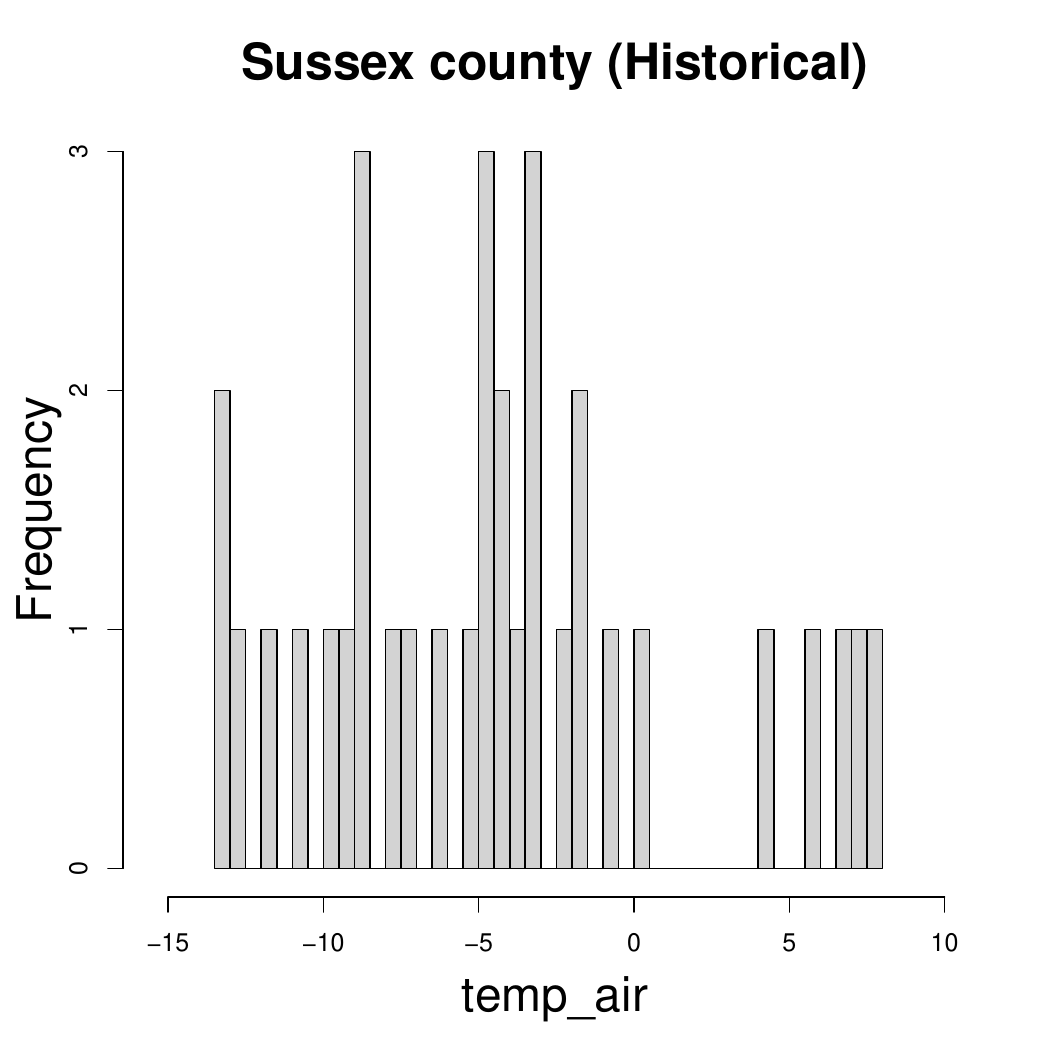}
    \includegraphics[scale=.28]{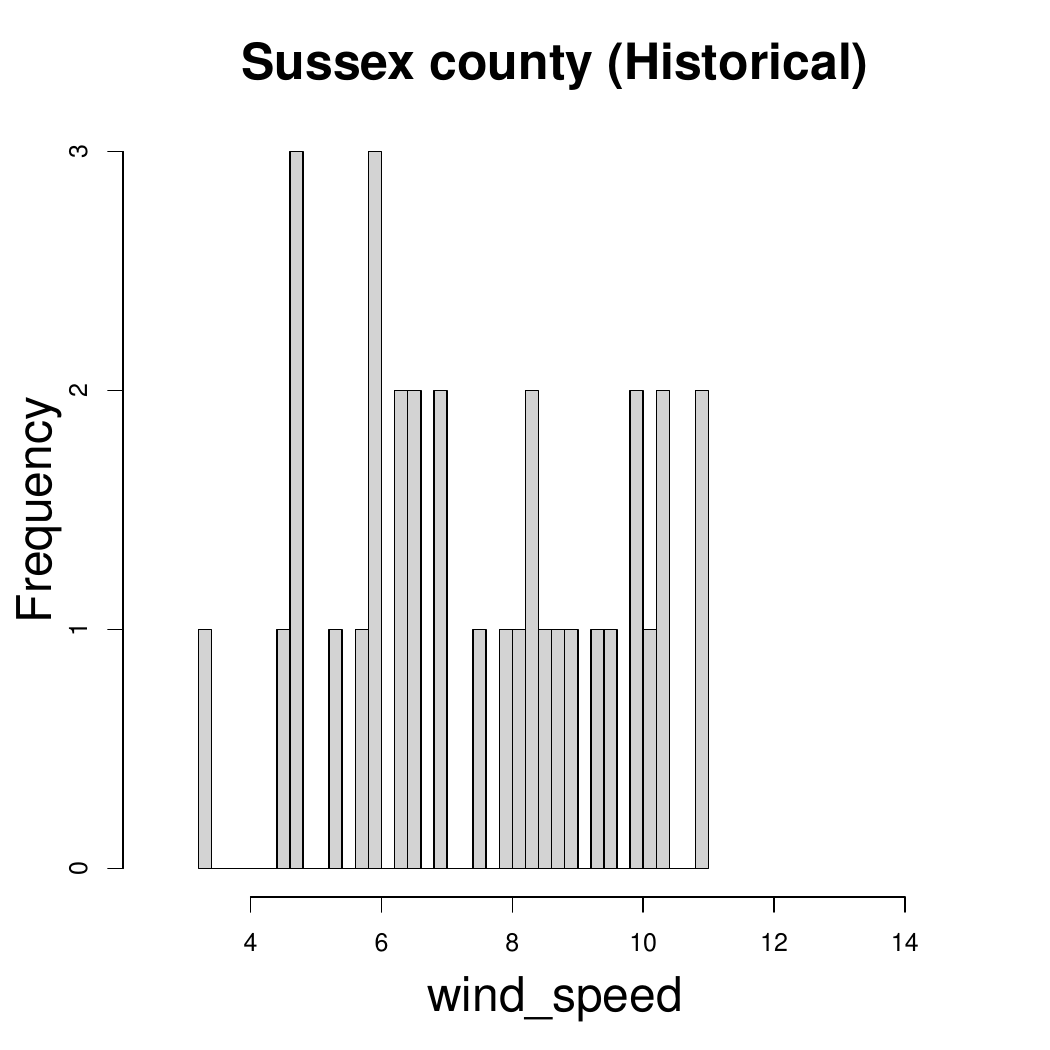}
    \includegraphics[scale=.28]{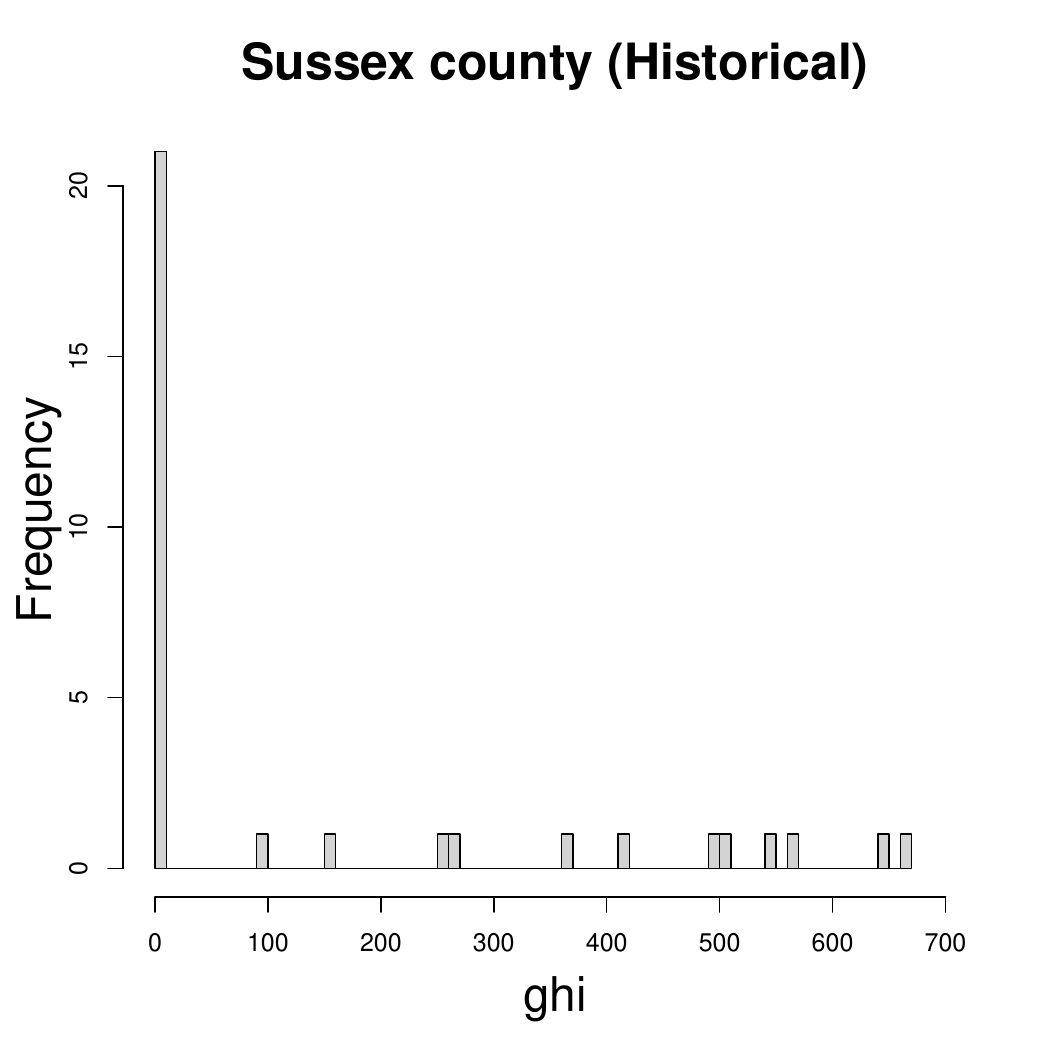}

    \includegraphics[scale=.28]{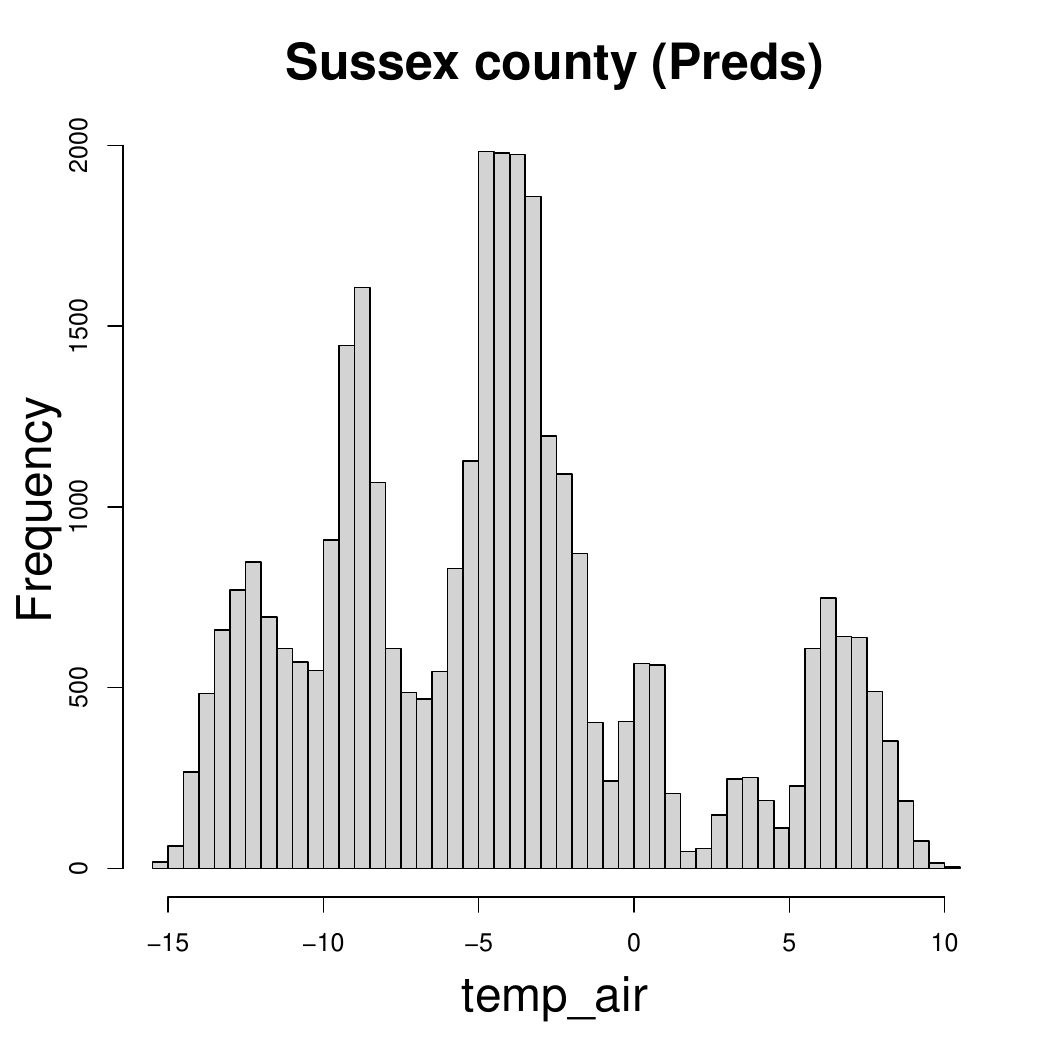}
    \includegraphics[scale=.28]{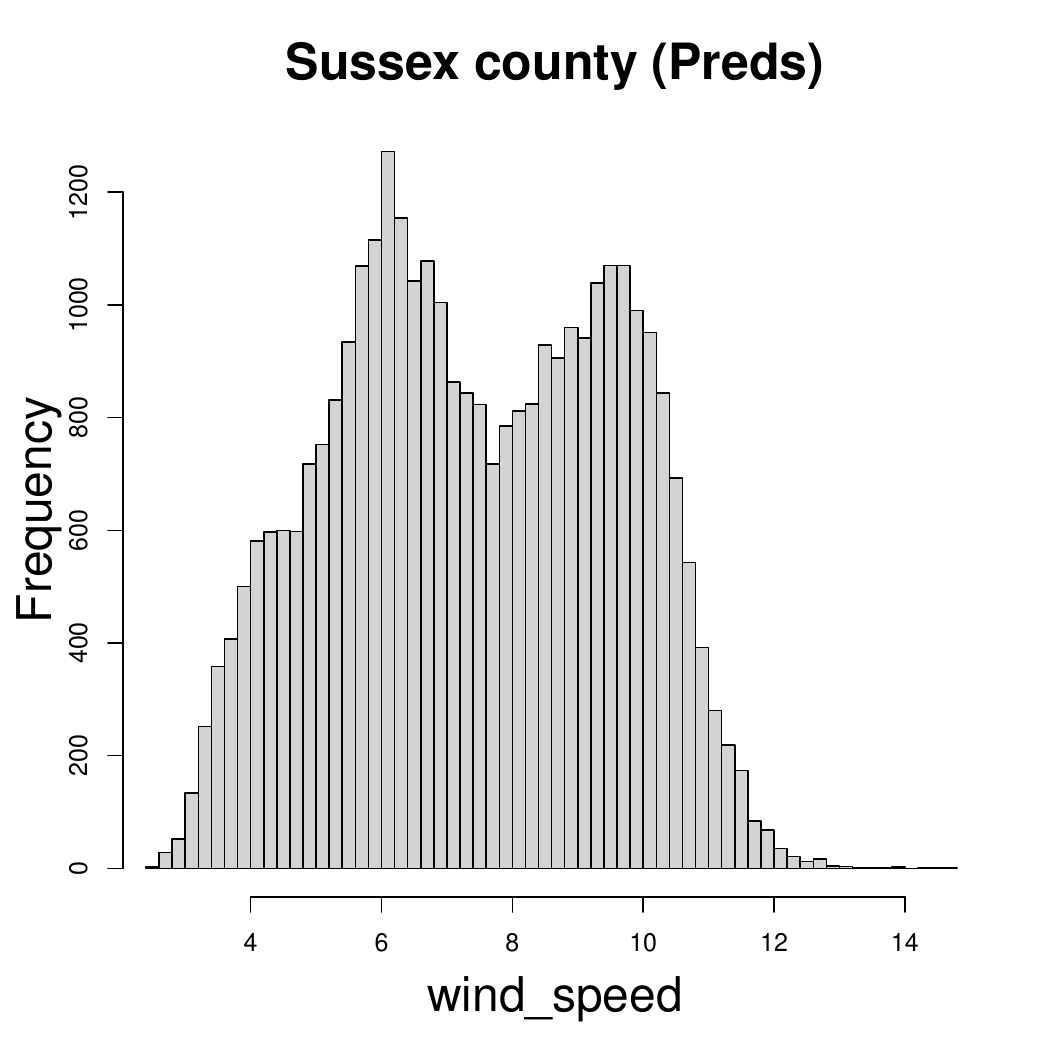}
    \includegraphics[scale=.28]{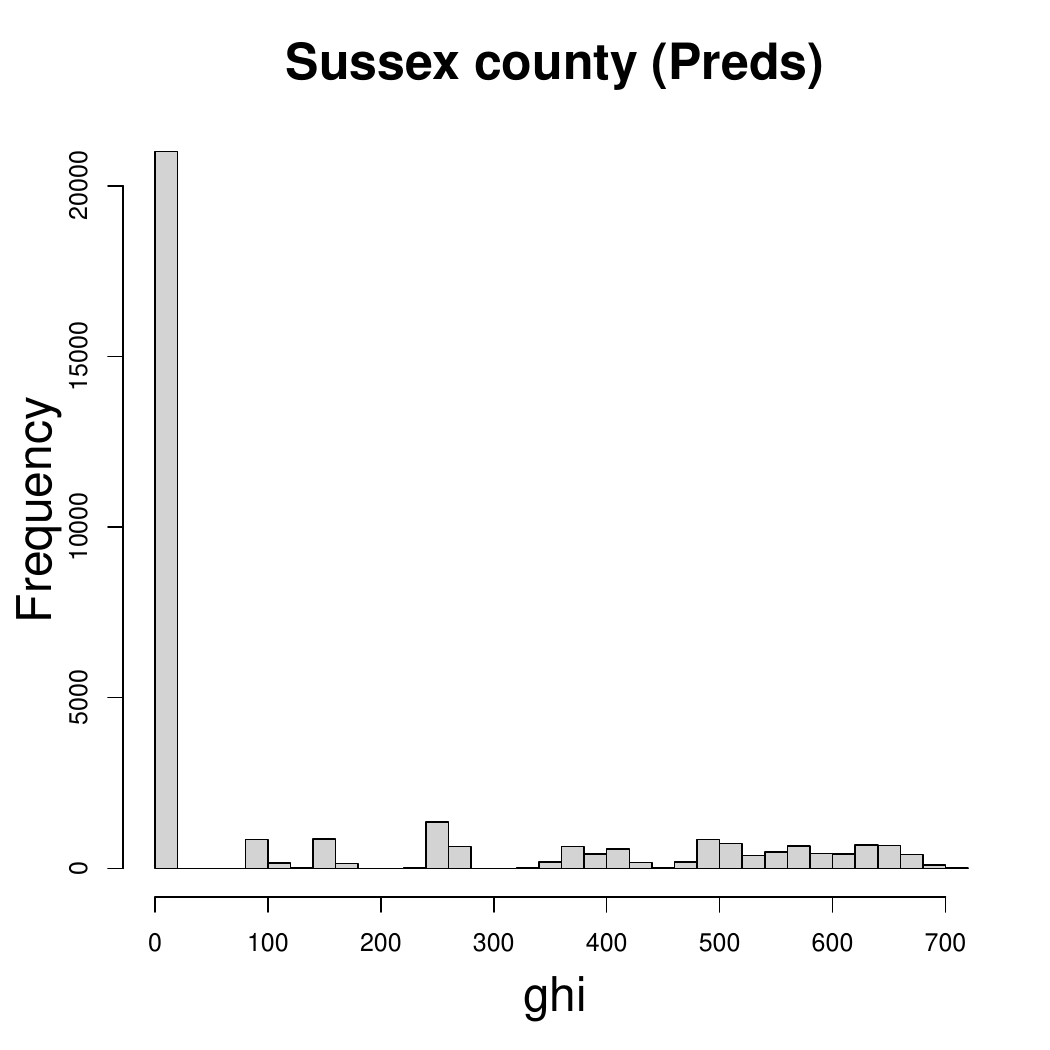}

    \includegraphics[scale=.28]{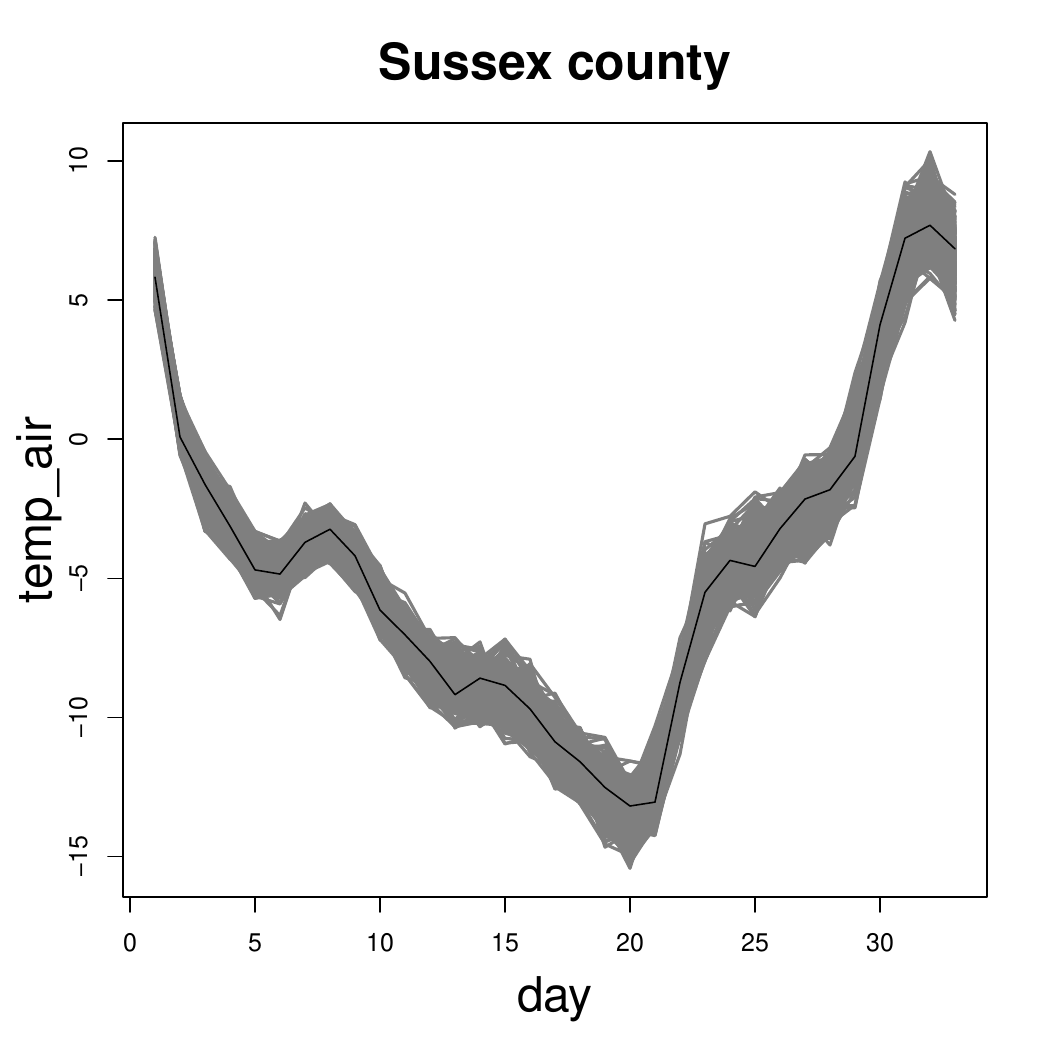}
    \includegraphics[scale=.28]{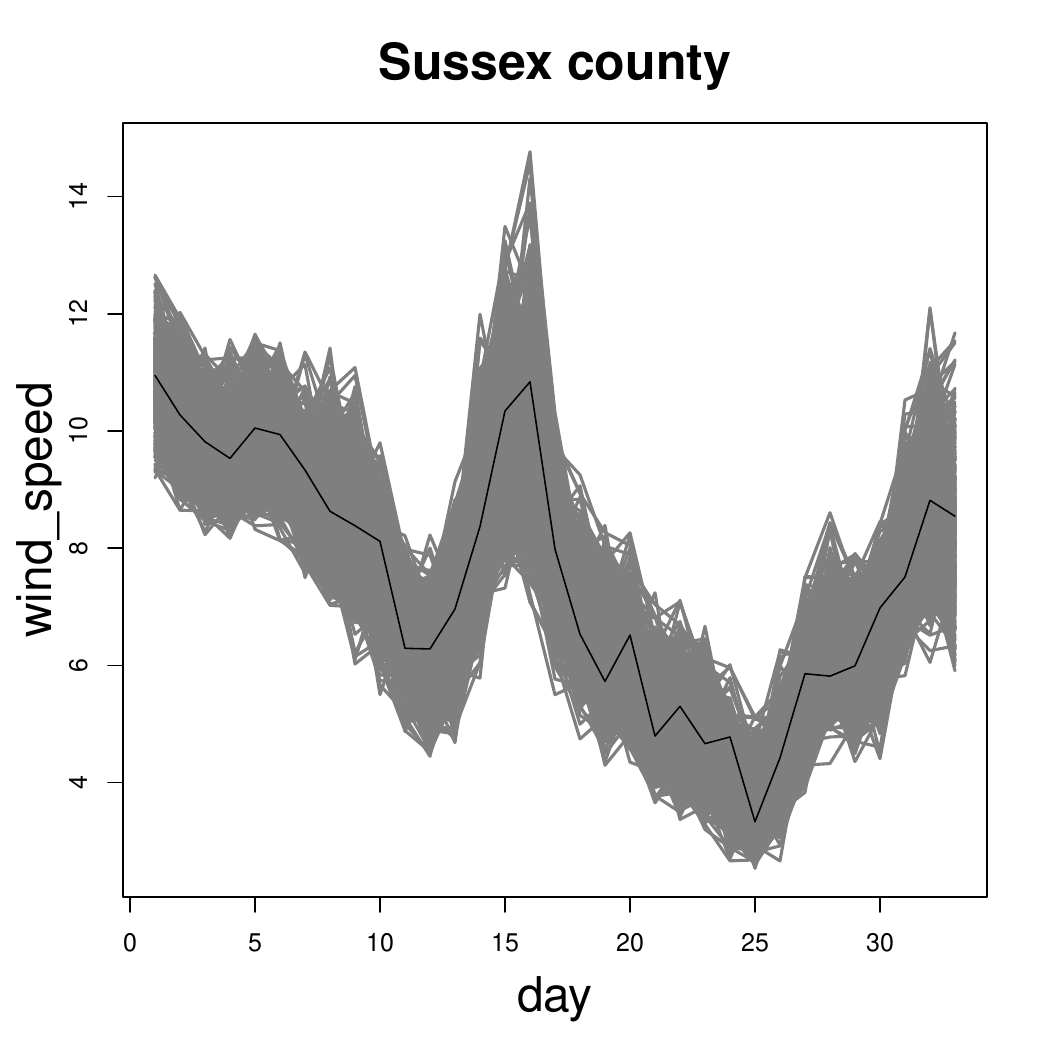}
    \includegraphics[scale=.28]{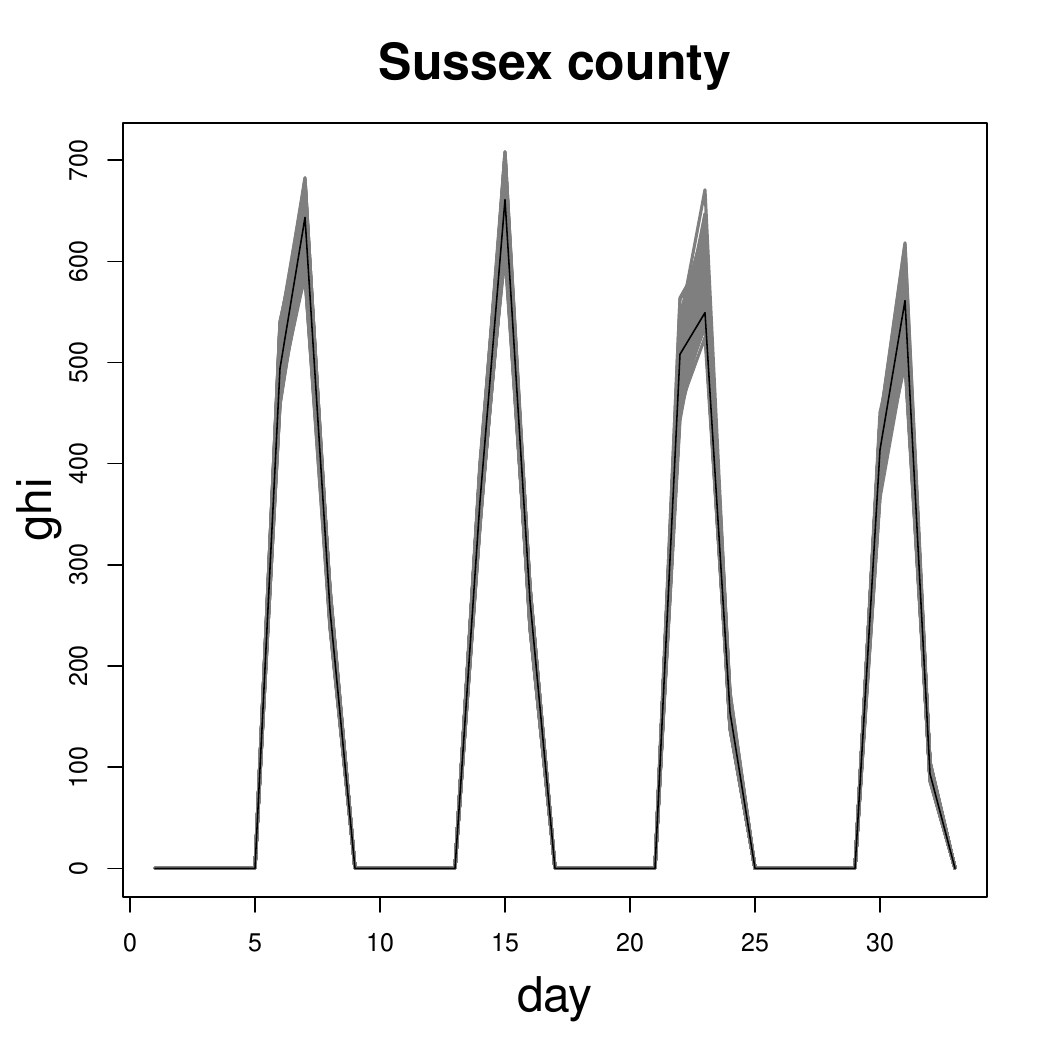}

    \caption{Assessing model performance at a single location. First and second rows show the historical and simulated histograms, respectively.
    The third row shows the historical time series as the black line and simulated time series in lighter gray.}
    \label{fig:locationwisecomparison}
\end{figure}

\begin{figure}[h!]
    \centering
    \includegraphics[scale=.38]{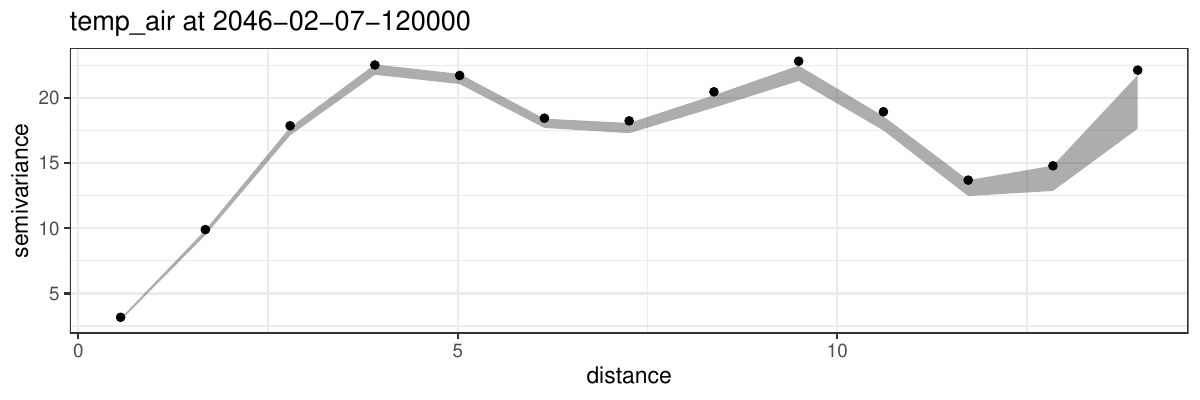} 
    \includegraphics[scale=.38]{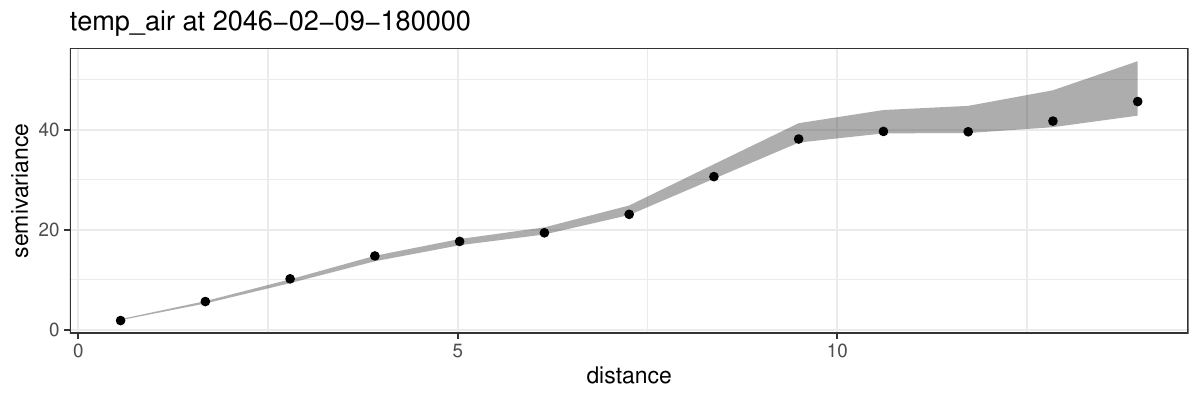} \\
    
    \includegraphics[scale=.38]{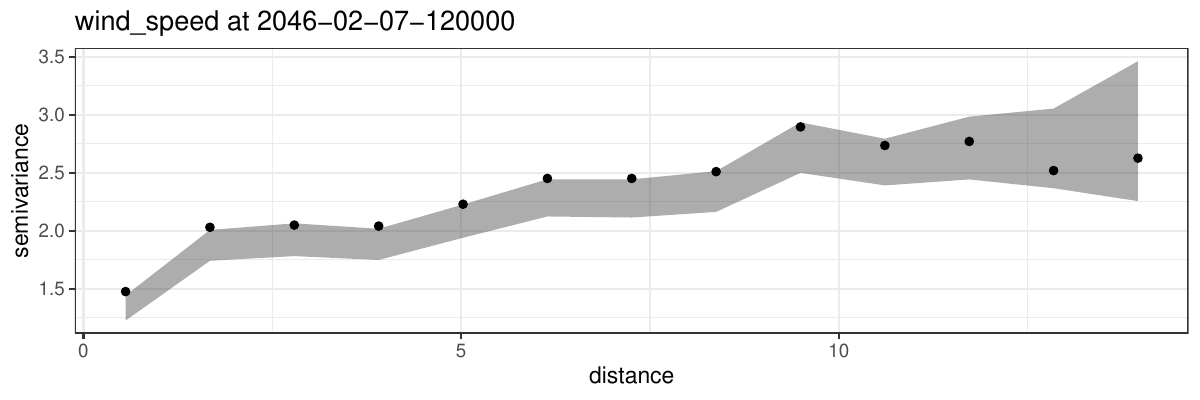} 
    \includegraphics[scale=.38]{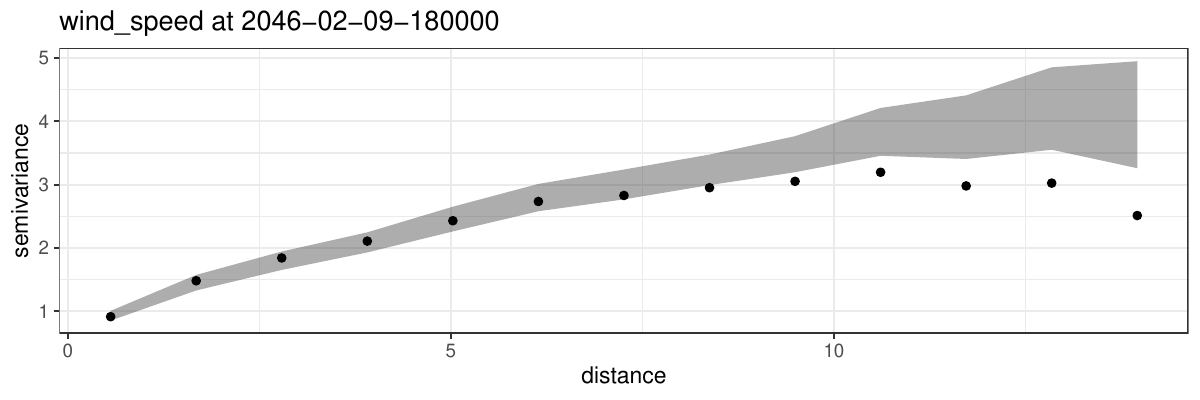} \\
    
    \includegraphics[scale=.38]{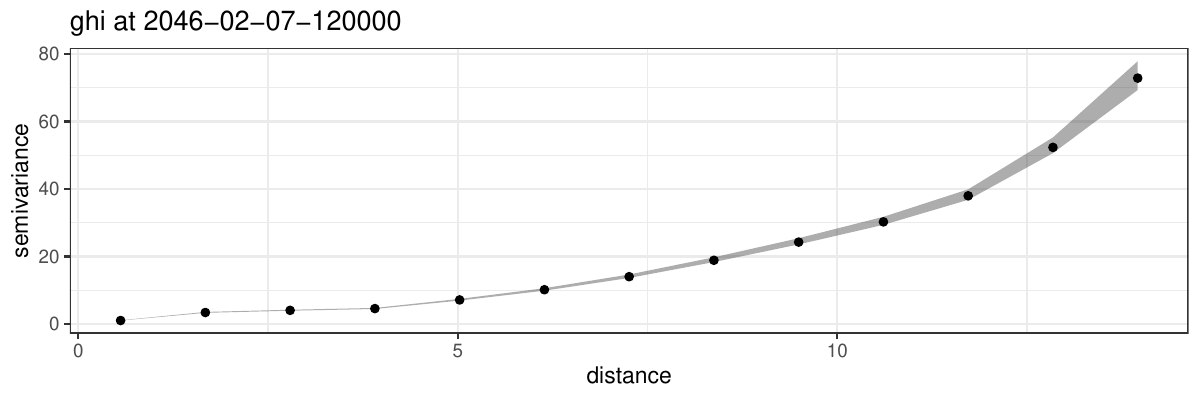} 
    \includegraphics[scale=.38]{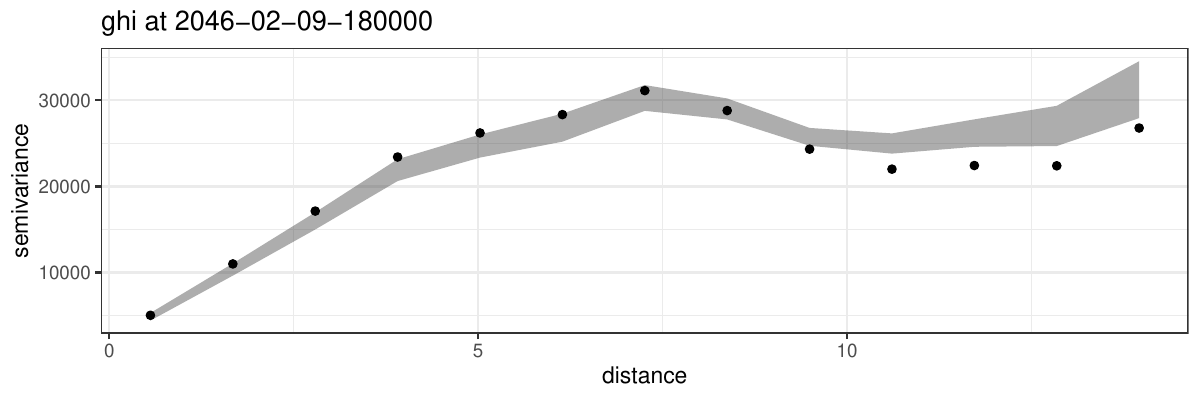} 

    \caption{Variogram plots for the three variables at the same dates as Figures \ref{fig:date1plot} and \ref{fig:date2plot}.
    Historical values are shown as black dots, while the gray lines correspond to BMW-GAM simulations.}
    \label{fig:variogram}
\end{figure}

For validation, \cite{economou2022} suggest visual checks to assess model quality.
Figures \ref{fig:date1plot} and \ref{fig:date2plot} correspond to two distinct times during the storm.
For each variable (column), posterior means (second row) and historical observations (first row) are hard to distinguish at a glance.
Posterior variances (third row) are reasonable, except in Figure \ref{fig:date2plot}, values of global horizontal irradiance around New York City and to the north appear quite large.
Autocorrelation and partial autocorrelation functions (acf and pacf) measure temporal dependence as a function of lags; a common usage is identifying the order of moving average and autoregressive time series models, respectively.
Figures \ref{fig:acfpacftemp}, \ref{fig:acfpacfwind}, and \ref{fig:acfpacfghi} display acf and pacf for historical versus predicted values, up to the first three lags.
Once again, there is strong consensus between observed and simulated values, and the same is true for higher order lags up to 16 (not shown).
The most noticeable discrepancy appears at the second lag for pacf of wind speed; simulations fail to take on positive values seen in the historical plot.
Figure \ref{fig:locationwisecomparison} shows several validation plots at a single location in Sussex County, Delaware.
The flexibility of BMW-GAM is apparent, as histograms match their historical counterparts, although one concern may be overfitting in the case of air temperature, since the simulated distribution appears highly multimodal.
In the last row, we show the observed time series versus simulated values to provide a measure of uncertainty.
Similarly, Figure \ref{fig:variogram} displays variograms for historical and simulated cases, indicating agreement in spatial correlation, except for minor deviations at large distances during the later date.

\section{Discussion and Conclusions}
\label{sec:discussion}
BMW-GAM \citep{economou2022} is enriched with a separable multivariate spatio-temporal correlation function.
The resulting Kronecker product structure of the Gaussian copula correlation matrix scales favorably to large data sets.
Moreover, marginal inference is embarrassingly parallel across different windows.
Enforcing the positive mean of select response variables with square-root link function is crucial.
In summary, BMW-GAM is robust to the data size and invariant to the practitioner's definition of extreme event,
and rapidly generates multivariate spatio-temporal fields for comprehensive model checking and uncertainty quantification.

Incorporating other global climate model simulations (ADDA or otherwise) introduces challenges beyond the scope of this paper \citep{knutti2010}.
We are aware of a few relevant papers, for example, \cite{subbian2013} combine multiple global climate models via regression with a Laplacian penalty ensuring spatial smoothness. 
\cite{singh2021} estimate joint return periods of warm-wet and warm-dry scenarios in Canada.
At each location, temperature and precipitation are modeled with bivariate copula whose parameter is inferred under a Bayesian framework, pooling data across 3 climate ensembles.
\cite{VAVRUS201510} study Northeast U.S.\ with 13 climate models to project changes in extreme heat, cold, and precipitation by the mid-21st century under various forcing scenarios.
They measure classical statistical metrics such as skewness and inter-model standard deviation, and derive uncertainty bounds from bootstrapping.

When interlinked with A-LEAF and NGTransient, BMW-GAM scenarios will be leveraged to assess resilience of critical energy systems during extreme weather threats.
BMW-GAM is a practical choice for non-Gaussian multivariate spatio-temporal Bayesian data analysis, and we anticipate its application in many areas beyond energy systems.
Selecting between different BMW-GAMs is difficult due to the large number of parameters and hyperparameters.
Besides performing automated model selection and including multiple climate models in BMW-GAM, we plan to investigate other climate variables and spatial domains.
Covariance separability is vital to BMW-GAM's computational tractability but not always a realistic assumption.

Ignoring the zeros of global horizontal irradiance is a weakness of our current approach.
Initially, we tried a BMW-GAM with Tweedie distribution and log link function to accommodate zeros; \cite{economou2022} modeled precipitation in this manner.
However, our simulated spatial fields and posterior variances were unrealistic.
Incorporating uncertainty about zeros of irradiance with specialized methodology \citep{Holtzhausen2024,MULLER2023869,euan2022,zhang2022,ZHANG2018370,zhang2019,upton2025subhourlyspatiotemporalstatisticalmodel,halder2021,dahl2019} is left for future research.

\clearpage

\section*{Acknowledgements}
The authors are especially grateful to Michael Stein for suggesting, at the 2025 Contemporary Advances in Statistics of Extremes workshop hosted by the University of Missouri, that we forego modeling the zero values of global horizontal irradiance.
High-performance computing resources were provided on the clusters Bebop and Improv operated by the Laboratory Computing Resource Center at Argonne National Laboratory.
Argonne National Laboratory's work was supported by the U.S. Department of Energy, Office of Science, Laboratory Directed Research and Development, Argonne National Laboratory, under contract DE-AC02-06CH11357.
The submitted manuscript has been created by UChicago Argonne, LLC, Operator of Argonne National Laboratory (“Argonne”). Argonne, a U.S. Department of Energy (DOE) Office of Science laboratory, is operated under Contract No. DE-AC02-06CH11357. 
The U.S. Government retains for itself, and others acting on its behalf, a paid-up nonexclusive, irrevocable worldwide license in said article to reproduce, prepare derivative works, distribute copies to the public, and perform publicly and display publicly, by or on behalf of the Government. 
The Department of Energy will provide public access to these results of federally sponsored research in accordance with the DOE Public Access Plan \url{http://energy.gov/downloads/doe-public-access-plan}.
This report was prepared as an account of work sponsored by an agency of the United States Government. Neither the United States Government nor any agency thereof, nor UChicago Argonne, LLC, nor any of their employees or officers, makes any warranty, express or implied, or assumes any legal liability or responsibility for the accuracy, completeness, or usefulness of any information, apparatus, product, or process disclosed, or represents that its use would not infringe privately owned rights. Reference herein to any specific commercial product, process, or service by trade name, trademark, manufacturer, or otherwise, does not necessarily constitute or imply its endorsement, recommendation, or favoring by the United States Government or any agency thereof. The views and opinions of document authors expressed herein do not necessarily state or reflect those of the United States Government or any agency thereof, Argonne National Laboratory, or UChicago Argonne, LLC.

\bibliography{LDRD2025.bib}

\end{document}